**Title:**

# Universal scaling of the critical temperature for thin films near the superconducting-to-insulating transition

**Authors:**

Yachin Ivry,[1*] Chung-Soo Kim,[1] Andrew E. Dane,[1] Domenico De Fazio,[1,2] Adam McCaughan,[1] Kristen A. Sunter,[1] Qingyuan Zhao[1] and Karl K. Berggren.[1*]

**Affiliations:**

[1] Research Laboratory of Electronics, Massachusetts Institute of Technology, 77 Massachusetts Avenue, Cambridge, Massachusetts 02139, USA.

[2] Department of Electronics and Telecommunications (DET), Polytechnic University of Turin, Corso Duca degli Abruzzi 24, 10129 Torino, Italy.

* Correspondence to: berggren@mit.edu, ivry@mit.edu.

Abstract:

**Thin superconducting films form a unique platform for geometrically-confined, strongly-interacting electrons. They allow an inherent competition between disorder and superconductivity, which in turn enables the intriguing superconducting-to-insulator transition and believed to facilitate the comprehension of high-$T_c$ superconductivity. Furthermore, understanding thin film superconductivity is technologically essential *e.g.* for photo-detectors, and quantum-computers. Consequently, the absence of an established universal relationships between critical temperature ($T_c$), film thickness ($d$) and sheet resistance ($R_s$) hinders both our understanding of the onset of the superconductivity and the development of miniaturised superconducting devices. We report that in thin films, superconductivity scales as $d \cdot T_c(R_s)$. We demonstrated this scaling by analysing the data published over the past 46 years for different materials (and facilitated this database for further analysis). Moreover, we experimentally confirmed the discovered scaling for NbN films, quantified it with a power law, explored its possible origin and demonstrated its usefulness for superconducting film-based devices.**

Relationships between low-temperature and normal-state properties are crucial for understanding superconductivity. For instance, the Bardeen-Cooper-Schrieffer theory (BCS) successfully associates the normal-to-superconducting transition temperature, $T_c$, with material parameters, such as the Debye temperature ($\Theta_D$) and the density of states at the Fermi level ($N(0)$). Hence, the BCS model allows us to infer superconducting characteristics (*i.e.* $T_c$) from properties measured at higher temperatures [1]. In the BCS framework, superconductivity occurs when attractive phonon-mediated electron-electron interactions overcome the Coulomb repulsion, giving rise to paired electrons (Cooper pairs) with a binding energy gap: $\Delta$. Moreover, within a superconductor, all Cooper pairs are coupled, giving rise to a collective electron interaction. Such a collective state is described by a complex global order parameter with real amplitude ($\Delta$) and phase ($\phi$): $\Psi = \Delta e^{i\phi}$.

Since superconductivity relies on a collective electron behaviour, the onset of superconductivity occurs when the number of participating electrons is just enough to be considered collective, .*i.e.* at the nanoscale [2–5]**.** Thus, it is known that the superconductivity-disorder interplay varies in thin films and is effectively tuned with the film thickness (*d*) or with the disorder in the system, which is represented by sheet resistance of the film at the normal state ($R_s$) [6–10]**.** The mechanism of superconductivity in thin films has been investigated since the 1930s [11]. The development of thin film growth methods in the late 1960s allowed Cohen and Abeles to demonstrate an increase in $T_c$ with decreasing thickness in aluminium films in a study that pioneered the currently ongoing research of thin superconducting films [12]. This enhancement of $T_c$, which is still not completely understood, was later confirmed by Strongin *et al*. [13], who reported also the more common behaviour of $T_c$–its suppression with reduced film thickness. Strongin *et al*. empirically examined different scaling options for the observed suppression of $T_c$ in lead and suggested that $T_c$ scales with $R_s$ better than it does with the other parameters, such as the film thickness. This suggestion, is still influential on the data analysis done in the field today and it encouraged the derivation of theoretical models to explain a dependence of $T_c$ on $R_s$. Indeed, Beasley *et al*. (followed by Halperin and Nelson) derived that $T_c$ depends only on $R_s$ for a Berezinsky-Kosterlitz-Thouless (BKT) transition, in which vortex-antivortex pairs, and not Cooper pairs, dominate the transition, which in turn is universal in nature [8,14]. In addition, the mathematical derivation of the interacting-boson treatment within the BKT framework was also used in a reminiscent framework that associates interacting paddles of Cooper pairs with a multiple-Josephson Junction array [15]. Likewise, Finkel'stein used renormalisation group tools to derive exactly a different expression for the dependence of $T_c$ on $R_s$ (with no direct dependence on the thickness). This derivation was based on a modified BCS equation, in which mean field theory was integrated with homogeneous disorder *i.e.* impurity scattering due to Coulomb and spin density interactions [9]. As opposed to these three models that claim that $T_c$ depends merely on $R_s$, competing models, such as the proximity effect [7] and the quantum size effect [10] theories suggest that $T_c$ depends on $d$ only, with no direct dependence on $R_s$. Nevertheless, none of these models is sufficient to explain the entirety of the accumulated experimental data [12,13,16–29] despite the long-standing attempt to do so either through a direct mathematical derivation as in the above model, or with the aid of empirical universal laws [30].

Relationships between $d$, $R_s$ and $T_c$ are significant even to a broader scope than thin superconducting films. That is, the dependencies $T_c(d)$ and $T_c(R_s)$ are important for superconducting films. However, the dependence of resistivity ($\rho \equiv d \cdot R_s$) on film thickness in thin metallic films has also occupied both scientists and technologists for many decades. Since above $T_c$ the superconducting films behave like normal metals, the relationship $R_s(d)$ is similar to that of normal metals. That is, presumably, the resistivity is expected to remain constant or to demonstrate a smooth minor monotonic increase with reduced film thickness. Therefore, it is not important if $T_c$ is expressed as a function of $d$ or of $R_s$, as presumably, one parameter can be replaced by the other straightforwardly. Theoretically, the relationship $\rho(d)$ is usually discussed in terms of derivatives of the Fuchs's theory [31], sometimes combined with Matthiessen's rule [32]. However, surprisingly, to-date, the existing theories encounter difficulties in fitting the experimental data which are often scattered when plotted on a $\rho(d)$ graph [33]. In addition to challenging our understanding of metallic thin films, such scatter prevents a smooth quantitatively valuable transition between descriptions of $T_c(R_s)$ and $T_c(d)$ in the case of superconductors.

A seminal experimental work by Goldman and co-authors [21] suggested that beyond certain film thickness or sheet resistance values $T_c$ is suppressed so much that practically, the material will never become superconducting. That is, the cooling curves of such thin films indicate that $R_s$ increases with decreasing temperature—a behaviour that is typical in insulators and not in metals. This observation began the race to understanding the superconducting-to-insulator phase transition, which is believed by many researchers to be of a quantum nature [5]. To date, although much data for thin film superconductivity have been accumulated [12,13,16–29] and the local disorder has already been observed directly [34], the mechanisms governing the collective behaviour close to the superconducting-to-insulating transition, or near the onset of superconductivity have remained elusive. That is, a model equivalent to the BCS but that is valid for thin films, is still missing. Specifically, the theories that suggest that $T_c$ varies with either $R_s$ or $d$ are material dependent, while for some materials, none of the existing theories agrees with the observations. The absence of a unified description of superconductivity in thin films is even more pronounced when bearing in mind that the onset of superconductivity in such geometries is believed to occur through a quantum phase transition, which is in principle universal. Moreover, understanding superconductivity in thin films is expected to clarify the behaviour of resistance in thin metallic films in general. Likewise, it has even been suggested that the superconductivity-disorder interplay in thin films is the key for understanding high-$T_c$ superconductivity [4]. Therefore, it is the goal of this paper to demonstrate a universal behaviour for $T_c$ in thin films as a function of both $d$ and $R_s$.

In addition to the scientific impact associated with understanding superconductivity in thin films, thin superconducting films are of a great technological significance as they are the basis for most miniaturised superconducting devices [35,36]**.** In particular, quantum-based technologies, such as computation, encryption and communication rely on such films. Similarly, the leading technology for sensing single photons fast [37] and at a broad spectral range [38]--superconducting nanowire single photon detectors (SNSPDs)--is also based on thin superconducting films [39]. Nevertheless, the lack of understanding of the underlying mechanism of superconductivity in thin films and the large scatter of the experimental data for the relationships between $T_c$, $R_s$ and $d$ typically lead to low confidence in the film growth process, encumbering the relevant technological developments. Specifically, the limited reproducibility and control of the film parameters impair both the yield and the size of devices made out of such films. For instance, the yield of SNSPDs made out of thin niobium nitride (NbN) films is low, while their active area is usually restricted, hindering the technological advances in the field. Hence, a universal scaling of the properties of thin superconducting films is expected to improve the control and reproducibility of the film properties, and therefore, to allow at last realisation of the potential of miniaturised superconducting devices.

We show that for a given material, the relationship between film thickness, sheet resistance and critical temperature scales as $d{\cdot}T_c(R_s)$. Moreover, this scaling typically follows a power law. We demonstrated the scaling on data gathered from some thirty different sets of materials published since 1968, which cover most of the literature. The materials studied included clean, dirty, granular, and amorphous superconductors. Some of these materials are type I in their bulk state, and some are type II. Most of these materials exhibited suppression of $T_c$ at reduced thicknesses, but some exhibited enhancement. The data in its entirety could not fit previous theories [7–10], but did fit the new power-law relationship across broad range of $T_c$, $d$ and $R_s$. We extracted the coefficient and exponent of the power law for each material, and demonstrated that the coefficient and the exponent are correlated. The power law fits the data from materials that fit also one of the previous models $T_c(d)$ or $T_c(R_s)$, as well as for materials that presumably are not BCS. We also examined our own new experimental data on NbN thin films. In these data, relating $d{\cdot}T_c$ to $R_s$ provided fits with reduced scatter relative to fits suggested by previous models [7–10,31–33]. Finally, we supply two possible explanations of the observed universal behaviour. We should note that the data gathered from the literature is available for further review in the Supplemental Material [40].

To illustrate the new scaling, we will start by examining our data on NbN films. We chose sputtered NbN as the material for this study for four main reasons: (a) it is widely researched, and experimental data collected for different growth methods and conditions are available; (b) there are contradicting reports about which of the existing models describes the $T_c$ suppression in NbN films. For instance, Kang *et al*. claimed that $T_c$ is suppressed due the quantum size effect [19], Wang and co-authors suggested that the suppression follows Finkel'stein's model [18], Semenov *et al*. determined that the suppression is governed by the proximity effect [20] and Koushik *et al.* argued that the transition is of a BKT type [41]; (c) the relatively high $T_c$ of NbN (16 K for a bulk NbN [42]) assists the experimental investigation; and (d) its properties make it useful for photodetectors [35,37,43,44].

Figures 1a and 1b show the dependence of $T_c$ on thickness and on sheet resistance for our NbN films, allowing a comparison of the data with the existing models [7–10,31–33]. Although a general trend can be seen in both $T_c(d)$ and $T_c(R_s)$, the scatter in these graphs is too large to allow confident fitting to any model of the form $T_c(d)$ or $T_c(R_s)$. Bearing in mind the metallic characteristic of the films, the resistivity of the grown films corresponds to their inverse mean free path and hence should increase monotonically with decreasing thickness [31]. However, Fig. 1c shows the dependence of resistivity on thickness in our films, revealing again, large scatter of the data points with only vaguely the expected trend (as a side remark we should note that the regime in which the scatter is the largest is when the thickness is around 6 nm, which is the nominal coherence length of NbN films [20]).

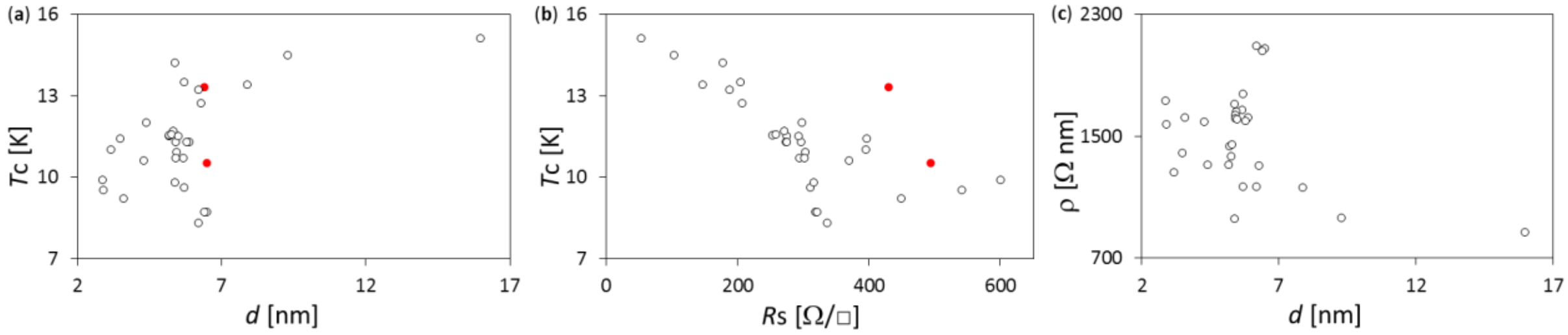

**Figure 1| Metallic and superconducting behaviour of thin NbN films**. (**a**) Critical temperature of NbN films as a function of the film thickness (*d*) and (**b**) sheet resistance ($R_s$) indicates no clear correlation with general functions of the form $T_c(R_s)$ or $T_c(d)$. (**c**) Resistivity ($\rho \equiv R_s \cdot d$) of the NbN films vs. thickness reveals a scattered data set. Chemical treatment of the substrate prior to deposition is suspected of influencing the properties of the red solid points here and in Fig. 3, while their high resistivity is outside the range presented in Fig. 1c but is discussed in the Supplemental Material [40].

One potential possible origin for the large scatter and for the deviation from a clear trend of the curves $T_c(d)$ or $T_c(R_s)$ in our NbN films is low material quality that might stem *e.g.* from extensive granularity, poor crystallinity, large strain etc. To avoid such effects and to obtain high material quality, we grew the NbN films on MgO substrates, with which the lattice mismatch is small (< 3.5%). Moreover, during the deposition, we heated the substrate to a nominal temperature of 800° C to further improve the crystallographic growth by relaxing the deposited film. We also used a system with a low base pressure ($4.5\text{-}9 \cdot 10^{-9}$ Torr), minimising the magnetic and other impurities in the films. To determine the quality of our NbN films, we demonstrated (Fig. 2a) with transition electron microscopy (TEM) that our NbN films are grown epitaxially on the MgO substrate, forming highly-orientated crystallinity, *i.e.* clear long range cubic structure with low level of granularity, if exists at all (we examined with TEM representative films, see Methods for details). In addition to the atomic resolution TEM imaging, the highly-orientated crystallinity of the NbN films and the good lattice matching between the NbN and the MgO substrate were observed also in the selective area electron diffraction (Fig. 2b). Furthermore, the measured lattice constant of the cubic NbN films is very close to the literature value, suggesting the films are relaxed.

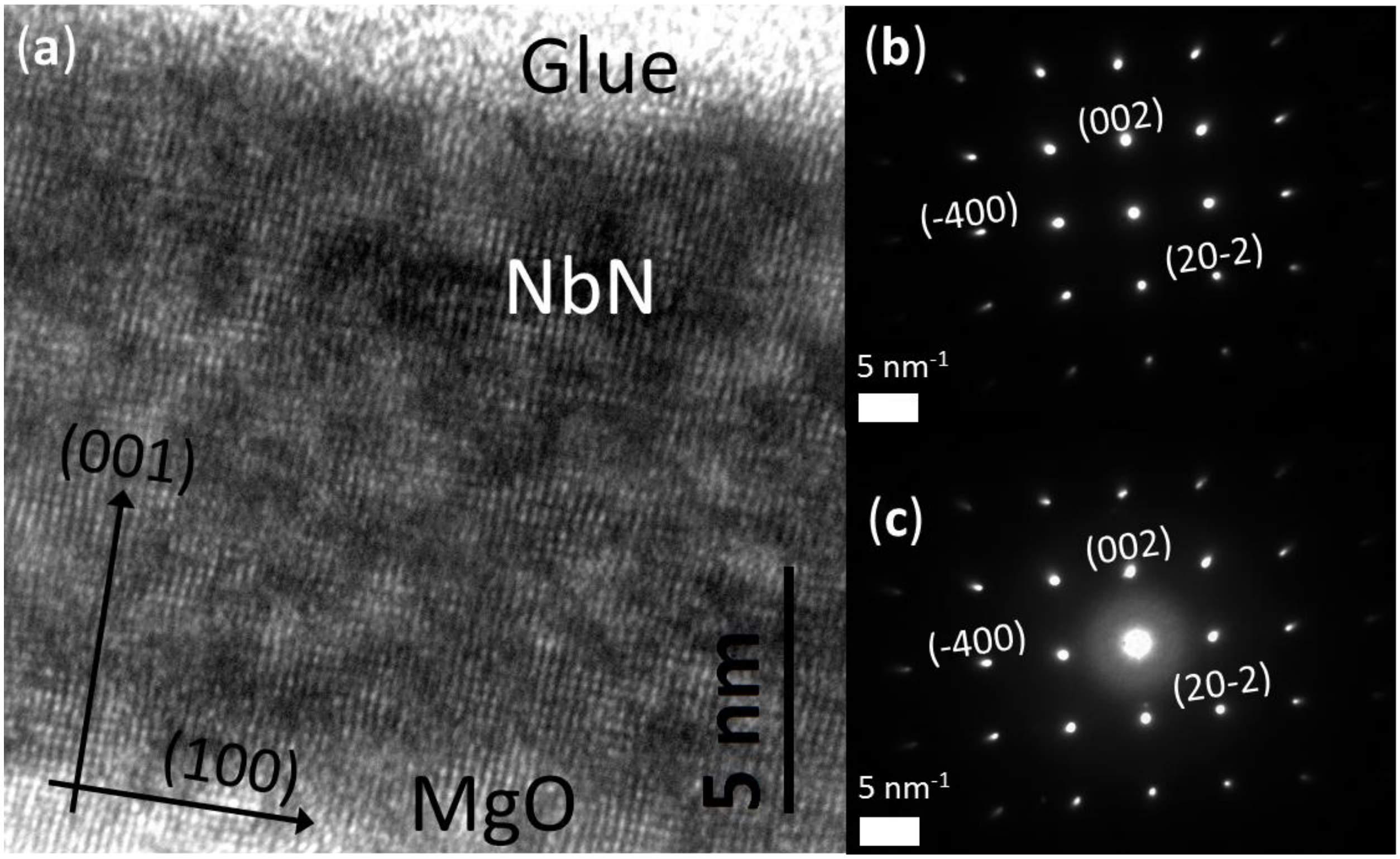

**Figure 2| TEM micrographs of epitaxial NbN film on an MgO substrate**. (**a**) Atomic structure of an NbN film on an MgO substrate, demonstrating epitaxial growth of long-range cubic structure and good lattice matching with the substrate. The lattice constants of both MgO and NbN are 4.35 ± 0.1 Å. (**b**) Selective area electron diffraction from an area within the MgO substrate only and (**c**) from an area that spans the MgO substrate, NbN film and the glue layer (that is used to protect the film from the top) demonstrates high crystallinity of both the MgO and NbN and a good lattice

matching between these substances. The bright spot in the centre of (c) is due to the amorphous glue.

Given that the epitaxial films were of high quality and that they were grown under similar conditions, the fact that Fig. 1 failed to show any clear correlations between the parameters $d$, $R_s$ and $T_c$ suggests that a different scaling method is required. Since $T_c$ is usually suppressed with reduced thickness and with the increase in disorder (*i.e.* with increasing $R_s$), we examined the relationship $d{\cdot}T_c$ as a function of $R_s$. Indeed, Fig. 3 shows that plotting $d{\cdot}T_c$ vs. $R_s$ reveals a clear trend, while the scatter was reduced significantly with respect to the traditional scaling curves that were presented in Fig. 1. This decrease in scatter is even more remarkable when taking into account that when multiplying two parameters that were measured independently (*i.e.* thickness and $T_c$) the statistical noise should increase, and not decrease.

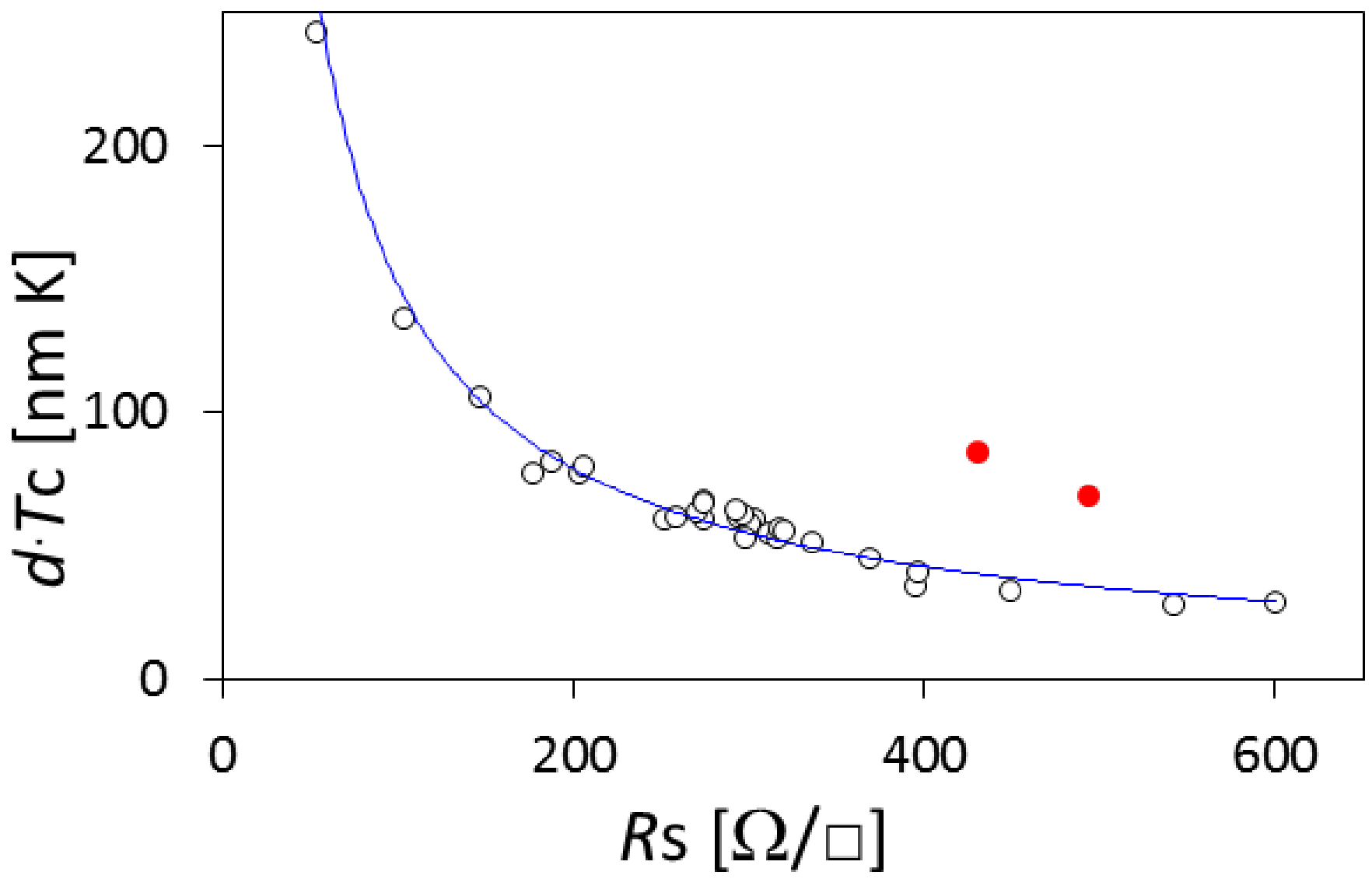

**Figure 3| Fitting our NbN on MgO data to the power law $d{\cdot}T_c(R_s)$.** (**a**) Plotting $d{\cdot}T_c$ vs. $R_s$ for the NbN films reduces the scattering significantly with respect to the curves in Fig. 1. The blue

line is the best fit to Eq. 1 ($A$ = 9448.1 and $B$ = 0.903). Red solid points are discussed in the Supplementary Material [40].

The blue solid line in Fig. 3 was added not only to guide the eye for the clear trend and reduced scatter with respect to Fig. 1, but this line is also the best fit of the data to the power law:

$d{\cdot}T_c = A{\cdot}R_s^{-B}$ **(Eq. 1)**

where $A$ and $B$ are fitting parameters and hereafter $d$, $T_c$ and $R_s$ are unitless when the appropriate values are given in nm, K and Ω/□. The exponent $B$ in Fig. 3 is close to unity ($B \approx 0.9$) so technically, one can approximate Eq. 1 to a reduced form: $\rho{\cdot}T_c \sim$ constant. Yet, when using Eq. 1 to predict the $T_c$ of a film [40], the exponent $B$ is needed.

We can suggest two approaches to explain the origin of Eq. 1. The first approach is based on the BCS-related models. Specifically, one can rewrite Eq. 1 as:

$T_c = (A/d){\cdot}e^{-B\ln(R_s)}$ **(Eq. 2a)**

Bearing in mind the BCS-based frameworks, Eq. 2a is written in a similar form to these equations *i.e.* $T_c$ equals to an amplitude ($A/d$) times an exponent that expresses the electron interactions ($B{\cdot}\ln(R_s)$). For instance, in the framework of the BCS-based McMillan equation [45,46] $\left(\boldsymbol{T_c = \frac{\Theta_D}{1.45} e^{-\frac{1.04(1+\lambda)}{\lambda-\mu^*(1+0.62\lambda)}}}\right)$, Eq. 2a implies that changes in $N(0)$ or in the interaction (λor μ) may scale as $B{\cdot}\ln(R_s)$ (where λ and μ are the electron-phonon coupling constant and the Coulomb repulsive interactions). This outcome is reminiscent also of Finkel'stein's derivation of $T_c(R_s)$ for homogenous superconductors where the interaction term was also rephrased in terms of the sheet resistance (while we recall that a logarithmic accuracy was claimed in that framework), but with the main difference that here $d$ appears explicitly [9].

The second approach to explain Eq. 1 relies on the fact that above $T_c$, conventional superconductors are normal metals. Thus, the relationship between $d$ and $R_s$ for the examined thin superconducting films is the same as that for metals in general. Hence, here, Eq. 1 implies a somewhat broader generalisation of the thickness dependence of resistance in thin metallic films. That is, one can isolate $R_s$ as a function of $d$ and $T_c$:

$$R_s = (A/d \cdot T_c)^{1/B} \qquad \textbf{(Eq. 2b)}$$

This manipulation is justified *e.g.* if $A$, $B$ and $T_c$ are representatives of simple material properties such as $\Theta_D$, mechanical strain, granularity, $N(0)$ etc. In this case, a power-law-form thickness dependence of these properties can also explain Eq. 1. We should emphasise here that although for thick materials, $R_s \sim 1/d$, this is usually not true for thin films (*i.e.* when $T_c$ is deviates significantly from its bulk value), as can be seen for instance in Fig. 1c. Therefore, the fit of our data to Eq. 1 (Fig. 3) cannot due to such simple relationships.

It is worth mentioning that Eq. 1, and more so its reduced form, resembles Homes's Law, which empirically relates $T_c$ through the superfluid density to the normal state conductivity in the case of high-$T_c$ superconductors [3]. However, thus far, we were not able to derive a direct relationship between the two laws.

To demonstrate the full range of applicability of Eq. 1, we showed that this equation fits data gathered from the literature for ~ 30 other superconductors studied over the past 46 years that summarise all of the reports from which we could extract $d$, $R_s$ and $T_c$ [12,13,16–29]. In some cases, we merged data reported in different publications by the same authors. We should note that, although $R_s$ can usually be measured rather accurately, the thickness, which is measured indirectly, is typically reported with a lower level of confidence [40]. Moreover, although there is an ongoing dispute of how to determine $T_c$ in thin films, usually the values for $T_c$ are measured in a consistent manner within a data set of a given material, allowing an examination of each data set at least with itself [47]. These data include NbN sets of films that were reported previously by other groups, each of which were reported to be in agreement with one of the models of the form $T_c(R_s)$ [18] or $T_c(d)$ [19,20]. Moreover, it includes some 'classical' examples such as the seminal Bi films by Goldman and co-workers [21] and the homogeneous αMoGe fimls of Graybeal *et al*. who were used for demonstrating Fnikelstein's model [9,48]. The data and analysis of each of these materials is discussed in detail on a linear scale in the Supplemental Material [40] (*e.g.* a detailed analysis of αMoGe films is brought in Section S7 in the Supplemental Material [40]). We should note that in addition to the contribution of the Supplemental Material [40] to the current work, this inclusive database is available also for readers who seek further investigation of superconducting and metallic behaviour in thin films.

Although detailed analysis of the individual materials is brought in the Supplemental Material [40], the most common method to present data points that follow Eq. 1 is by linearity on a log-log scale of $d{\cdot}T_c$ vs. $R_s$. Indeed, the linearity of the data in Fig. 4a-b clearly validates Eq. 1 for a broad range of $T_c$, $d$ and $R_s$ (we divided the data between Fig. 4a and 4b arbitrarily to spread data that otherwise would have been too crowded to distinguish). To eliminate the possibility that the scaling of Eq. 1 is due to *e.g.* an inverse proportionality between $R_s$ and $d$ which by chance fits with a power law relation for $T_c(R_s)$ or for $T_c(d)$, in Fig. 4c we presented the resistivity as a function of thickness for these materials. Likewise, in Fig. 4d and 4e we showed the dependence of $T_c$ on thickness and on sheet resistance. The non-linearity and non-uniformity of the data in Fig. 5c-e emphasise the universality presented in Fig. 5a-b (we should note that the set of $\alpha Nb_3Ge$ films reported by Kes and Tsuei [24] and the thicker films of Wang and co-authors [16–18] are brought as examples for thick films, in which both the resistivity and $T_c$ are rather constants over the entire range of thickness reported for these films. However, it is clear from Fig. 4 that this is not the case for all the other data sets). In addition, the complete presentation of the individual sets of data on a linear scale in the Supplementary Material [40], demonstrates that Eq. 1 quantitatively fits well the data from each material. This exhaustive list that surveys thin-film superconductors and presents their properties also includes some superconductors that require more gentle treatment, for instance, superconductors that only qualitatively agree with the scaling $d{\cdot}T_c$ vs. $R_s$ (*e.g.* $MgB_2$ [49]) as well as the few material sets that do not exhibit convincing agreement with this scaling (Ga [50], Sn [13], $Nb_3Sn$ and $V_3Si$ [51]).

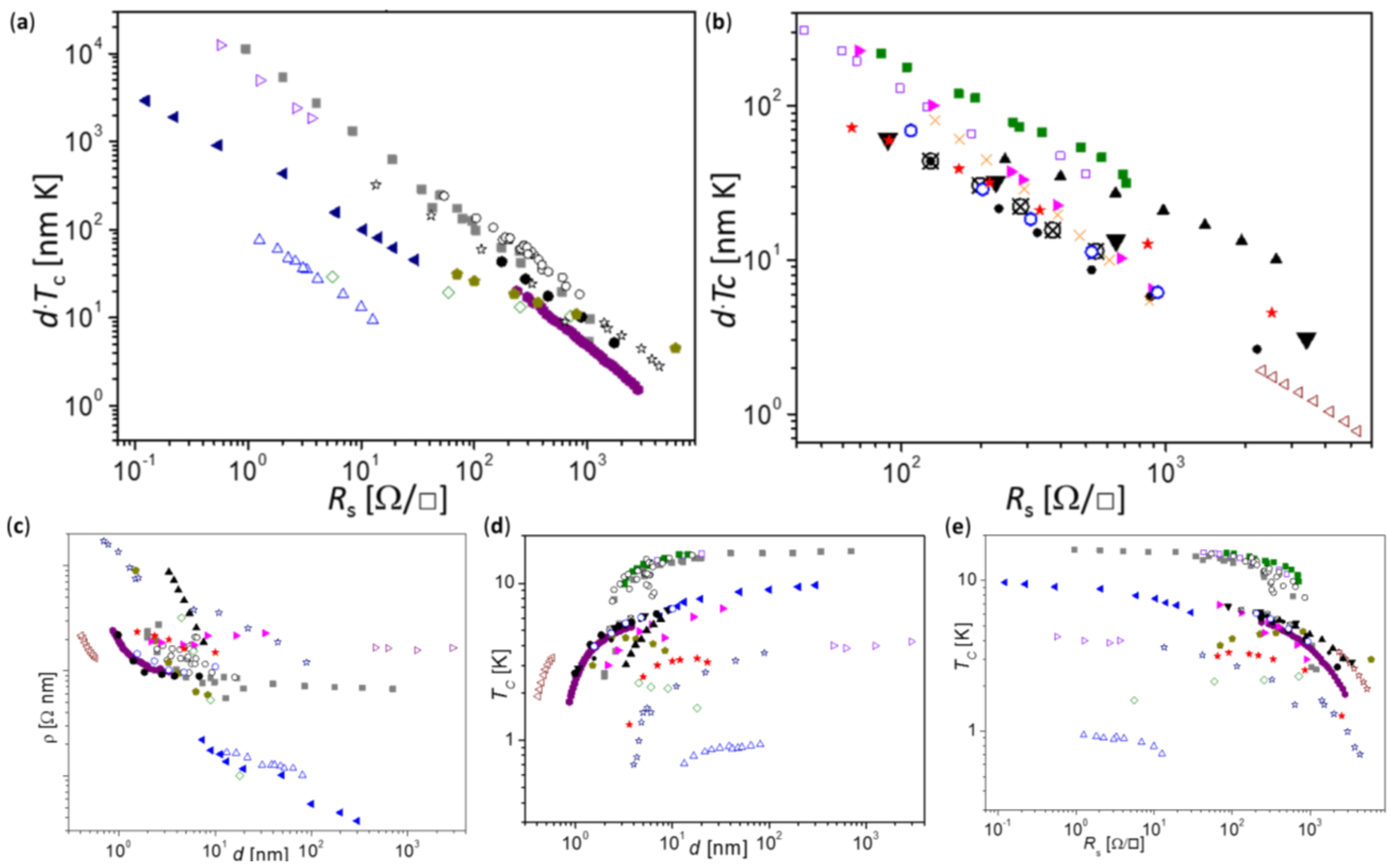

**Figure 4| Scaling of superconducting and metallic properties in thin films**. (**a**) and (**b**) show the dependence of $d{\cdot}T_c$ on $R_s$ for various superconductors (data sets are arbitrarily split to two panels to prevent indistinguishability). The linearity of the data on a log-log scale is in good

agreement with Eq. 1. (**c**) Resistivity dependence on thickness for representative materials (log-log scale). (**d**) $T_c$ as a function of thickness and (**e**) of sheet resistance of these materials on log-log scales. The following symbols were used in (a-f): ■-NbN from Wang and co-authors [16–18]; ■-NbN by Semenov *et al.* [20]; □-NbN by Kang *et al.* [19]; ○-our NbN films (from Fig. 1); △-Mo [52]; ⬣-Bi and ◁-Pb by Haviland, Liu and Goldman [21]; ◇-Al from Cohen and Abeles [12]; ◀-Nb [53]; ☆-disordered TiN by Klapwijk and co-authors [22,23]; ★- disordered TiN by Baturina and co-authors [54]; ▷-α$Nb_3Ge$ [24]; ×-αMoGe from Graybeale and Beasley [55]; ▶-αMoGe from Graybeal and co-authors [25–27]; ◆-αMoGe by Yazdani and Kapitulnik [28]; ✱-αReW by Raffy *et al.* [29]; while ⬟ is Al, and ▼, ▲, ⌖, ●, • and ○ are Pb films, corresponding to the same symbols used in Strongin *et al*. [13]. A complete list of the data is given in the Supplemental Material [40].

To allow further examination of the universality presented in Fig. 4a-b, we plotted in Fig. 5a the intercepts of the different curves as a function of their slopes (*A* vs. *B* in Eq. 1). In this way, each material is represented by a single data point, allowing a comparison between the different superconductors. Figure 5a shows that the data points follow a general trend, so that *A* and *B* are correlated. It is interesting to note that the materials at the two extreme points of this curve are aluminium (in which $T_c$ is enhanced in thin films) and αMoGe, implying that *A* and *B* may be determined by the granularity of the superconductor. In fact, since the interaction in Eq. 2a is reminiscent of Finkelstein's model, which is turn had no implicit dependence on thickness and is valid for homogeneous (amorphous) superconductors, Fig. 5a suggests that the thickness-dependent coefficient is more significant to granular films, while for amorphous films, the $R_s$-based interaction is dominating. Further discussion about the potential relationship between *A* and *B* can be found in the Supplemental Material (mainly in Sections S7.1 and in S17) [40].

Independently, the data aggregation around $B$ = 1 indicates that $\rho{\cdot}T_c$ ~ constant is a reasonable approximation in several cases. More specifically, the histogram in Fig. 5b suggests that $B \approx 0.9$ to 1.1 is a universal exponent that represents the scaling of Eq. 1. In fact, a correlation between the coefficient and the exponent such as the one observed in Fig. 5a indicates that a logarithmic correction to the power law may support universality of the exponent $B \approx 0.95$. We should note that a universal value of $B$ (e.g. $B$ = 0.95) means that this may help describing superconductivity in thin films in general, but it does not mean that such a value is a good approximation for any of the specific materials.

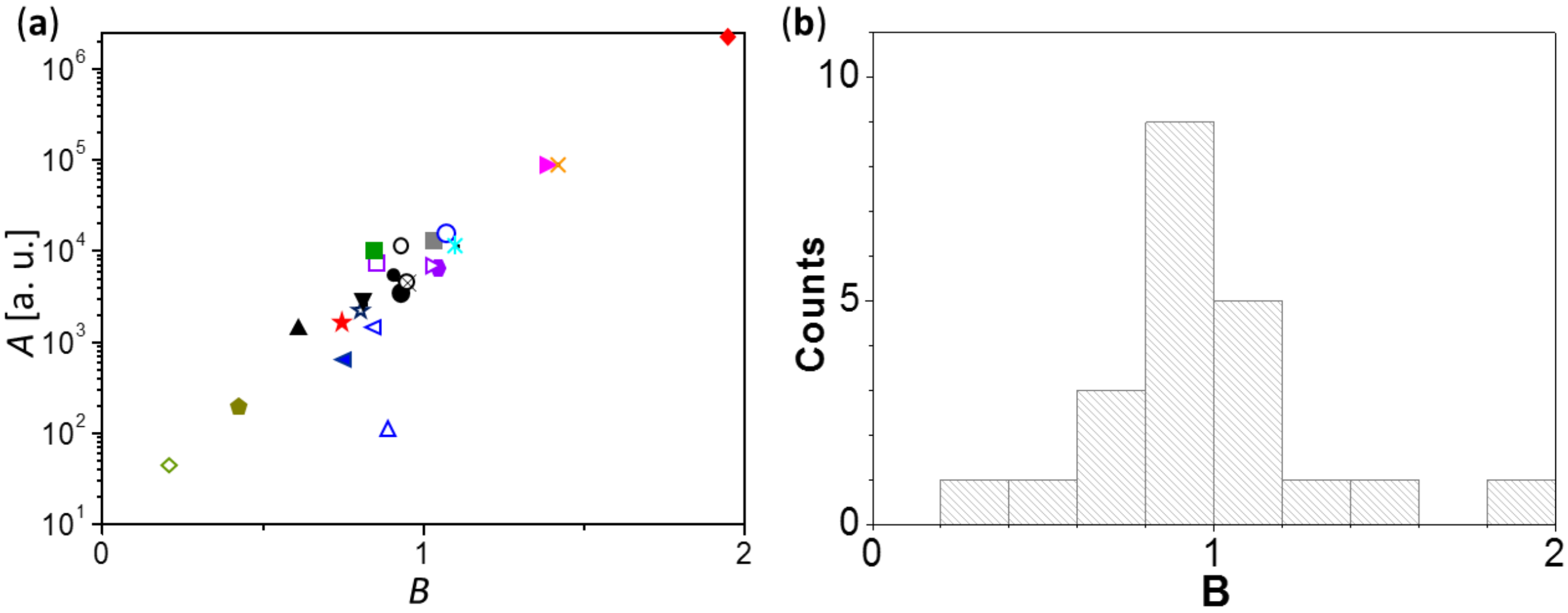

**Figure 5| Universality of the scaling $d{\cdot}T_c(R_s)$.** (**a**) Intercept versus slope ($A$ vs. $B$ with respect to Eq. 1) of the best linear fits for the graphs in Fig. 4a-b suggests that the parameters $A$ and $B$ are correlated (for details, see S7 and S17 in [40]). Legend corresponds to Fig. 4, while the only material outside the trend (△) is molybdenum. (**b**) histogram of the exponents $B$ with a mean value at $B \approx 0.95$.

It often occurs that one or more films in a data set are different than the others *e.g.* due to faults in the growth process. In many cases, it is difficult to identify such a film in a $T_c(R_s)$ or in a $T_c(d)$ curve. Therefore, the confidence in determining whether the growing system is stable or not is low. This lack of confidence obstructs the relevant scientific studies. Furthermore, it affects very badly the ability to reliably fabricate miniaturised superconducting devices. The solid red points in our own data in Figures 1 and 3 and two films from Semenov *et al*. [20] are examples, and are discussed in detail in the Supplementary Material [40] (Sections S11.5 and S11.2). Since such films stand out on a $d{\cdot}T_c(R_s)$ curve, we propose to use this scaling as a practical method to assess the quality of superconducting films. Moreover, once the parameters $A$ and $B$ in Eq. 1 are determined for a specific set of films, $T_c$ can be derived from measurements that are done in the normal state (*e.g.* at room temperature). Indeed, using the scaling law of Eq. 1 we were able to better control and evaluate our growth system. This control helped us in increasing the yield of SNSPDs made in our group. Moreover, the improvement of control of the film properties allowed us to make larger area, and hence more advanced nano-superconducting devices [56,57]. In addition, the analysis done with Eq. 1 also proved useful to predict the behaviour of thin TiN films as discussed in Section S15.1 in the Supplemental Material [40].

In conclusion, we showed that in thin superconducting films, close to the superconducting-to-insulating transition, the scaling $d{\cdot}T_c(R_s)$ describes the relationships between the film properties. We demonstrated this scaling for the films grown by us and showed that it fits our data better than the previously proposed scaling for $T_c$ in thin films. Moreover, by examining the data existing in the literature, we demonstrated the universality of this scaling. We quantified the scaling with a power law and supplied possible explanations of its origin. Furthermore, using this scaling, we presented a method to evaluate the quality of a grown film as well as to estimate its $T_c$, assisting the control of thin superconducting films, and hence expediting the development of miniaturised superconducting devices and the research of superconductivity at low dimensions. In addition, because existing theories of metallic thin films relating $R_s$ (or $\rho$) and $d$ do not involve $T_c$, while our finding does, our result may help understanding better metallic thin films more generally. Finally, the inclusive database formed to allow quantitative analysis of the existing data from the literature can be used for further investigation of the universality of superconductivity in thin films.

**Acknowledgements:**

The authors would like to thank Terry P. Orlando, Mehran Kardar, Teunis M. Klapwijk, Bertrand I. Halperin, Patrick A. Lee, Leonid S. Levitov, Yosef Imry, Biaobing Jin, Karen Michaeli, Eduard Driessen, Pieter Jan C. J. Coumou, Emanuele G. Dalla Torre, Baruch Barzel and Richard G. Hobbs for very useful discussions. Moreover, they thank the Center for Excitonics, an Energy Frontier Research Center funded by the U.S. Department of Energy, Office of Science, Office of Basic Energy Sciences under award number DE-SC0001088 for funding the project, and Y.I.; A.E.D. and K.A.S. as well as the test apparatus and growth were supported by IARPA (award number FA8650-11-C-7105), while A.M.C. acknowledges fellowship support from the NSF iQuISE program, award number 0801525. The TEM experiment was supported by the U.S. Office of Naval Research (contract #N00014-13-1-0074).

**Authors contribution:**

Y.I. initiated the study, analysed the data, gathered the published data, grew and characterised NbN films, helped with the TEM study and wrote the paper; CSK helped prepare and conducted the TEM study; A.E.D. helped characterise NbN films and helped gather published data; D.D.F. grew and characterised NbN films; A.MC. developed the electrical characterisation systems for the NbN films; K.A.S. developed the thickness measurement system for the NbN films and contributed to the paper writing; QZ helped grow NbN films and K.K.B. initiated the study, supervised the research and helped write the paper. All authors contributed to the discussion and provided feedback on the manuscript.

## Materials and Methods

NbN films were sputtered with an ATC ORION Sputtering System, EMOC-380 Power Distribution, and SHQ-15A PID Heater Controller from AJA International Inc. Sputtering conditions were nominal temperature of 800° C; a total pressure of 1.5-6 mTorr; Ar and $N_2$ flows of 26.5 sccm and 3-7.5 sccm, respectively; a sputtering current of 400 mA; and a target-sample distance of 47 mm. The sputtering time ranged from 45 sec to 300 sec. We used 2" diameter × 0.25" thick Nb 99.95% ExTa targets from Kurt J. Leskor and 10 × 10 × 0.5 mm$^3$ <100> MgO substrates with both sides polished. $R_s$ was extracted at ambient conditions from standard 4-probe measurements with a Remington Test LCC stage and a Keithley 2400 SourceMeter. $T_c$ was determined as the temperature at which $R_s = (0.9 \cdot R_S(@20K)+0.1 \cdot R_s(@20K))/2$, where $R_s$(@20K) is the measured sheet resistance at 20 K. $T_c$ was measured in liquid He with an in-house made dipstick (Omegalux KHLV-102/10 flexible heater, a DT-670A LakeShore temperature sensor, and a cry·con34 temperature controller). Finally, $d$ values were measured with an in-house made reflectometer (photodiodes: ThorLabs DET 36A Biased Detectors 350-1100 nm wavelength; LED: 470 nm HI VIS TO-5 IDX:1 OptoDiode Corp. 00-469L-ND; LED Driver: ThorLabs LEDD1B; and Hewlett Packard 34401A Multimeter). TEM images were taken with JEM 2010F by JEOL with a 200 kV beam for several samples from different locations from two films of 14.2 nm and 2.5 nm, all were found to share a similar structure. The selective area electron diffraction images were in great agreement with the fast Fourier transform of the atomic images taken from the same areas. Data presented here are from a film with $d$ = 14.2 nm as measured with the reflectometer (14.6 nm extracted from the TEM image), $R_s$ = 75.69 Ω/□, and $T_c$ = 14.24 K. The film was sputtered while being held with a thinner sample holder than that used for the set of films presented in Figures 1 and 3, potentially giving rise to a slightly higher substrate temperature.

Previously published data were collected with DataThief III version 1.6 [58]. Whenever the data were collected from several different sources (*i.e.* from different figures within the same work), cross-checking was done, and data points were discarded when the inconsistency was large. Moreover, data were also discarded when it was stated clearly by the authors that the films were too thin to be continuous or to allow reliable measurements (some of the published data were sent by the authors of these publications). A complete list of the collected and presented data, as well as the cross-checking and discussion of each data set are given in the Supplementary Material [40].

**References:**

[1] M. Tinkham, *Introduction to Superconductivity*, 2nd editio (Dover Publications, New York, 2004), p. 454.

[2] S. Chakravarty, A. Sudbø, P. W. Anderson, and S. Strong, Science **261**, 337 (1993).

[3] C. C. Homes, S. V Dordevic, M. Strongin, D. A. Bonn, and R. Liang, **430**, 539 (2004).

[4] Y. Dubi, Y. Meir, and Y. Avishai, Nature **449**, 876 (2007).

[5] V. F. Gantmakher and V. T. Dolgopolov, Phys. Uspekhi **180**, 3 (2010).

[6] A. Shalnikov, Nature **142**, 74 (1938).

[7] L. N. Cooper, Phys. Rev. Lett. **6**, 689 (1961).

[8] M. Beasley, J. Mooij, and T. Orlando, Phys. Rev. Lett. **42**, 1165 (1979).

[9] A. M. Finkel'stein, Phys. B Condens. Matter **197**, 636 (1994).

[10] Y. Guo, Y.-F. Zhang, X.-Y. Bao, T.-Z. Han, Z. Tang, L.-X. Zhang, W.-G. Zhu, E. G. Wang, Q. Niu, Z. Q. Qiu, J.-F. Jia, Z.-X. Zhao, and Q.-K. Xue, Science **306**, 1915 (2004).

[11] A. Shalnikov, Nature **142**, 74 (1938).

[12] R. W. Cohen and B. Abeles, **109**, 444 (1967).

[13] M. Strongin, R. Thompson, O. Kammerer, and J. Crow, Phys. Rev. B **1**, 1078 (1970).

[14] B. Halperin and D. Nelson, Phys. Rev. Lett. **41**, 121 (1978).

[15] M. W. Johnson and A. M. Kadin, Phys. Rev. **57**, 3593 (1998).

[16] Z. Wang, A. Kawakami, Y. Uzawa, and B. Komiyama, J. Appl. Phys. **79**, 7837 (1996).

[17] S. Miki, Y. Uzawa, A. Kawakami, and Z. Wang, Electron. Commun. Japan (Part II Electronics) **85**, 77 (2002).

[18] S. Ezaki, K. Makise, B. Shinozaki, T. Odo, T. Asano, H. Terai, T. Yamashita, S. Miki, and Z. Wang, J. Phys. Condens. Matter **24**, 475702 (2012).

[19] L. Kang, B. B. Jin, X. Y. Liu, X. Q. Jia, J. Chen, Z. M. Ji, W. W. Xu, P. H. Wu, S. B. Mi, a. Pimenov, Y. J. Wu, and B. G. Wang, J. Appl. Phys. **109**, 033908 (2011).

[20] a. Semenov, B. Günther, U. Böttger, H.-W. Hübers, H. Bartolf, a. Engel, a. Schilling, K. Ilin, M. Siegel, R. Schneider, D. Gerthsen, and N. Gippius, Phys. Rev. B **80**, 054510 (2009).

[21] D. B. Haviland, Y. Liu, and A. M. Goldman, Phys. Rev. Lett. **62**, 2180 (1989).

[22] E. F. C. Driessen, P. C. J. J. Coumou, R. R. Tromp, P. J. de Visser, and T. M. Klapwijk, Phys. Rev. Lett. **109**, 107003 (2012).

[23] P. C. J. J. Coumou, E. F. C. Driessen, J. Bueno, C. Chapelier, and T. M. Klapwijk, Phys. Rev. B **88**, 180505(R) (2013).

[24] H. Kes and C. C. Tsuei, Phys. Rev. B **28**, 5126 (1983).

[25] H. Tashiro, J. Graybeal, D. Tanner, E. Nicol, J. Carbotte, and G. Carr, Phys. Rev. B **78**, 014509 (2008).

[26] S. Turneaure, T. Lemberger, and J. Graybeal, Phys. Rev. B **63**, 174505 (2001).

[27] S. Turneaure, T. Lemberger, and J. Graybeal, Phys. Rev. B **64**, 179901 (2001).

[28] A. Yazdani and A. Kapitulnik, Phys. Rev. Lett. **74**, 3037 (2000).

[29] H. Raffy, R. Laibowitz, P. Chaudhari, and S. Maekawa, Phys. Rev. B **28**, 6607 (1983).

[30] S. Wolf, D. Gubser, and Y. Imry, Phys. Rev. Lett. **42**, 324 (1979).

[31] K. Fuchs and H. H. Wills, Math. Proc. Cambridge Philos. Soc. **34**, 100 (1938).

[32] P. A. Badoz, A. Briggs, E. Rosencher, F. A. D'Avitaya, and C. D'Anterroches, Appl. Phys. Lett. **51**, 169 (1987).

[33] T. Sun, B. Yao, A. P. Warren, K. Barmak, M. F. Toney, R. E. Peale, and K. R. Coffey, Phys. Rev. B **81**, 155454 (2010).

[34] B. Sacépé, C. Chapelier, T. Baturina, V. Vinokur, M. Baklanov, and M. Sanquer, Phys. Rev. Lett. **101**, 157006 (2008).

[35] G. N. Gol'tsman, O. Okunev, G. Chulkova, A. Lipatov, A. Semenov, K. Smirnov, B. Voronov, A. Dzardanov, C. Williams, and R. Sobolewski, Appl. Phys. Lett. **79**, 705 (2001).

[36] J. Clarke and F. K. Wilhelm, Nature **453**, 1031 (2008).

[37] F. Najafi, F. Marsili, E. Dauler, R. J. Molnar, and K. K. Berggren, Appl. Phys. Lett. **100**, 152602 (2012).

[38] F. Marsili, F. Bellei, F. Najafi, A. E. Dane, E. A. Dauler, R. J. Molnar, and K. K. Berggren, Nano Lett. **12**, 4799 (2012).

[39] G. N. Gol'tsman, O. Okunev, G. Chulkova, A. Lipatov, A. Semenov, K. Smirnov, B. Voronov, A. Dzardanov, C. Williams, and R. Sobolewski, Appl. Phys. Lett. **79**, 705 (2001).

[40] Supplemental Material is avaulable online at [URL will be inserted by Publisher].

[41] R. Koushik, S. Kumar, K. R. Amin, M. Mondal, J. Jesudasan, A. Bid, P. Raychaudhuri, and A. Ghosh, Phys. Rev. Lett. **111**, 197001 (2013).

[42] B. Matthias, T. Geballe, and V. Compton, Rev. Mod. Phys. **35**, 1 (1963).

[43] K. M. Rosfjord, J. K. W. Yang, E. A. Dauler, A. J. Kerman, V. Anant, B. M. Voronov, G. N. Gol'tsman, and K. K. Berggren, Opt. Express **14**, 527 (2006).

[44] F. Marsili, F. Najafi, E. Dauler, F. Bellei, X. Hu, M. Csete, R. J. Molnar, and K. K. Berggren, Nano Lett. **11**, 2048 (2011).

[45] W. McMillan, Phys. Rev. **167**, 331 (1968).

[46] P. B. Allen, Phys. Rev. B **12**, 905 (1975).

[47] T. I. Baturina, S. V. Postolova, A. Y. Mironov, A. Glatz, M. R. Baklanov, and V. M. Vinokur, EPL (Europhysics Lett. **97**, 17012 (2012).

[48] J. M. Graybeal and M. R. Beasley, Phys. Rev. B **29**, 4167 (1984).

[49] A. V. Pogrebnyakov, J. M. Redwing, J. E. Jones, X. X. Xi, S. Y. Xu, Q. Li, V. Vaithyanathan, and D. G. Schlom, Appl. Phys. Lett. **82**, 4319 (2003).

[50] H. M. Jaeger, D. B. Haviland, B. G. Orr, and A. M. Goldman, Phys. Rev. B **40**, 182 (1989).

[51] T. Orlando, E. McNiff, S. Foner, and M. Beasley, Phys. Rev. B **19**, 4545 (1979).

[52] L. Fàbrega, a Camón, I. Fernández-Martínez, J. Sesé, M. Parra-Borderías, O. Gil, R. González-Arrabal, J. L. Costa-Krämer, and F. Briones, Supercond. Sci. Technol. **24**, 075014 (2011).

[53] A. Gubin, K. Il'in, S. Vitusevich, M. Siegel, and N. Klein, Phys. Rev. B **72**, 064503 (2005).

[54] T. I. Baturina and S. V. Postolova, in Int. Work. Strongly Disord. Supercond. Supercond. Insul. Transit. (2104), Villard de Lans, France.

[55] J. M. Graybeal, Phys. B+C **135**, 113 (1985).

[56] Q. Zhao, A. McCaughan, F. Bellei, F. Najafi, D. De Fazio, A. Dane, Y. Ivry, and K. K. Berggren, Appl. Phys. Lett. **103**, 142602 (2013).

[57] Q. Zhao, A. McCaughan, A. Dane, F. Najafi, F. Bellei, D. De Fazio, K. Sunter, Y. Ivry, and K. K. Berggren, arXiv:1408.1124 (2014).

[58] DataTheif III version 1.6, Http://datathief.org, B. Tummers, (2006).

# Supplemental Material to:

# "Universal scaling of the critical temperature for thin films near the superconducting-to-insulating transition"

**Introduction to the Supplemental Material**:

The following sections give an inclusive list of the values of $d$, $T_c$ and $R_s$ for some 35 sets of experiments on thin superconducting films in the form of tables and figures. This allows a careful examination of the data, mainly with respect to the scaling presented in the main text.

We present the values of our measurements as well as those collected from published works. These include all the data presented in Fig. 4, along with a few datasets that were not included in Fig. 4.

**What is included below**?

The datasets for the different superconductors are presented in alphabetic order. Wherever there are multiple sources for a given material, an additional chronological order was used.

Each material is presented with a very brief relevant overview. We then present the raw data for $d$, $T_c$ and $R_s$ as collected either with DataThief [1] or directly (if applicable) while specifying the actual source (figure/table number in the original report). We also include the data of the films measured and characterized in our lab. For each dataset, we presented the values that we found for $A$ and $B$ by fitting the data to Eq. 1 ($d{\cdot}T_c = A/R_s^B$). We then fed back these values for $A$ and $B$ and also input the thickness and sheet resistance values to extract the corresponding calculated value of $T_c$ for each film, and we present this recalculated value for $T_c$ ($T_{c_RC}$). Finally, the error of $T_{c_RC}$ with respect to the measured $T_c$ is also presented (‘Err $T_{c_RC}$%’). The statistical coefficients of determination, $R^2$, of the fitting curves to Eq. 1 are also added. In addition to the table of the raw data, each dataset is displayed graphically in four panels. In the first two panels, $T_c$ is shown as a function of thickness and of the normal state sheet resistance. The third one includes the 3D resistivity, ρ, as a function of thickness. Finally, the scaling of $d{\cdot}T_c$ vs. $R_s$ is presented, while the best fit to Eq. 1 is also drawn when applicable.

As can be qualitatively deduced from Fig. 4, the data agree rather well with the scaling. However, one should bear in mind that the original publications from which the data were extracted did not consider Eq. 1 at all. Thus, despite the good agreement, there are several factors that in some cases made the quantitative analysis somewhat difficult.

**Sources of error in the data**

**(a) Data extraction**

In some cases, the data points were taken from several different sources, *e.g.*, from several figures in the original publication, each of which contains only partial data. For instance, the data points were sometimes reconstructed from two independent figures in the original publication, one graph of $T_c(d)$ and one of $T_c(R_s)$. Hence, errors have already accumulated in the data extraction process. We evaluated this error in the following way. After extracting the values for $d$, $T_c$ and $R_s$ from the two sources, the error in the values that were extracted more than once were calculated. Bearing in mind the example of data extracted from two independent graphs of $T_c(d)$ and of $T_c(R_s)$, the difference in the values of $T_c$ can be calculated to evaluate the error in the data extraction process. We presented these values below, designated by "Err $d$%", "Err $T_c$%" and "Err $R_s$%". In these cases, if the cross-checked values resulted in a low level of confidence for some data points, these data points were presented (and highlighted) but were not taken into consideration for the quantitative analysis.

**(b) Thickness measurements**

Another error source is the low certainty accompanied with the determination of the measured values of ultrathin films. In particular, in most cases, the thickness is measured indirectly. Typically, this is done by measuring the frequency shift on a quartz oscillator that is located close to the substrate and then calculating the volume, and hence the thickness of the deposited material, given that the density of the material and the relevant geometrical factors are known.

**(c) Films thinner than a single unit cell**

In some cases, the reported thickness of the films was smaller than a single unit cell of that material. Hence, for these sets of films, we considered only films that are thicker than one unit cell size.

**(d) Inhomogeneity of films within a given set**

In addition, for the analysis we did, we assumed that films that belong to the same dataset are similar in nature. Nevertheless, in many cases, films within the same dataset vary according to different parameters that are not always reported or able to be detected, such as stoichiometry, strain, grain size, dopant concentration, physical dimensions, etc. Therefore, such unknown variations between films within the same dataset are likely to encumber the quantitative analysis of the data. In fact, in some cases, it was specified explicitly that the growth conditions were changed from one film to another for different reasons (*e.g.*, for optimizing $T_c$)

(**e**) **Homogeneity within a given film**

Another potential source of error in the values reported for the measured values (mainly for the thickness) is the homogeneity of the material within a given film. In some cases, it is impractical to examine the homogeneity of the superconductor with respect to, *e.g.*, the granularity of the material, the distribution of chemical or magnetic contaminations, and the homogeneity of the thickness (*e.g.*, films reported to be thinner than a unit cell cannot have constant thickness across the samples). Hence, this might lead to inconsistency in the results.

(**f**) **Reported in low level of confidence**

In several cases, the authors who report the values of the film they measured add a note regarding to certain films (usually the thinnest films), in which they suggest that the values measured for these films should be considered in cautious. For instance, this can be due to one or more of the reasons specified above. In such cases, we specified these values below. However, we did not consider these films for the quantitative analysis.

(**g**) **Incomplete or irrelevant data**

Since we wanted to examine the relations between $R_s$, $T_c$ and $d$, we looked only at the data sets in which all these three values were measurements for enough films from that set. We give some examples below for paper that we omitted from the current survey, as some of the reported values were nominal, rather than measured. For instance, several studies that explored superconductivity in Bi films reported the values measured for $T_c$ and $R_s$, while the values reported for *d*, were nominal based on the assumption that the resistivity is constant for all of the films.  Hence, we did not consider these papers.

Likewise, some work on thin film superconductivity deals with 'unconventional' superconductors. For instance, this can include a layered material, which is composed of two different materials. Examples for this are the work by Strunk *et al*. on layered Nb/Gd films [2], and the work by Kapitulnik and co-authors on layered $MgB_2$/MgO films [3].

The fact that most of the datasets agree both quantitatively and qualitatively with the universal scaling despite these hurdles strengthens the fact that the reported power law is universal. The above factors can usually allow a sufficient explanation wherever there is a deviation from the scaling of Eq. 1. Below, we analyzed each of the datasets and specified the factors that may have affected the quantitative analysis, wherever applicable.

**1. Aluminum**.

The scaling of superconductivity in thin Al films is unique in the sense that $T_c$ is enhanced rather than suppressed. That is, unlike in most other superconductors, the reported $T_c$ in thin aluminum films usually exceeds its bulk value ($T_c$ of the bulk: ~1.2 K [4]). Moreover, the value of $T_c$ in Al generally increases with decreasing thickness. Hence, the fact that the scaling of Eq. 1 fits the data for Al is presumably a surprise. We present here data collected on superconductivity in aluminum by three independent studies, and it should be noted that the third dataset is one of the only cases where Eq. 1 does not fit the data very well.

It is interesting to point out that the exponents for the two cases where Eq. 1 fits the data are of the smallest values among the examined materials ($B < 0.5$).

**1.1. Aluminum (Cohen and Abeles [5], ◇ in Fig. 4).**

An early report (1968) by Cohen and Abeles [5] demonstrated a continuous enhancement of $T_c$ with decreasing thickness in aluminum. This increase resulted in $T_C$ higher than that of the bulk value ($T_C$ of the bulk: ~1.2 K [4]).

**Table S1.1. Superconductivity in aluminum films (Cohen and Abeles [5]), ◇ in Fig. 4.** $d$, $T_c$, and $R_s$ (the latter was measured at 4.2 K) reproduced from Table 1 of Cohen and Abeles [5], as well as the values calculated for $A$, $B$, $T_{c_RC}$, and Error in $T_{c_RC}$ %.

| | | | | |
|---|---|---|---|---|
| | | | $A$ | 42.653 |
| | | | $B$ | 0.212 |
| $d$ [nm] | $T_c$ [K] | $R_s$@4.2K [Ω/□] | $T_{c_RC}$ [K] | Err $T_{c_RC}$% |
| 18 | 1.6 | 5.56 | 1.647 | 2.96 |
| 9 | 2.13 | 58.89 | 1.997 | -6.23 |
| 6 | 2.18 | 255 | 2.196 | 0.73 |
| 4.5 | 2.31 | 711.11 | 2.356 | 1.98 |

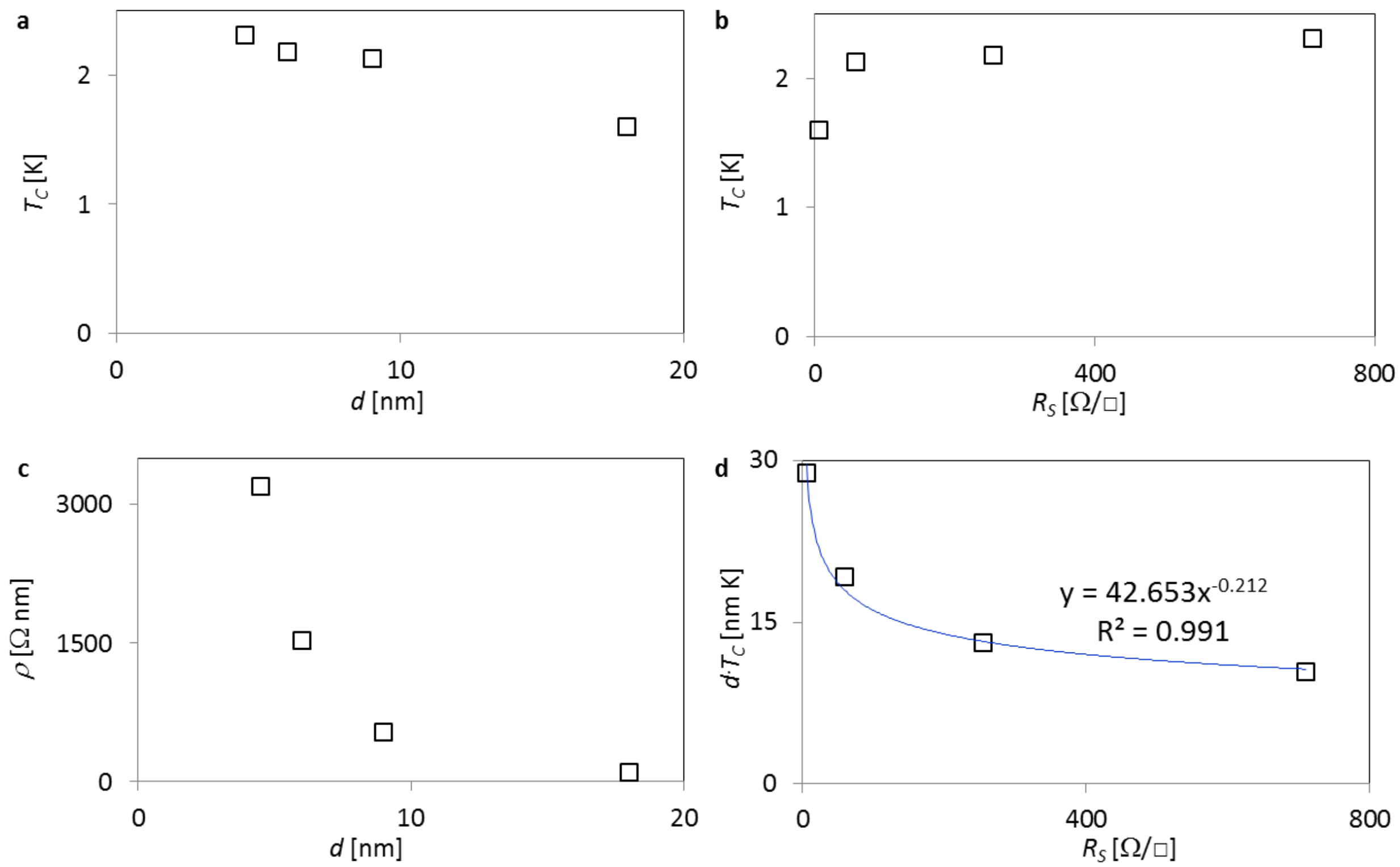

**Figure S1.1. Superconductivity in aluminum films (Cohen and Abeles** [5]**), ◊ in Fig. 4.** Critical temperature as a function of (**a**) thickness and (**b**) sheet resistance indicates enhancement of $T_C$ with decreasing thickness. (**c**) Resistivity as a function of thickness. (**d**) $d \cdot T_c$ vs. $R_s$ fits to Eq. 1 with $A$ = 42.653, and $B$ = 0.212.

**1.2. Aluminum (Strongin *et al.*,** [6] **⬟ in Fig. 4).**

Following Cohen and Abeles' experiments, Strongin *et al*. observed a similar increase in $T_c$ for thinner film, but here $T_C$ started to drop down below $d$ = 2.4 nm (which corresponds to $R_s$ = 803Ω/□). Yet, in all films, the measured $T_c$ was higher than the bulk value. Despite this dissimilarity, this dataset also agrees with Eq. 1.

**Table S1.2. Superconductivity in aluminum films (Strongin *et al.*** [6]**), ⬟ in Fig. 4.** $d$, $T_c$, and $R_s$ from the two panels of Fig. 1 from Strongin *et al*. [6] as well as the values calculated for $A$, $B$,

$T_{c_RC}$, and Error in $T_{c_RC}$%. Since the data points were matched through a common thickness value, the difference in *d* extracted from the two panels is also added (‘Error in *d*%’)

| Taken from Fig. 1 Top | Taken from Fig. 1 Top | Taken from Fig. 1 Bottom | Taken from Fig. 1 Bottom | | *A* | 192.43 |
|---|---|---|---|---|---|---|
| | | | | | *B* | 0.431 |
| *d* [nm] | $T_c$ [K] | *d* [nm] | $R_s$ [Ω/□] | Err *d*% | $T_{c_RC}$ [K] | Err $T_{c_RC}$% |
| 1.503 | 2.989 | 1.527 | 5996 | 1.59 | 3.01 | 1.38 |
| 2.389 | 4.591 | 2.412 | 803 | 0.98 | 4.51 | -3.42 |
| 3.299 | 4.496 | 3.321 | 364 | 0.69 | 4.59 | 2.84 |
| 4.202 | 4.416 | 4.233 | 226 | 0.73 | 4.43 | 0.31 |
| 6.324 | 4.110 | 6.313 | 100 | -0.18 | 4.18 | 1.1 |
| 8.400 | 3.716 | 8.4 | 70.114 | -0.01 | 3.67 | -0.56 |

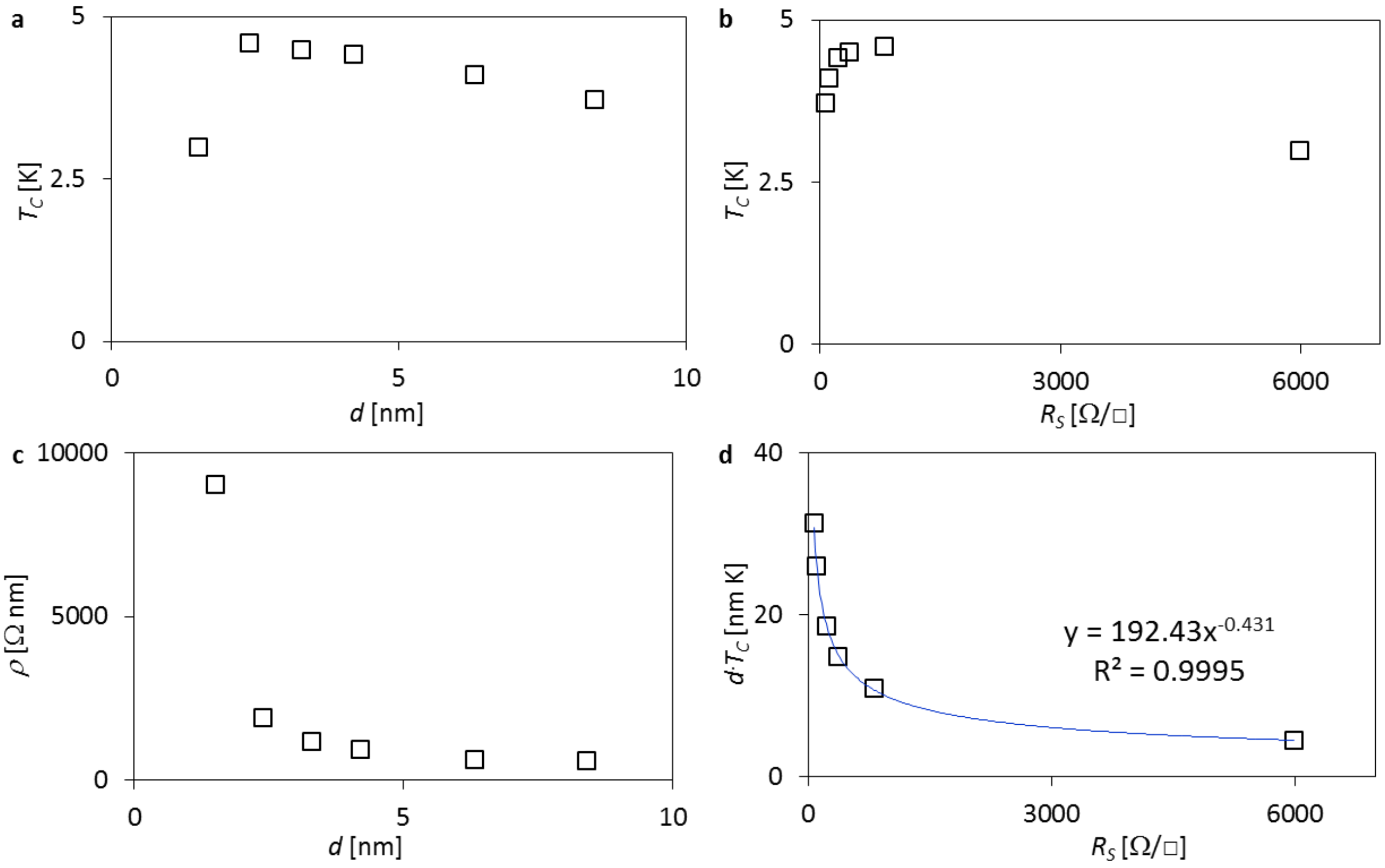

**Figure S1.2. Superconductivity in aluminum films (Strongin *et al.*** [6]**), ⬟ in Fig. 4.** Critical temperature as a function of (**a**) thickness and (**b**) sheet resistance indicates enhancement of $T_c$ with decreasing thickness, while the trend changes at $d$ = 2.4 nm ($R_S$ = 803 Ω/□). (**c**) Resistivity as a function of thickness. (**d**) $d{\cdot}T_c$ vs. $R_s$ fits to Eq. 1 (blue curve) with $A$ = 192.43, and $B$ = 0.431.

**1.3. Aluminum (Liu *et al.*** [7]**)**

A later paper by Goldman and co-authors repeated the previous measurements but with samples grown in somewhat cleaner conditions. Similarly to the case of Strongin *et al.*, although $T_c$ assumed values larger than the bulk, it started decreasing at a certain thickness ($d$ = ~4.2 nm), while here it even went below the bulk $T_c$ (at $d$ < ~2.5 nm). On the other hand, the dependence of $T_c$ on $R_s$ was found to be rather linear. When plotting $d \cdot T_c$ as a function of $R_s$, one obtains a smooth monotonic function as in the other superconductors. However, this function decreases much more slowly than a power law (the decrease is approximately logarithmic) and does not agree with Eq. 1. A possible partial reason for this is the large uncertainty in the values of $d$, $R_s$ and $T_c$. Another possible deviation of this dataset from the framework of Eq. 1 might be related to the thickness measurement or more likely, to the Ge substrate used for these films as was discussed by the authors. Specifically, the proximity effect could have played a significant role in changing the superconducting properties of the Al films grown on the Ge substrate. Although the scaling of Eq. 1 was found to agree well with the data of Bi and Pb films grown in the same method (Fig. S2.1 and S12.7), the proximization may have influenced the Al films over a larger thickness scale. A more insightful discussion about these thin aluminum films is given in Section 17.2.

**Table S1.3. Superconductivity in aluminum films (Liu *et al.*** [7]**).** $d$, $T_c$, and $R_s$ from the two panels of Fig. 9 from Liu *et al.* [7] Since the data points were matched through a common $T_c$ value, the difference in $T_c$ extracted from the two panels is also added ('Error in measured $T_{c_RC}$%'), which was found to be relatively large.

| Taken from Fig. 9 Top | Taken from Fig. 9 Top | Taken from Fig. 9 bottom | Taken from Fig. 9 bottom |
|---|---|---|---|
| $T_c$ [K] | $1/d$ [1/Å] | $T_c$ [K] | $R_s$@14K [Ω/□] |
| 0.544 | 0.0415 | 0.535 | 16889.76 |
| 0.942 | 0.0402 | 0.936 | 15291.92 |
| 1.209 | 0.0398 | 1.202 | 13989.39 |
| 1.408 | 0.0393 | 1.4 | 12881.26 |
| 1.434 | 0.0374 | 1.471 | 12478.31 |
| 1.476 | 0.039 | 1.425 | 11083.42 |
| 1.495 | 0.0371 | 1.486 | 10486.34 |
| 1.936 | 0.0369 | 1.932 | 9808.783 |
| 2.068 | 0.0368 | 2.064 | 9182.692 |
| 2.151 | 0.0352 | 2.167 | 8901.873 |
| 2.167 | 0.0357 | 2.163 | 8686.155 |
| 2.270 | 0.0348 | 2.144 | 8549.596 |
| 2.360 | 0.0342 | 2.259 | 8254.349 |
| 2.408 | 0.0335 | 2.355 | 7815.351 |
| 2.534 | 0.0333 | 2.407 | 7398.089 |
| 2.617 | 0.0322 | 2.525 | 6930.246 |
| 2.669 | 0.0313 | 2.615 | 6591.946 |
| 2.788 | 0.0308 | 2.66 | 6260.998 |
| 2.858 | 0.0287 | 2.785 | 5836.278 |
| 2.878 | 0.0297 | 2.875 | 5426.068 |
| 3 | 0.0285 | 2.859 | 5059.385 |
| 3.121 | 0.0282 | 2.997 | 4699.338 |
| 3.147 | 0.0275 | 3.122 | 4396.865 |
| 3.26 | 0.027 | 3.147 | 4130.706 |
| 3.337 | 0.026 | 3.263 | 3813.886 |
| 3.439 | 0.0165 | 3.34 | 3525.969 |
| 3.442 | 0.02 | 3.333 | 3159.251 |
| 3.459 | 0.0248 | 3.452 | 2899.948 |
| 3.481 | 0.0242 | 3.478 | 2669.743 |
| 3.507 | 0.0226 | 3.535 | 2439.424 |
| 3.516 | 0.0165 | 3.535 | 1605.269 |
| 3.533 | 0.0211 | 3.451 | 1346.694 |
| 3.545 | 0.0193 | 3.55 | 1159.369 |
| 3.546 | 0.0236 | | |
| 3.578 | 0.022 | | |

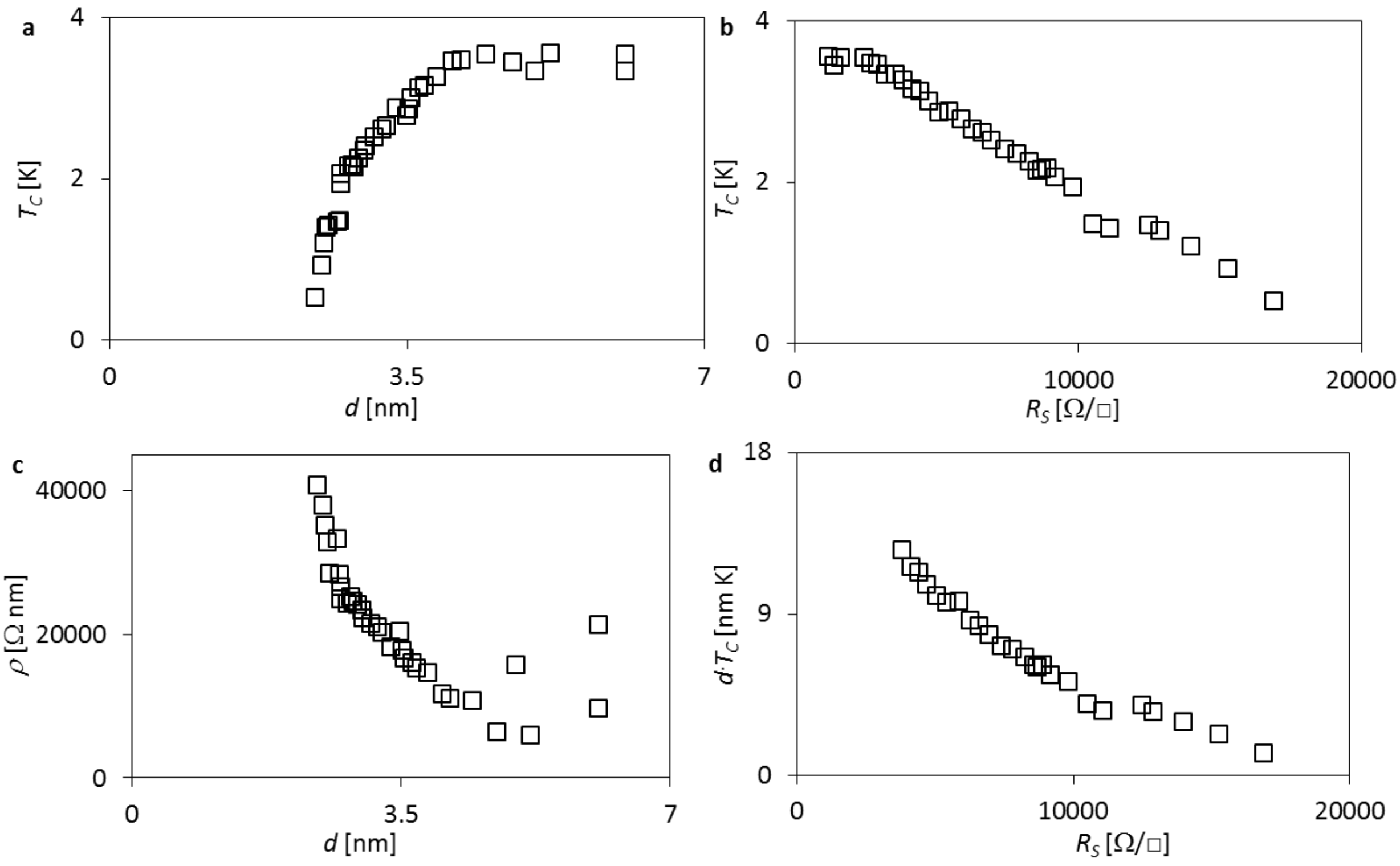

**Figure S1.3. Superconductivity in aluminum films (Liu *et al*.** [7]**).** (**a**) Critical temperature as a function of thickness indicates a slow decrease of $T_c$ with decreasing thickness, while the value of $T_c$ falls to the bulk value at $d$ = ~2.5 nm. (**b**) $T_c$ vs. sheet resistance demonstrates a rather linear dependence. (**c**) Resistivity as a function of thickness. (**d**) $d{\cdot}T_c$ vs. $R_s$ decreases monotonically but much more slowly than the power law of Eq. 1.

**2. Bismuth**.

A classic example for superconducting-insulator quantum transition, Bi is an interesting material for testing the scaling of Eq. 1. Unfortunately, most published data on Bi are either incomplete or irrelevant. For instance, Landau *et al.* tried to proximitize Bi in a special way [8], while Silverman only assumed the values for the thickness without actually measuring it in two papers [9,10] and the data published by Naugle *et al*. [11] did not contain all the relevant values (the two datasets of Silverman fit Eq. 1 perfectly, even if merged together). Yet, it is possible to examine the data by Haviland, Liu and Goldman, in which the quantum phase transition was reported [12]. It should be noted that the exponent is very close to unity ($B$ = ~1) for the data reported by both Silverman [9,10] and Haviland, Liu and Goldman [12], suggesting that the datasets are also fit well by $\rho \cdot T_c$ = constant.

**2.1. Bismuth- extracted from Haviland, Liu and Goldman** [12] **(⬢ in Fig. 4).**

Haviland, Liu and Goldman found that the quantum superconducting-insulator transition in their Bi films is around 6.5 kΩ/□. This allowed them to report on superconducting films up to < ~5 kΩ/□. Since the error in data extraction increased for $d$ < 0.85 nm, only data points with $d$ > 0.85 nm were considered to examine our model. It should be noted that in the thinnest films, superconductivity may have been influenced by the proximization with the Ge substrate, hence changing the trend from a power law to a more complex form. For a more thorough discussion about these thin bismuth films, please see Section 17.2.

**Table S2.1. Superconductivity in bismuth films (Haviland, Liu and Goldman** [12]**), ⬢ in Fig. 4.** $d$, $T_c$, and $R_s$ reproduced from Fig. 2 and Fig. 3 of Haviland *et al.* [12]. The error in the values extracted for $T_C$ in the two figures is presented as an indication of the data collection error. The values calculated for $A$, $B$, $T_{c_RC}$, and Error in $T_{c_RC}$% are also presented.

| | | | | | | |
|---|---|---|---|---|---|---|
| | | | | | $A$ | 6402.8 |
| | | | | | $B$ | 1.043 |
| Taken from Fig. 1 $d$ [nm] | Taken from Fig. 1 $T_c$ [K] | Taken from Fig. 3 $R_s$@14K [Ω/□] | Taken from Fig. 3 $T_c$ [K] | Err $T_c$% | $T_{c_RC}$ [K] | Err $T_{c_RC}$ % |
| 7.407 | 5.67 | | | | | |
| 6.41 | 5.61 | 154.468 | 5.614 | -0.02 | 5.19 | -7.53 |
| 5.376 | 5.53 | 168.837 | 5.521 | 0.07 | 5.64 | 2.08 |
| 4.405 | 5.40 | 204.76 | 5.399 | 0.07 | 5.63 | 4.2 |
| 3.773 | 5.29 | 237.09 | 5.28 | 0.12 | 5.64 | 6.69 |
| 3.322 | 5.15 | 294.567 | 5.1438 | 0.11 | 5.11 | -0.75 |
| 2.95 | 5.02 | 330.489 | 5.016 | 0.05 | 5.1 | 1.62 |
| 2.703 | 4.91 | 387.966 | 4.905 | 0.05 | 4.71 | -4.01 |
| 2.519 | 4.83 | 402.335 | 4.825 | 0.1 | 4.87 | 0.83 |
| 2.326 | 4.68 | 438.258 | 4.677 | 0.03 | 4.82 | 3.02 |
| 2.11 | 4.53 | 495.734 | 4.534 | -0.03 | 4.67 | 3.02 |
| 1.992 | 4.39 | 560.395 | 4.397 | -0.04 | 4.35 | -1.02 |
| 1.901 | 4.29 | 596.318 | 4.288 | 0.07 | 4.27 | -0.5 |
| 1.757 | 4.16 | 668.163 | 4.153 | 0.06 | 4.11 | -1.11 |
| 1.704 | 4.09 | 711.271 | 4.087 | 0.12 | 3.97 | -2.99 |
| 1.605 | 3.91 | 772.34 | 3.91 | 0.12 | 3.87 | -1.14 |
| 1.529 | 3.84 | 819.039 | 3.833 | 0.11 | 3.82 | -0.46 |
| 1.46 | 3.75 | 876.515 | 3.738 | 0.25 | 3.72 | -0.73 |
| 1.408 | 3.65 | 933.992 | 3.643 | 0.1 | 3.61 | -1 |
| 1.357 | 3.54 | 1009.43 | 3.537 | 0.17 | 3.46 | -2.35 |
| 1.316 | 3.43 | 1066.906 | 3.429 | 0.16 | 3.37 | -1.87 |
| 1.258 | 3.30 | 1153.121 | 3.297 | 0.23 | 3.25 | -1.64 |
| 1.218 | 3.18 | 1235.743 | 3.17 | 0.23 | 3.12 | -1.79 |
| 1.163 | 3.03 | 1350.696 | 3.03 | 0.13 | 2.98 | -1.77 |
| 1.124 | 2.87 | 1426.134 | 2.868 | 0.12 | 2.91 | 1.33 |
| 1.093 | 2.79 | 1544.679 | 2.787 | 0.29 | 2.75 | -1.6 |
| 1.067 | 2.70 | 1645.263 | 2.691 | 0.19 | 2.64 | -2.1 |
| 1.045 | 2.60 | 1738.662 | 2.599 | 0.18 | 2.54 | -2.45 |
| 1.009 | 2.47 | 1878.761 | 2.467 | -0.04 | 2.43 | -1.45 |
| 0.978 | 2.31 | 2047.598 | 2.306 | 0.26 | 2.29 | -0.94 |
| 0.945 | 2.16 | 2216.435 | 2.153 | 0.25 | 2.18 | 1.03 |
| 0.922 | 2.04 | 2396.049 | 2.039 | 0.23 | 2.06 | 0.8 |
| 0.895 | 1.92 | 2593.624 | 1.915 | 0.2 | 1.96 | 2.14 |
| 0.868 | 1.75 | 2794.791 | 1.754 | 0.02 | 1.87 | 6.59 |
| 0.845 | 1.60 | 3053.435 | 1.596 | 0.29 | | |
| 0.822 | 1.45 | 3297.71 | 1.44 | 0.63 | | |
| 0.802 | 1.32 | 3545.577 | 1.312 | 0.82 | | |
| 0.786 | 1.19 | 3779.075 | 1.179 | 0.62 | | |
| 0.767 | 1.03 | 4073.642 | 1.024 | 0.6 | | |
| 0.752 | 0.90 | 4343.062 | 0.892 | 0.59 | | |
| 0.74 | 0.77 | 4605.299 | 0.771 | 0.23 | | |
| 0.725 | 0.65 | 4892.681 | 0.652 | -0.28 | | |

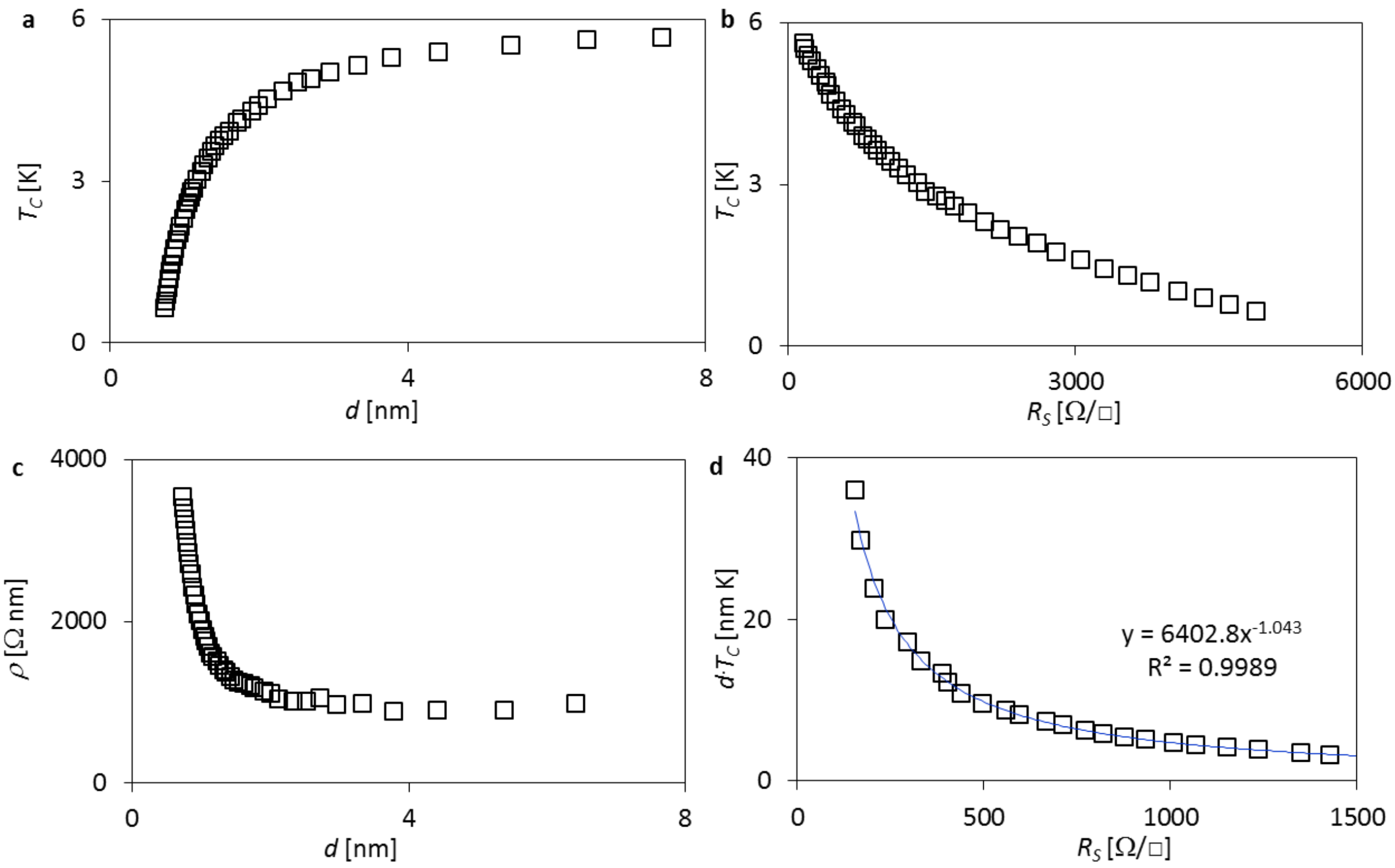

**Figure S2.1. Superconductivity in bismuth films (Haviland *et al*.** [12]**), ⬢ in Fig. 4.** Critical temperature as a function of (**a**) thickness and (**b**) sheet resistance. (**c**) Resistivity as a function of thickness. (**d**) $d{\cdot}T_c$ vs. $R_s$ fits Eq. 1 with $A$ = 6402.8 and $B$ = 1.043.

**3. $CoSi_2$.**

Badoz *et al*. investigated the interplay between conductivity and superconductivity in thin (~1.5 – 70 nm thick) $CoSi_2$ films [13]. In particular, they were looking at the relation between the deviation from the theory of Fuchs for conductivity in thin metallic films and the suppression of $T_c$ observed in their materials. The authors explained this abrupt change in both superconductivity and metallic behavior with the proximity effect. Unfortunately, we could not extract the data in a way that allows for a reliable examination of the data with Eq. 1, as the mismatch between the extracted data was >>1%. Yet, we present below the data we extracted. Yet, Fig. S15 suggests that the data should agree with the found scaling.

**Table S3. Superconductivity in $CoSi_2$ films .**

Thickness, $R_s$ and critical temperature values extracted from Badoz *et al*. for $CoSi_2$ films [13] (Figures 2, inset from 2 and Fig. 3 therein). Large errors in the data extraction process (>>1%) do not allow quantitatively analysis of the data.

| Fig. 2 Inset | | Fig2 | | Fig3 | | | | |
|---|---|---|---|---|---|---|---|---|
| *d* [nm] | $R_s$ [Ω] | *d* [nm] | $\rho_0/\rho_\infty$ | *d* [nm] | $T_c$[K] | Err $d_{2inset2}$% | Err $d_{23}$% | Err $d_{2inset3}$% |
| 70.922 | 0.311 | 60.041 | 0.94 | 59.713 | 1.118 | 18.122 | 0.55 | 18.772 |
| 23.095 | 1.028 | 20.022 | 1.112 | 20.139 | 1.041 | 15.347 | -0.579 | 14.679 |
| 15.038 | 2.251 | 14.166 | 1.483 | 14.085 | 0.928 | 6.156 | 0.573 | 6.764 |
| 12.484 | 2.299 | 12.092 | 1.226 | 12.048 | 1.013 | 3.243 | 0.364 | 3.619 |
| 10.549 | 3.611 | 10.1 | 1.636 | 10.141 | 0.852 | 4.437 | -0.396 | 4.023 |
| | | | | 10.121 | 0.787 | | | |
| 8.203 | 5.4 | 7.766 | 1.961 | 7.896 | 0.358 | 5.634 | -1.644 | 3.898 |
| 5.133 | 12.167 | 4.644 | 2.706 | 4.823 | 0.22 | 10.535 | -3.713 | 6.431 |
| 4.098 | 31.679 | 4.23 | 4.726 | | | -3.101 | | |
| 4.08 | 26.498 | 4.218 | 5.314 | | | -3.279 | | |
| 3.559 | 31.126 | 3.546 | 3.496 | 3.737 | 0.02 | 0.356 | -5.101 | -4.763 |
| 3.549 | 21.699 | 3.417 | 5.029 | | | 3.846 | | |
| 2.535 | 132.625 | 2.747 | 14.826 | | | -7.686 | | |
| 1.401 | 1523.5 | | | | | | | |

## 4. Ga (from Naugle *et al.* [11] and Ga and In by Jaeger *et al.* [14] ).

The data published by Naugle *et al.* [11] on Ga were not sufficiently complete for us to test the scaling of Eq. 1. Moreover, although Jaeger *et al*. [14] reported sets of In and Ga films for granular films (relying in part on Naugle's data), in which the mechanism governing the transition might be considered unusual and different than that of normal films, we wanted to test the scaling of Eq. 1 for these data sets as well. Unfortunately, we encountered technical difficulties in extracting these datasets (the data points are too crowded on the given scales, leading to a large inconsistency error in the extracted data and therefore are not presented here). Yet, we should state here that, although we have low confidence in the data extracted for Ga from Jaeger *et al*. [14], the data we did extract did not fit Eq. 1 very well. However, at this point, we are uncertain whether this disagreement is due to the error in data extraction or due to the fact that the mechanism governing the transition is unique.

**5. $MgB_2$- extracted from Pogrebnyakova *et al*.** [15].

Magnesium diboride is often considered a conventional superconductor, even though it is not quite so, due to, *e.g.*, its high transition temperature and the coexistence of two types of Cooper pairs. Moreover, experimental difficulties hinder the understanding of superconductivity in $MgB_2$. For instance, to date, growing a thin film of high-quality superconducting $MgB_2$ is still a challenge. In addition, most electrical properties in $MgB_2$ are anisotropic, which leads to difficulties in obtaining reproducible results. Yet, since $MgB_2$ is considered a conventional superconductor, we wanted to examine whether the scaling of Eq. 1 can potentially assist in understanding the electronic properties of this superconductor. To do so, we extracted data for thin films (~80-430 nm thick) of $MgB_2$ from Pogrebnyakova *et al*. [15].

Since the reported $T_C$ in this dataset spanned a small range (~41-41.8 K), it was impossible to examine the scaling of Eq. 1 quantitatively with respect to its prediction of $T_c$. Nevertheless, we did find out that the scaling $d \cdot T_c$ vs $R_s$ reduces the scattering that exists in other dependencies. Pogrebnyakova *et al*. [15] reported several different datasets of $MgB_2$ that are distinguishable by their preparation conditions (a flow of $B_2H_6$ gas with rates ranging from 50 to 250 sccm). Here, we demonstrate the reduction in scattering for one dataset only (200 sccm), in which the effect can be demonstrated qualitatively, but the effect also occurs in most other datasets of this report, though with a lower degree of confidence.

**Table S5. Superconductivity in $MgB_2$ (Pogrebnyakova *et al*.** [15]**) for 200 sccm $B_2H_6$ gas flow.** $d$, $T_c$, and $R_s$ reproduced from Fig. 4 of Pogrebnyakova *et al*. (full squares in the top and bottom panels) [15]. The error in the values extracted for $d$ in the two figures is presented as an indication

of the data collection error. The scaling $d{\cdot}T_c$ vs. $R_s$ qualitatively demonstrates a reduction in the data scattering.

| Bottom panel | | Small panel | | | Top panel | | | | |
|---|---|---|---|---|---|---|---|---|---|
| $d_1$ [nm] | $\rho$@14K [μΩ·nm] | $d_2$ [nm] | $\Delta\rho$ [μΩ·nm] | Err $d_{12}$% | $d_3$ [nm] | $T_c$ [K] | Err $d_{13}$% | Err $d_{23}$% | $R_s$@300K [Ω/□] |
| 80.7 | 0.84 | 80.69 | 9.54 | -0.02 | 80.87 | 41 | 0.2 | 0.22 | 12.85 |
| 151.86 | 0.37 | 150.57 | 8.14 | -0.86 | 151.97 | 41.3 | 0.07 | 0.93 | 5.61 |
| 225.1 | 0.28 | 224.4 | 7.62 | -0.31 | 225.41 | 41.7 | 0.14 | 0.45 | 3.51 |
| 230.08 | 0.36 | 229.78 | 7.92 | -0.13 | 230.83 | 41.19 | 0.33 | 0.46 | 3.6 |
| 337.4 | 0.6 | 337.75 | 11.48 | 0.1 | 336.89 | 41.82 | -0.15 | -0.25 | 3.58 |
| 364.39 | 0.33 | 364.09 | 9.32 | -0.08 | 363.71 | 41.42 | -0.19 | -0.1 | 2.65 |
| 428.41 | 0.31 | 428.89 | 9.59 | 0.11 | 428.16 | 41.47 | -0.06 | -0.17 | 2.31 |

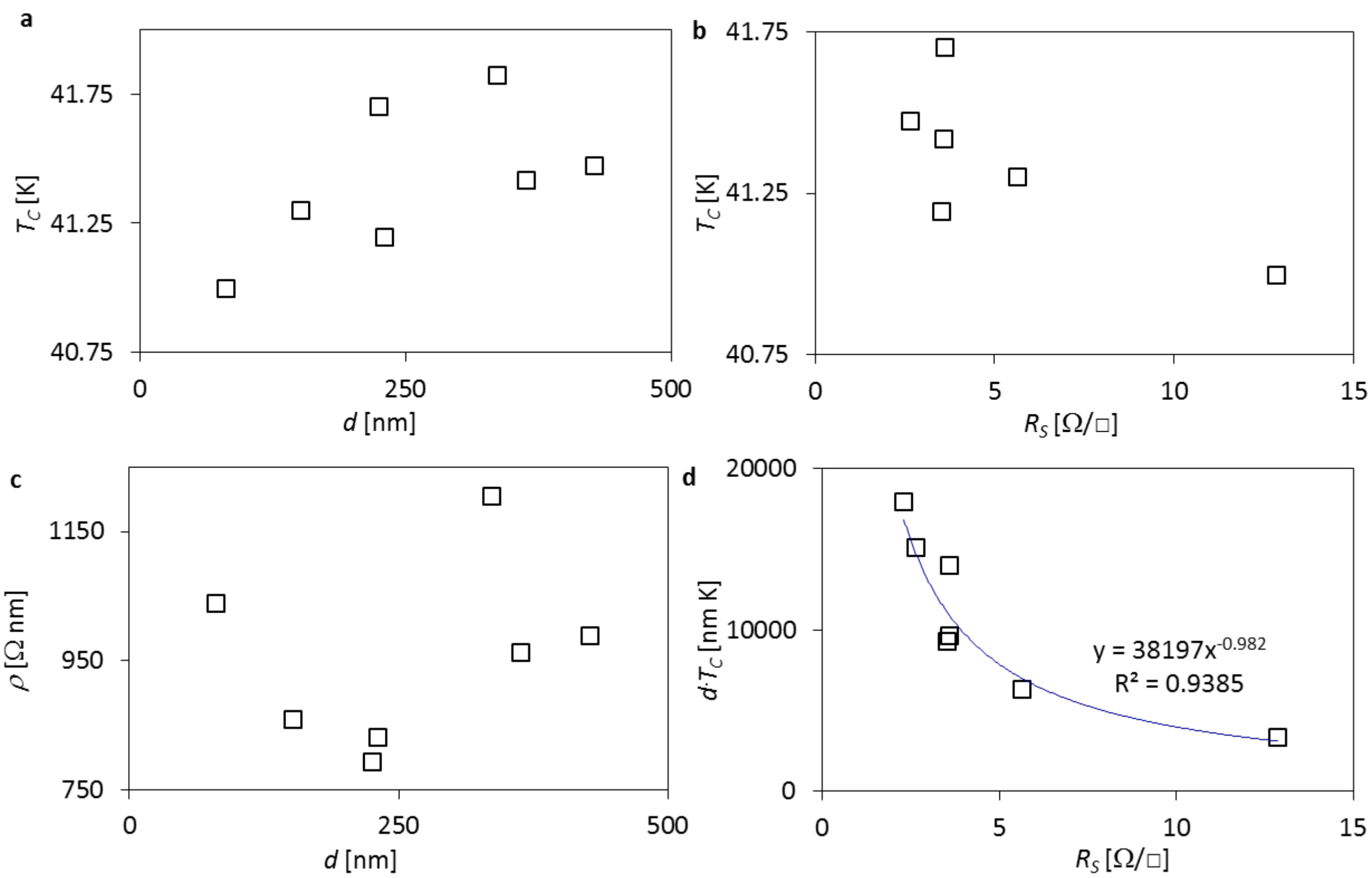

**Figure S5.1. Superconductivity in $MgB_2$ films (Pogrebnyakova *et al.*** [15]**) for 200 sccm $B_2H_6$ gas flow.** Critical temperature as a function of (**a**) thickness and (**b**) sheet resistance. (**c**) Resistivity as a function of thickness. (**d**) $d{\cdot}T_c$ vs. $R_S$ demonstrates qualitative agreement with the scaling of Eq. 1, in which the trend is more obvious than in (a-c).

**6. Molybdenum (Fàbrega *et al*,** [16]**, △ in Fig. 4).**

Similar, *e.g.*, to Al, bulk Mo is a type I superconductor (while thin films of conventional superconductors are effectively type II). The data points presented below were extracted from the work by Fàbrega *et al*. [16], who characterized two-dimensional Mo films. Although these data are in agreement with Eq. 1, the relation between the fitting coefficients *A* and *B* do not follow the general trend that seems to appear for the other superconductors examined, as shown in Fig. 5c.

**Table S6. Superconductivity in molybdenum films (Fàbrega *et al*.** [16]**), △ in Fig. 4.** $d$, $T_c$, the resistivity at 4.2K and the residual resistivity ratio ($RRR = R_S$@300K/$R_T$@4.2K) taken from Figures 1 and 3 by Fàbrega *et al*., [16] as well as the values calculated for $A$, $B$, $T_{c_RC}$, and Err $T_{c_RC}$%. Since the data points were matched through a common $T_c$ value and through a common thickness value, the difference in $T_c$ and $d$ extracted from the two panels (three datasets) is also added ('Err $T_c$%').

| | |
|---|---|
| *A* | 99.661 |
| *B* | 0.892 |

| From Fig. 1 | From Fig. 1 | From Fig. 1 | From Fig. 1 | | From Fig. 3 | From Fig. 3 | | | |
|---|---|---|---|---|---|---|---|---|---|
| $d$ [nm] | ρ[Ωnm]@4.2K | $d$ [nm] | $T_c$ [K] | Err $d$% | $T_c$ [K] | *RRR* | *Err* $T_c$% | $T_{c_RC}$ [K] | *Err* $T_{c_RC}$% |
| 13.252 | 100.297 | 13.252 | 0.707 | 0 | 0.709 | 1.68 | 0.28 | 0.777 | 9.91 |
| 16.4 | 95.21 | 16.4 | 0.79 | 0 | 0.794 | 1.74 | 0.57 | 0.771 | -2.35 |
| 21.681 | 82.493 | 21.579 | 0.837 | -0.47 | 0.844 | 1.82 | 0.76 | 0.821 | -1.95 |
| 28.281 | 78.847 | 28.281 | 0.865 | 0 | 0.873 | 1.879 | 0.84 | 0.803 | -7.15 |
| 41.483 | 60.958 | 41.483 | 0.875 | 0 | 0.883 | 2.089 | 0.89 | 0.882 | 0.81 |
| 30.82 | 64.265 | 30.922 | 0.885 | 0.33 | 0.894 | 1.979 | 1 | 0.909 | 2.71 |
| 47.576 | 59.432 | 47.677 | 0.888 | 0.21 | 0.896 | 2.1 | 0.98 | 0.883 | -0.5 |
| 52.247 | 55.108 | 52.247 | 0.895 | 0 | 0.903 | 2.149 | 0.87 | 0.918 | 2.54 |
| 38.639 | 62.06 | 38.639 | 0.909 | 0 | 0.918 | 2.049 | 1.03 | 0.89 | -2.05 |
| 64.737 | 56.295 | 64.737 | 0.915 | 0 | 0.925 | 2.099 | 1.08 | 0.899 | -1.81 |
| 47.677 | 66.978 | 47.677 | 0.916 | 0 | 0.925 | 1.958 | 1.05 | 0.846 | -7.57 |
| 80.579 | 50.784 | 80.68 | 0.93 | 0.13 | 0.94 | 1.999 | 1.04 | 1.004 | 7.91 |

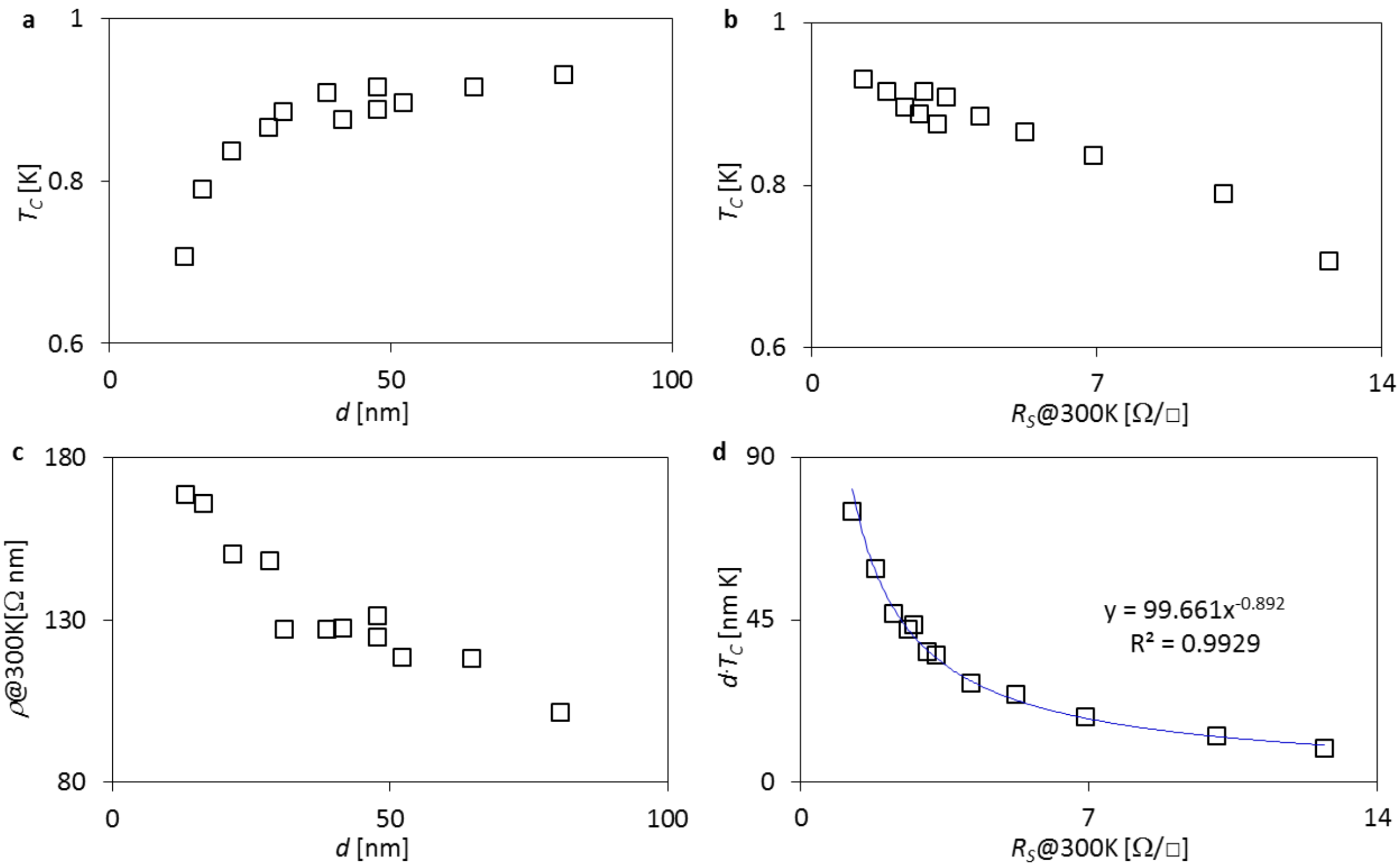

**Figure S6. Superconductivity in molybdenum films (Fàbrega *et al*.** [16]**), △ in Fig. 4.** Critical temperature as a function of (**a**) thickness and (**b**) sheet resistance. (**c**) Resistivity as a function of thickness. (**d**) $d{\cdot}T_C$ vs. $R_S$ fits Eq. 1 with $A$ = 99.661 and $B$ = 0.892.

## 7. Amorphous MoGe.

The significance of amorphous MoGe films to the understanding of superconductivity has been reported as two-fold by Graybeal and co-authors in 1984 [17], 1985 and in three more occasions, [18–20] as well as by Yazdani and Kapitulnik [21]. First, understanding superconductivity in thin α-MoGe films is interesting from the superconducting-insulator transition perspective. Second, thin α-MoGe films exhibit a deviation from conventional superconductivity that is still not understood. Hence, examining the scaling of Eq. 1 for the case of thin α-MoGe films may address these two issues.

In addition to the importance of these films as was identified by Graybeal and co-author, these films were used to demonstrate the validity of the model for homogeneous superconductors by Finkel'stein [22]. Hence, a comparison of the fitting of the data to the power law presented in this work and to Finkel'stein's model is also brought here, demonstrating that the fitting to Eq. 1 is at least comparable to the fitting of the data to Finkel'stein's model, at least for these films. Moreover, to complete the analysis, we also used these data to examine the universal behavior discussed below in Eq. S2b and we demonstrated that such a universal behavior indeed may exist.

Bearing in mind the universality of Fig. 5a in the main text, α-MoGe is at the extreme where the exponent $B$ (and the coefficient $A$) has its highest value. Hence, since the other edge of this curve is defined by aluminum, it is possible that the position of a superconductor on this curve is determined by its homogeneity.

Surprisingly, unlike most other materials, the resistivity of several of the data sets of α-MoGe was found to decrease with decreasing thickness. Yet, the α-MoGe data fit Eq. 1 with a very good agreement. Hence, this may strengthen the hypothesis that Eq. 1 encompasses a broader relation between resistance and thickness in metals in general. This may be surprising when bearing in mind that, so far, the electric properties of α-MoGe have been considered as unique and unexplained.

**7.1. α-MoGe (Graybeal and Beasley 1984** [17]**), × in Fig. 4.**

Graybeal and Beasley reported the suppression of $T_c$ in thin (> 2 nm) homogeneous films of α-MoGe with reduced thickness [17]. In fact, they reported that the suppression correlates very well the increase in $R_s$, as suggested by some of the theories at that time that laid the grounds to the later derivation of the dependence of $T_c$ on $R_s$ by Finkel'stein. [22] Yet, as shown below, we found a very good agreement of the data with the power law of Eq. 1. We should also note that these films were also discussed in a later paper by Graybeal and co-authors [23].

Since this data set is considered a 'classical' example for the well-established Finkel'stein's model, we performed to this data set a somewhat deeper analysis in comparison to the other materials. Specifically, similarly to the analysis done with other materials, in Table S7.1.1 and Fig. S7.1.1 we introduced the analysis of the data with respect to the power law discussed in this work (Eq. 1). However, unlike the analysis done with other materials, here, in Table S7.1.2, we introduced also the values calculated for $T_c$ based on Finkesltein's model (Eq. 13 in Ref. [22]). For calculating these values that are designated by $T_c$_F we used the values reported for $R_s$ by Graybeal and co-authors as well as $\gamma$ = -1/8.2 and $T_{c0}$=7.2 K ($\gamma$ and $T_{c0}$ correspond to the notation used in Ref. [22]). To allow a visual comparison between the different models, we presented in Fig. S7.1.2 the values of the measured $T_c$, the re-calculated values for $T_c$ with the fitting to Eq. 1 ($T_{c_RC}$) and the values for $T_c$ calculated while fitting to Finkel'stein's model ($T_c$_F) as a function of $R_s$. Moreover, in Table S7.1.2 we also calculated the error at percent in $T_c$ for each film: Err $T_c$_F%=100·($T_c$_F – $T_c$)/$T_c$ . This estimation of the error allows us to quantitatively compare the accuracy of the two models (Finkel'stein's model and the power law in and Eq. 1) by comparing the errors in the fitting, *i.e.* to compare the value |Err $T_c$%| with the value |Err_$T_c$_F%| (lower value suggests better accuracy). As shown in both Table S7.1.2 and Fig. S7.1.2 the merit of the fitting of the data of these homogeneous films is at least comparable to the quality of the fit to Finkel'stein's model for this set of films.

We should remind the reader that presumably, Finkel'stein's model is used for the correlation between two values that are measured independently ($R_s$ and $T_c$), while Eq. 1 is using three of such values ($R_s$, $T_c$ and $d$). Moreover, presumably, Finkel'stein's model requires no fitting parameters, as $T_{c0}$ can, in principle be measured directly and $\gamma=1/\ln(T_{c0}\cdot\tau)$ can also be estimated if the relaxation time constant, $\tau$ is measured independently. Nevertheless, typically, both $T_{c0}$ and $\gamma$ are extracted as fitting parameters from the curve $T_c(R_s)$. Likewise, the two parameters $A$ and $B$ in Eq. 1 are also extracted as fitting parameters.

In addition to the comparison to Finkel'stein's model, we used the data of these α-MoGe films to evaluate the fitting of Eq. 1 to the data while using also the possible correlation between the parameters $A$ and $B$ that is introduced below in Eq. S2b. Specifically, it has been proposed that a universal relationship exists between $A$ and $B$. In such a case, $B$ is the only fitting parameter and is the proportionality factor between $\ln(d\cdot T_c/13.7)$ and $\ln(R_s/464)$. Table S7.1.3 presents the fitting value for $B$ with Eq. S2b, as well as the values for the critical temperature that were calculated by feeding back this value of $B$ that was extracted from fitting the data to Eq. S2b as given in Fig. S7.1.3a. We designated these calculated critical temperature values: $T_c$_RC_1Par. In Fig. S7.1.3b we compared these values to the values calculated from Eq. 1, to the values calculated with Finkel'stein's model and to the values reported by Graybeal *et al*. as the raw data [23]. Finally, we calculated the error at percent (ErrT$_c$_RC_1Par%) between the value of the re-calculated critical temperature and the measured value as reported by Graybeal *et al*. These values of the error that are presented in Table S7.1.3 demonstrate a reasonable fitting to such a universal behavior (to Eq. S2b). We would like however to remind the reader that despite the agreement of the one-parameter universal description of Eq. S2b with the data here, such a universal approach should describe the data in general and is not necessarily accurate for each of the individual materials. Yet, we should note that although the scatter in Fig. 5a is larger for materials with $B$ close to unity, where the data is more crowded, we do expect the universality of Eq. S2b to describe the materials with $B$ values at the extrema, *i.e.* materials that are relatively more homogeneous or more granular. The α-MoGe films discussed here are an example for rather homogeneous films. Hence, it is not surprising that their behaviour has a good agreement with Eq. S2b.

**Table S7.1.1. Superconductivity in α-MoGe films (Graybeal and Beasley 1984** [17]**), × in Fig. 4.**

*d*, $T_c$, and $R_\square$ of α-MoGe films of the stoichiometry: 79:21 extracted from Graybeal and Beasley [17]. We present the values calculated for *A*, *B*, $T_{c_RC}$, and Error in $T_{c_RC}$% for these films, as well as the error in the data extraction process. One can notice that the three thickest samples had large mismatch between their extracted values, while the thinnest film was reported by the authors to be inhomogeneous (these four films are highlighted in red). Hence, these films were omitted from the calculations.

| | |
|---|---|
| *A* | 88643 |
| *B* | 1.421 |

| From Figure 2: $R_s$ [Ω/□] | $T_c$ [K] | From Figure 1: $R_s$ [Ω/□] | *d* [nm] | Err $R_s$% | $T_{c_RC}$ [K] | Err $T_{c_RC}$% |
|---|---|---|---|---|---|---|
| 8.575 | 7.35 | 6.901 | 234.545 | -24.254 | | |
| | | 14.374 | 111.289 | | | |
| 30.088 | 6.831 | 29.459 | 55.321 | -2.134 | | |
| 61.718 | 6.708 | 63.993 | 25.31 | 3.554 | | |
| 133.932 | 6.512 | 134.685 | 12.324 | 0.559 | 6.833 | 4.917 |
| 165.948 | 5.994 | 164.752 | 10.133 | -0.726 | 6.128 | 2.233 |
| 209.84 | 5.607 | 209.822 | 7.955 | -0.009 | 5.592 | -0.258 |
| 291.444 | 5.015 | 292.471 | 5.741 | 0.351 | 4.858 | -3.118 |
| 391.014 | 4.463 | 391.71 | 4.391 | 0.178 | 4.184 | -6.24 |
| 472.499 | 3.992 | 472.795 | 3.595 | 0.063 | 3.905 | -2.192 |
| 612.593 | 3.429 | 611.156 | 2.897 | -0.235 | 3.351 | -2.266 |
| 865.614 | 2.465 | 863.818 | 2.201 | -0.208 | 2.698 | 9.488 |
| 1382.976 | 1.235 | 1382.477 | 1.499 | -0.036 | | |

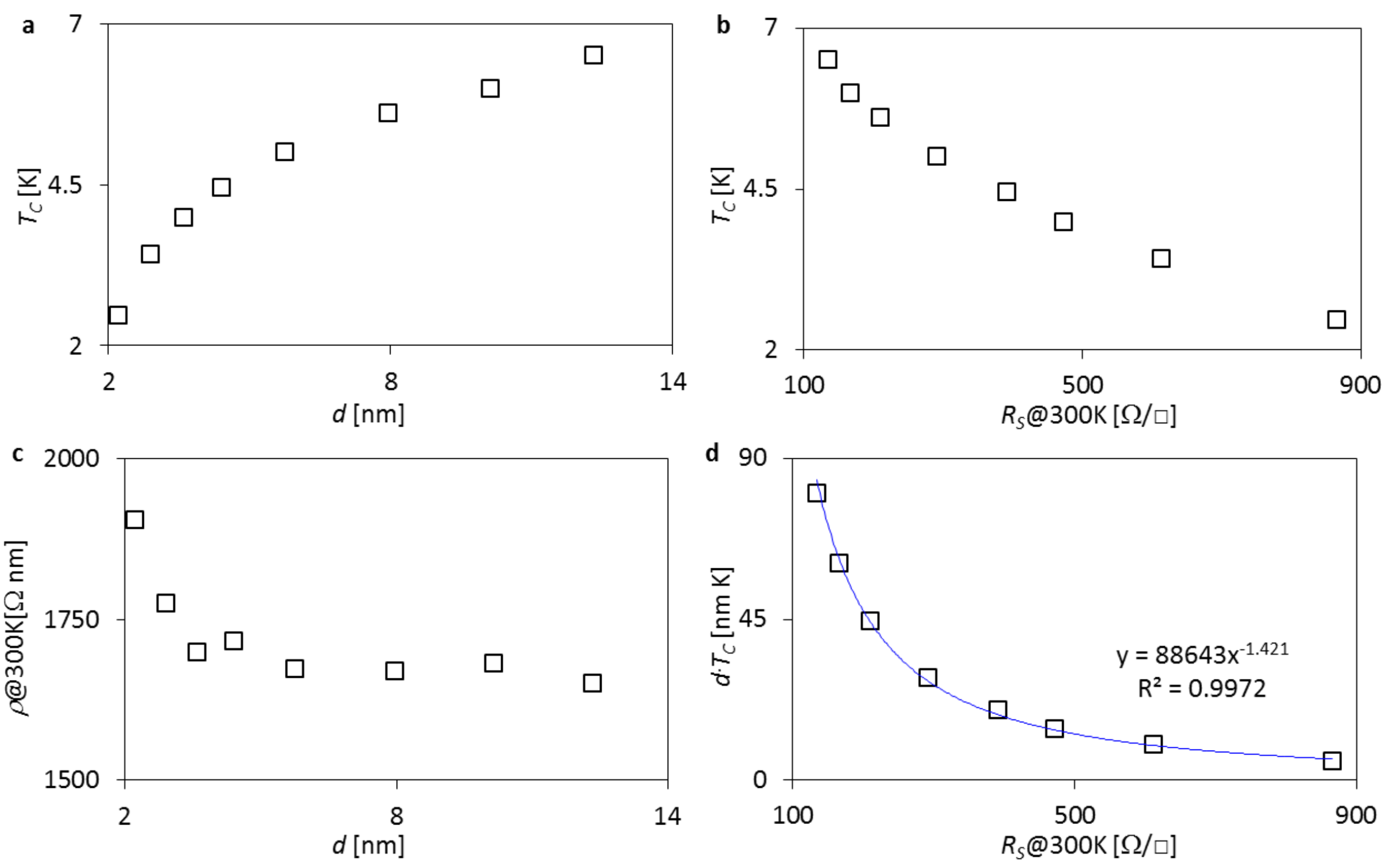

**Figure S7.1.1. Superconductivity in α-MoGe films (Graybeal and Beasley** [17]**), ✕ in Fig. 4.** Critical temperature as a function of (**a**) thickness and (**b**) sheet resistance. (**c**) Resistivity as a function of thickness. (**d**) $d{\cdot}T_c$ vs. $R_s$ with a best fit to Eq. 1 with $A$ = 88643 and $B$ = 1.421.

**Table S7.1.2. Comparison of Eq. 1 with the model reported in Ref.** [22] **for α-MoGe films (Graybeal and Beasley 1984** [17]**), ✕ in Fig. 4.**

$d$, $T_c$, and $R_{\square}$ of α-MoGe films of the stoichiometry: 79:21 extracted from Graybeal and Beasley [17]. We present a comparison of the discovered scaling law to the analysis of the data with Finkel'stein's model. We calculated the $T_c$ value with Finkel'stein's model ($T_c$_F) by using $T_{c0}$=7.2 K and γ=-1/8.2, as specified in Ref. [22]. We also presented the error at percent between the fit value for the critical temperature with this model ($T_c$_F) and the actual data that was extracted from Ref. [17] and we designated this error by: Err$T_c$_F%. Therefore, a quantitative comparison between Eq. 1 and the model presented in Ref. [22] can be done by comparing the errors in Err Tc_RC% and in ErrTc_F%

| Raw data | | | | Fit to Eq. 1 | | Fit to Ref. [22] | |
|---|---|---|---|---|---|---|---|
| | | | | A= | 88643 | Tc0= | 7.2 K |
| | | | | B= | 1.421 | γ= | -1/8.2 |
| $R_s$ [Ω/□] | d [nm] | $T_c$ [K] | Err $R_s$% | $T_{c_RC}$ [K] | Err $T_c$_RC% | $T_c$_F | Err$T_c$_F% |
| 8.575 | 234.545 | 7.35 | -24.254 | | | | |
| | 111.289 | | | | | | |
| 30.088 | 55.321 | 6.831 | -2.134 | | | | |
| 61.718 | 25.31 | 6.708 | 3.554 | | | | |
| 133.932 | 12.324 | 6.512 | 0.559 | 6.833 | 4.917 | 6.316 | -3.016 |
| 165.948 | 10.133 | 5.994 | -0.726 | 6.128 | 2.233 | 6.119 | 2.085 |
| 209.84 | 7.955 | 5.607 | -0.009 | 5.592 | -0.258 | 5.855 | 4.43 |
| 291.444 | 5.741 | 5.015 | 0.351 | 4.858 | -3.118 | 5.38 | 7.282 |
| 391.014 | 4.391 | 4.463 | 0.178 | 4.184 | -6.24 | 4.83 | 8.234 |
| 472.499 | 3.595 | 3.992 | 0.063 | 3.905 | -2.192 | 4.405 | 10.335 |
| 612.593 | 2.897 | 3.429 | -0.235 | 3.351 | -2.266 | 3.723 | 8.585 |
| 865.614 | 2.201 | 2.465 | -0.208 | 2.698 | 9.488 | 2.652 | 7.605 |
| 1382.976 | 1.499 | 1.235 | -0.036 | | | | |

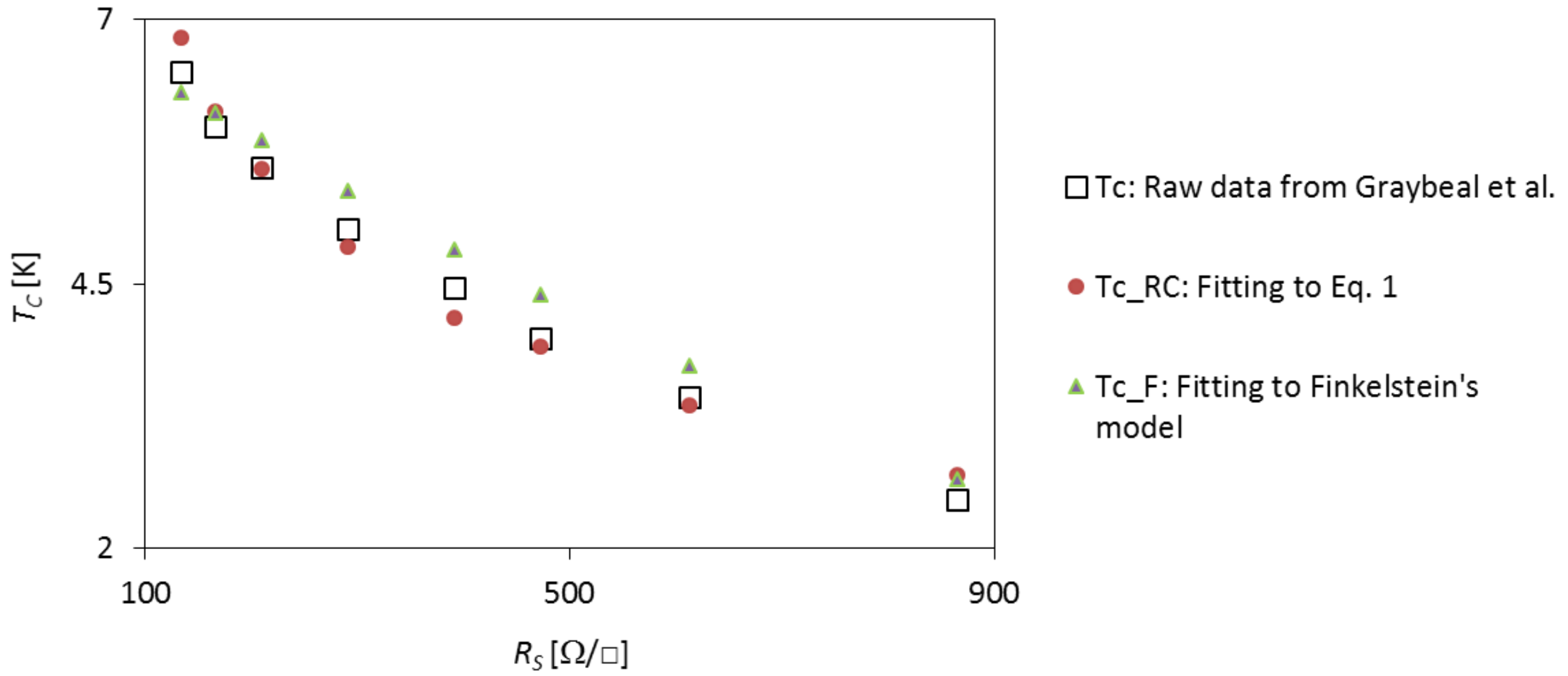

**Figure S7.1.2. Comparison between Eq. 1 and Finkel'stein's model for α-MoGe films (Graybeal and Beasley** [17]**), ✕ in Fig. 4.**

Comparing the power law in Eq. 1 to Finkel'stein's model [22] by presenting the $T_c$ values calculated with each model (*i.e.* $T_{c_RC}$ and $T_c$_F from Table S.7.1.2), as well as the raw data for $T_c$ as was extracted from Graybeal et al. The comparison suggests that the fitting of the data to Eq. 1 is at least comparable to the fitting of the data to the model of Ref. [22].

**Table S7.1.3. Fitting the data of α-MoGe films (Graybeal and Beasley 1984** [17], **× in Fig. 4) to the one-free-parameter universal formula given in Eq. S2b.**

Extracting the value $B$=1.5946 from fitting the proportional terms: ln($d{\cdot}T_c$/13.7) and ln($R_s$/464) as discussed in Eq. S2b and using this value to estimate the values of the critical temperature calculated from this model based on this value of $B$ ($T_c$_RC_1Par). The error at percent in the estimated critical temperature is also presented (Err$T_c$_RC_1Par%), demonstrating a reasonable fitting of the data to the universal behaviot of Eq. S2b. Nevertheless, we refer the reader to our remark in the head of this section (S7.1) with regard to the accuracy of this analysis for the general set of data. Red points are omitted from the fitting as explained above.

| **B=** | **1.4574** | Fit to Eq. S2b | |
|---|---|---|---|
| ln(dTc/13.7) | ln(Rs/464) | TcRC1Par | ErrTcRC1Par% |
| 4.83 | -3.99 | 19.61 | 166.81 |
| 3.32 | -2.74 | 13.35 | 95.44 |
| 2.52 | -2.02 | 10.24 | 52.65 |
| 1.77 | -1.24 | 6.8 | **4.42** |
| 1.49 | -1.03 | 6.05 | **0.93** |
| 1.18 | -0.79 | 5.47 | **-2.44** |
| 0.74 | -0.47 | 4.7 | **-6.28** |
| 0.36 | -0.17 | 4 | **-10.37** |
| 0.05 | 0.02 | 3.71 | **-7.07** |
| -0.32 | 0.28 | 3.15 | **-8.13** |
| -0.93 | 0.62 | 2.51 | **1.84** |
| -2 | 1.09 | 1.86 | 50.63 |

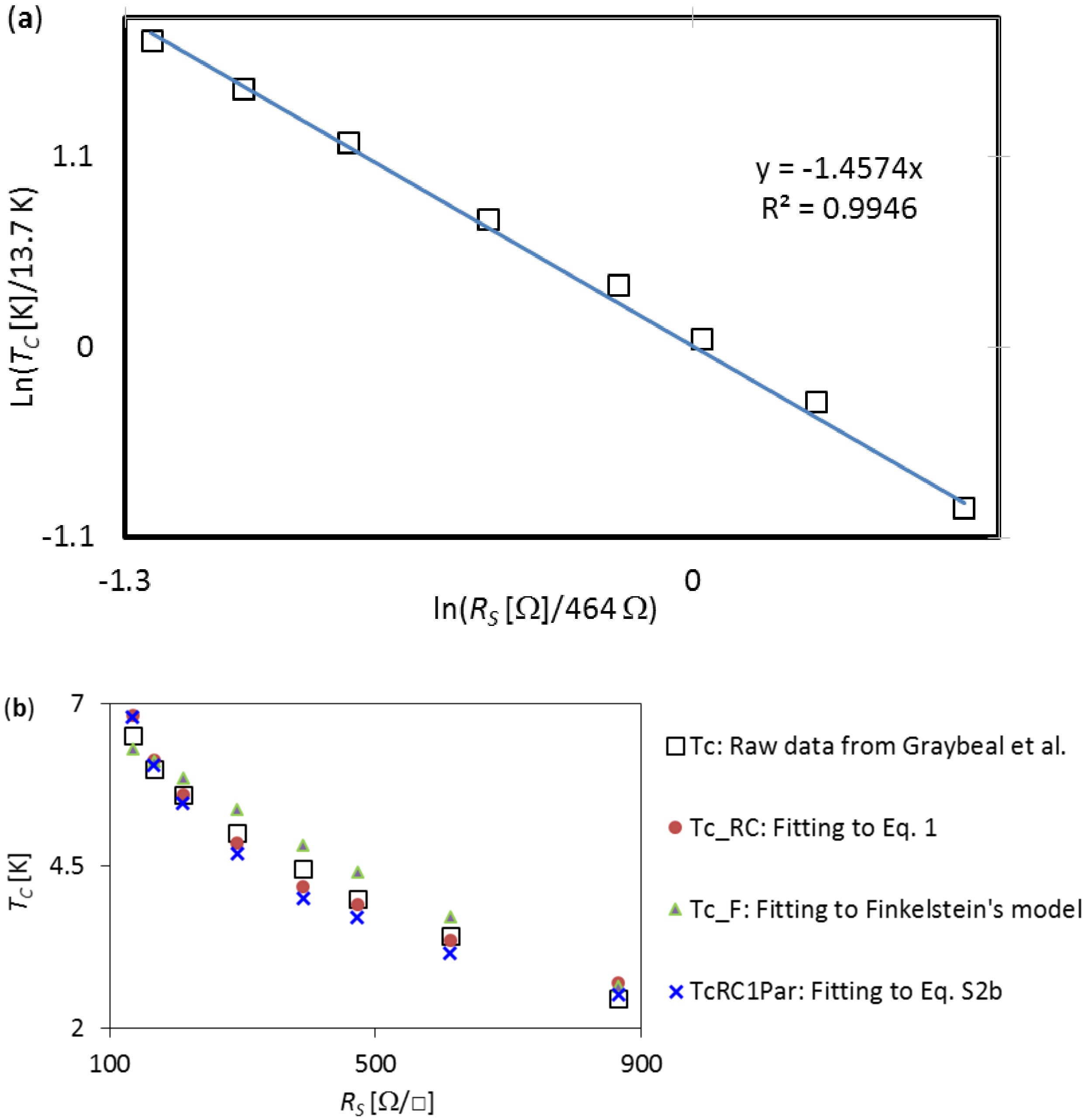

**Figure S7.1.3. Analyzing α-MoGe films (Graybeal and Beasley** [17], ✕ **in Fig. 4) with the different models.**

(**a**) Extracting *B* from Eq. S2b from a ln($T_c$ [k]/13. 7 K) vs. ln($R_s$ [Ω]/464 Ω) plot. (**b**) Comparing the values obtained by the different fittings (Eq. 1, Finkel'stein's model and Eq. S2b, as designated in the legend) to the values reported by Graybeal *et al*.

## 7.2. α-MoGe (Yazdani and Kapitulnik [21]) (◆ in Fig. 4).

Yazdani and Kapitulnik suggested that the properties of α-MoGe films are sample dependent [21]. They reported two sets of films with different stoichiometries. However, neither of these sets included enough films for our analysis (2 films of α-$Mo_{21}$Ge and three films of α-$Mo_{43}$Ge were reported). We combined these two sets, bearing in mind that the scattering may be large. Yet, we found that the data fit Eq. 1 rather well and much better than the other scaling options $T_c(d)$, $T_c(R_s)$ or $\rho(d)$ (Fig. S7.2). It is interesting to note that the values for $A$ and $B$ are consistent with the fact that αMoGe is at the extreme right of the curve in Fig. 5a. In fact, these values for $A$ and $B$ are also consistent with the linearity in Fig. 5a, strengthening the assumption that $A$ and $B$ are correlated so that each material may be defined by one free parameter only. Finally, we used $A$ and $B$ to calculate the resistance as a function of $T_C$ and $d$ and found the fitting rather valuable ($R_{N_RC}$ in Table S7.1). The error in this re-calculated value for the resistance with respect to the reported resistance (Err $R_{N_RC}$) is also presented in Table S6.1.

We should note that here, the dependence of $T_c$ and of ρ on film thickness is rather peculiar. We are not sure about the origin of this behavior. However, one possible explanation that can be considered is that the origin of this behavior stems from variations in the film stoichiometry.

**Table S7.2. Superconductivity in α-MoGe films from Yazdani and Kapitulnik** [21] **(◆ in Fig. 4).**

$d$, $T_c$, and $R_s$ of α-MoGe films with different stoichiometries extracted from Yazdani and Kapitulnik [21]. We also present the values calculated for $A$, $B$, $T_{c_RC}$, and Error in $T_{c_RC}$%.

| | | | | *A* | 2192805 | | |
|---|---|---|---|---|---|---|---|
| | | | | *B* | 1.96 | | |
| | $d$ [nm] | $R_N$ [Ω/□] | $T_c$ [K] | $T_{c_RC}$ [K] | *Err* $T_{c_RC}$% | $R_{N_RC}$ [K] | Err $R_{s_RC}$% |
| α-Mo21Ge | 7 | 1980 | 0.1 | 0.108 | 8.25 | 2060.946 | 4.088 |
| α-Mo21Ge | 8 | 1710 | 0.15 | 0.126 | -15.83 | 1566.05 | -8.418 |
| α-Mo43Ge | 3 | 1400 | 0.5 | 0.498 | -0.35 | 1397.532 | -0.176 |
| α-Mo43Ge | 4 | 951 | 1.01 | 0.797 | -21.04 | 842.997 | -11.357 |
| α-Mo43Ge | 6 | 658 | 1.02 | 1.094 | 7.28 | 682.025 | 3.651 |

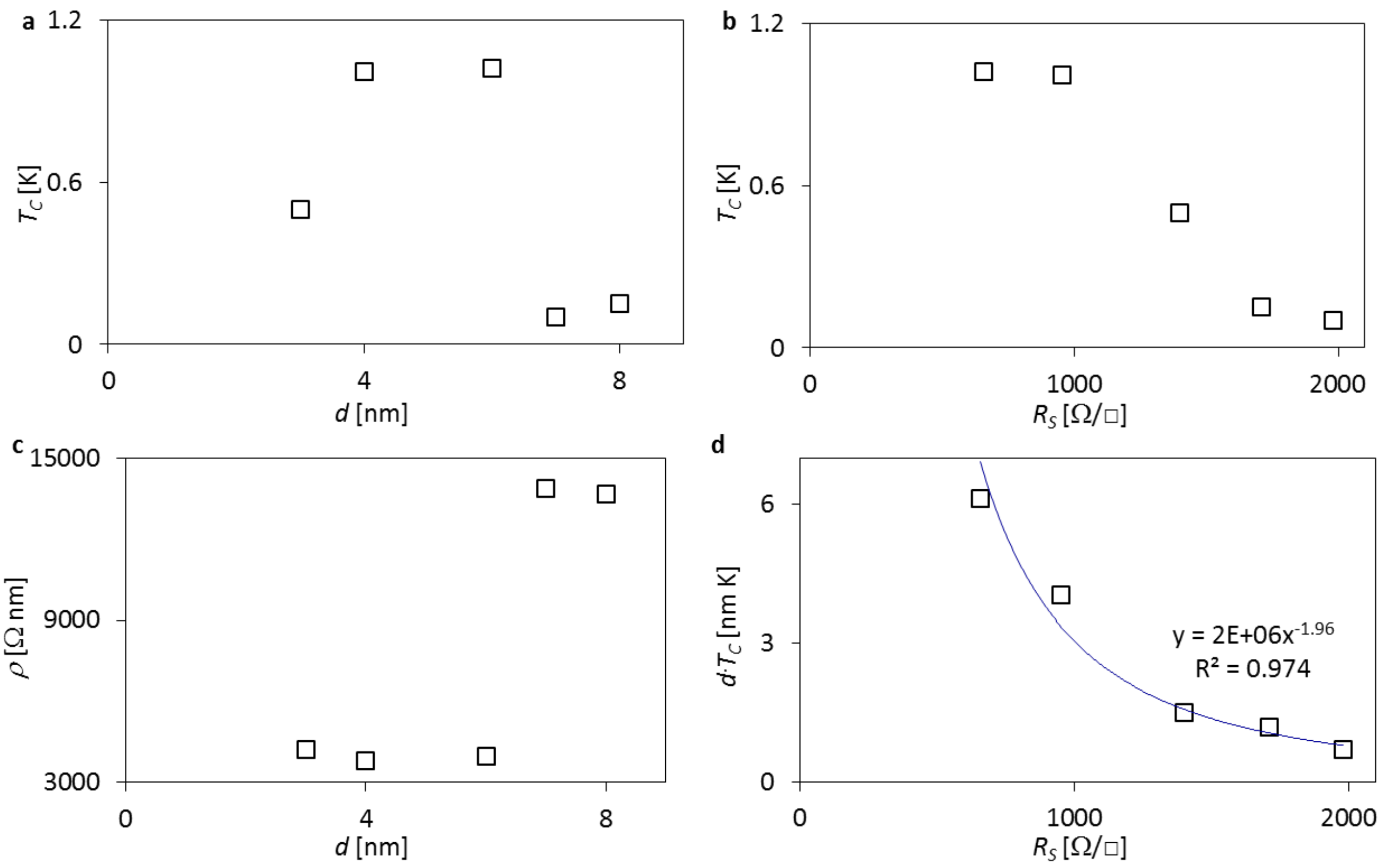

**Figure S7.2. Superconductivity in α-MoGe films from Yazdani and Kapitulnik** [21] **(◆ in Fig. 4).**

Critical temperature as a function of (**a**) thickness and (**b**) sheet resistance indicates enhancement of $T_C$ with decreasing thickness. (**c**) Resistivity as a function of thickness demonstrates a decrease in resistivity with decreasing film thickness. (**d**) $d{\cdot}T_c$ vs. $R_s$ with a best fit to Eq. 1 with $A$ = 2192805 and $B$ = 1.9601.

**7.3. Amorphous MoGe- extracted from Graybeal and co-authors** [18–20] **(▶ in Fig. 4).**

Below we present the data for α-MoGe films collected from different reports by Graybeal and co-authors [18–20], which demonstrate excellent agreement with Eq. 1.

**Table S7.3. Superconductivity in α-MoGe films (Graybeal and co-authors** [18–20]**), ▶ in Fig. 4.** *d*, $T_c$, and $R_s$ of α-MoGe films extracted from Graybeal and co-authors [18–20] as well as the values calculated for *A*, *B*, $T_{c_RC}$, and Error in $T_{c_RC}$%. Since the data points were matched through a common $T_c$ value and through a common thickness value, the difference in $T_c$ and *d* extracted from the two panels (three datasets) are also added ('Err $T_c$%' and 'Err *d*%'). The table includes data merged from the tables in two different publications by Graybeal and co-authors [18,19], while considering their complimentary information [20].

| | | | | |
|---|---|---|---|---|
| | | | *A* | 85126 |
| | | | *B* | 1.389 |
| *d* [nm] | $T_c$ [K] | $R_s$ [Ω/□] | $T_{c_RC}$ | Err $T_{c_RC}$% |
| 6.1 | 5.442 | 287 | 5.379 | 1.15 |
| 4.6 | 4.92 | 387 | 4.709 | 4.279 |
| 2.75 | 3.734 | 674 | 3.645 | 2.378 |
| 2.15 | 2.999 | 885 | 3.194 | -6.498 |
| 8.3 | 4.5 | 260 | 4.535 | -0.78 |
| 16.5 | 6.1 | 131 | 5.911 | 3.092 |
| 33 | 6.9 | 69 | 7.2 | -4.36 |

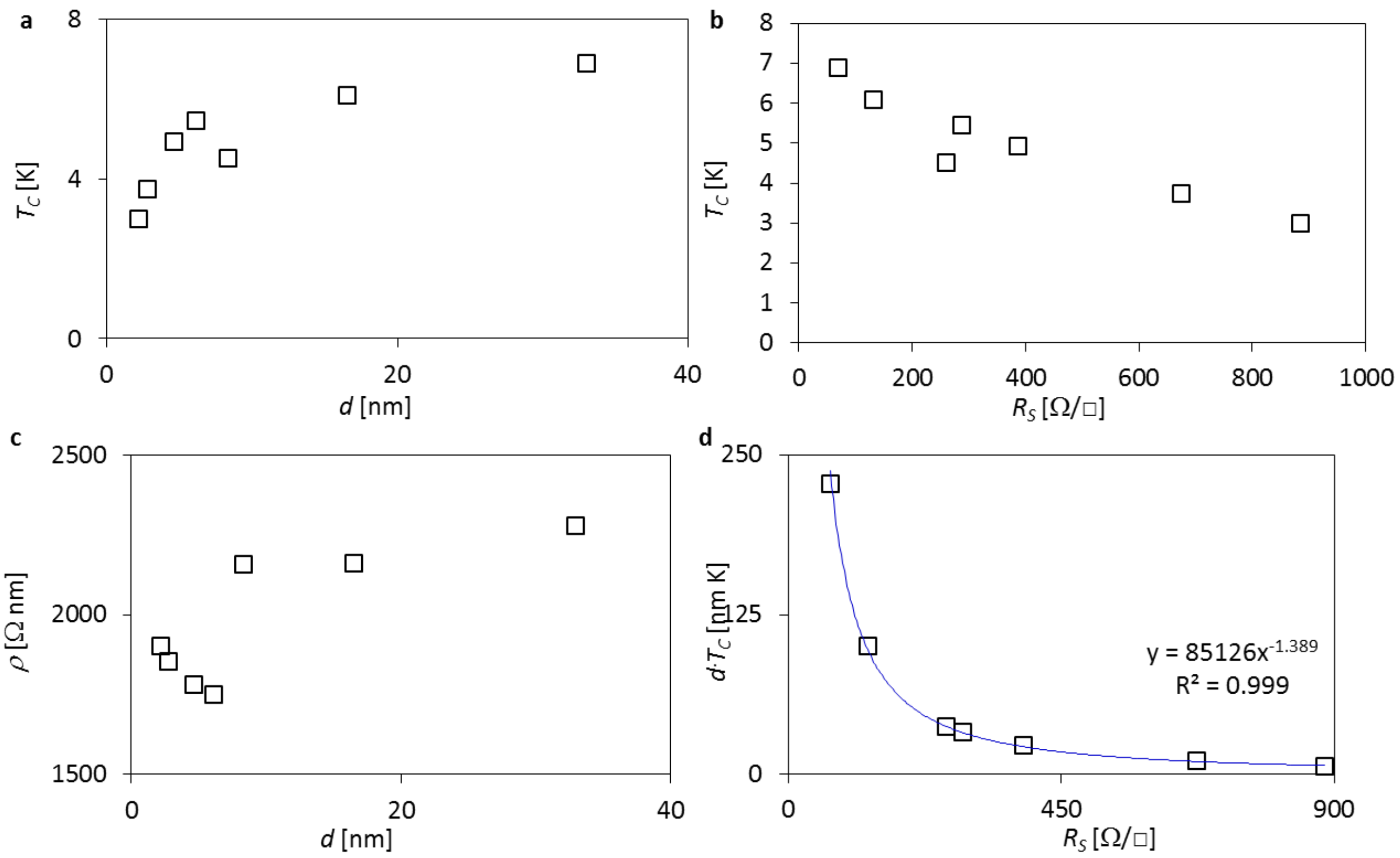

**Figure S7.3. Superconductivity in α-MoGe films (Graybeal and co-authors** [18–20]**), ▶ in Fig. 4.** Critical temperature as a function of (**a**) thickness and (**b**) sheet resistance indicates enhancement of $T_C$ with decreasing thickness. (**c**) Resistivity as a function of thickness demonstrates a decrease in resistivity with decreasing film thickness. (**d**) $d{\cdot}T_c$ vs. $R_s$ fits Eq. 1 with $A$ = 85126 and $B$ = 1.389.

**8. Nb- extracted from Gubin *et al.*** [24] **(◀ in Fig. 4).**

Gubin *et al*. studied the suppression of superconductivity in thin Nb films on Si substrates with and without the influence of an external magnetic field, suggesting that the mechanism governing superconductivity in these materials is the proximity effect ($T_c = T_c(d)$) [24]. We extracted their data and found that thin Nb films also fit the empirical power law of Eq. 1. However, looking carefully at the data, one can see that, although the error in the fit is low for most of the data points, it is exceptionally high for one specific data point (highlighted in red in Table S7). Hence, for the sake of data analysis, we fitted the data twice, once with this data point and once without it. This film also varies from the trend when looking at the thickness dependence of the resistivity (Fig. S8c), suggesting that it is different than the other films in this set. The recalculated values were found to be significantly improved when this data point was not considered for the fitting. Here we present the two fitting options in Table S7, while in Fig. S7 we present the fitting curve that included all the data points.

**Table S8. Superconductivity in Nb films (Gubin *et al.*** [24]**), ◀ in Fig. 4.** $d$, $T_c$, and $R_s$ of Nb films extracted from Gubin *et al.* [24] (Fig. 1 therein) as well as the values calculated for $A$, $B$, $T_{c_RC}$, and Error in $T_{c_RC}$%. Since the data points were matched through a common a common thickness value, the difference in the extracted $d$ is also added ('Error in $d$%'). The data analysis was done with and without the data point highlighted in red, and both calculations are presented.

| | | | | | With Red Data point | | Without Red Data point | |
|---|---|---|---|---|---|---|---|---|
| | | | | | $A$ | 611.38 | $A$ | 594.13 |
| | | | | | $B$ | 0.761 | $B$ | 0.759 |
| $d$ [nm] | $T_c$ [K] | $d$ [nm] | $R_s$ [Ω/□] | Err $d$% | $T_{c_RC}$ | Err $T_{c_RC}$% | $T_{c_RC}$ | Err $T_{c_RC}$% |
| 7.405 | 6.134 | 7.387 | 29.799 | -0.24 | 6.236 | 1.67 | 6.103 | -0.52 |
| 9.067 | 6.841 | 9.173 | 19.295 | 1.17 | 7.089 | 3.63 | 6.93 | 1.31 |
| 11.334 | 7.127 | 11.28 | 14.199 | -0.47 | 7.162 | 0.5 | 6.997 | -1.82 |
| 13.148 | 7.593 | 13.23 | 10.397 | 0.63 | 7.827 | 3.08 | 7.642 | 0.64 |
| 19.646 | 7.944 | 18.961 | 5.925 | -3.49 | 8.035 | 1.15 | 7.836 | -1.36 |
| 49.266 | 8.802 | 49.781 | 2.061 | 1.04 | 7.157 | -18.69 | 6.965 | -20.87 |
| 99.892 | 9.096 | 99.774 | 0.541 | -0.12 | 9.773 | 7.44 | 9.485 | 4.28 |
| 199.633 | 9.477 | 199.66 | 0.224 | 0.01 | 9.574 | 1.02 | 9.276 | -2.12 |
| 300.13 | 9.723 | 300.152 | 0.124 | 0.01 | 9.9835 | 2.68 | 9.661 | -0.63 |

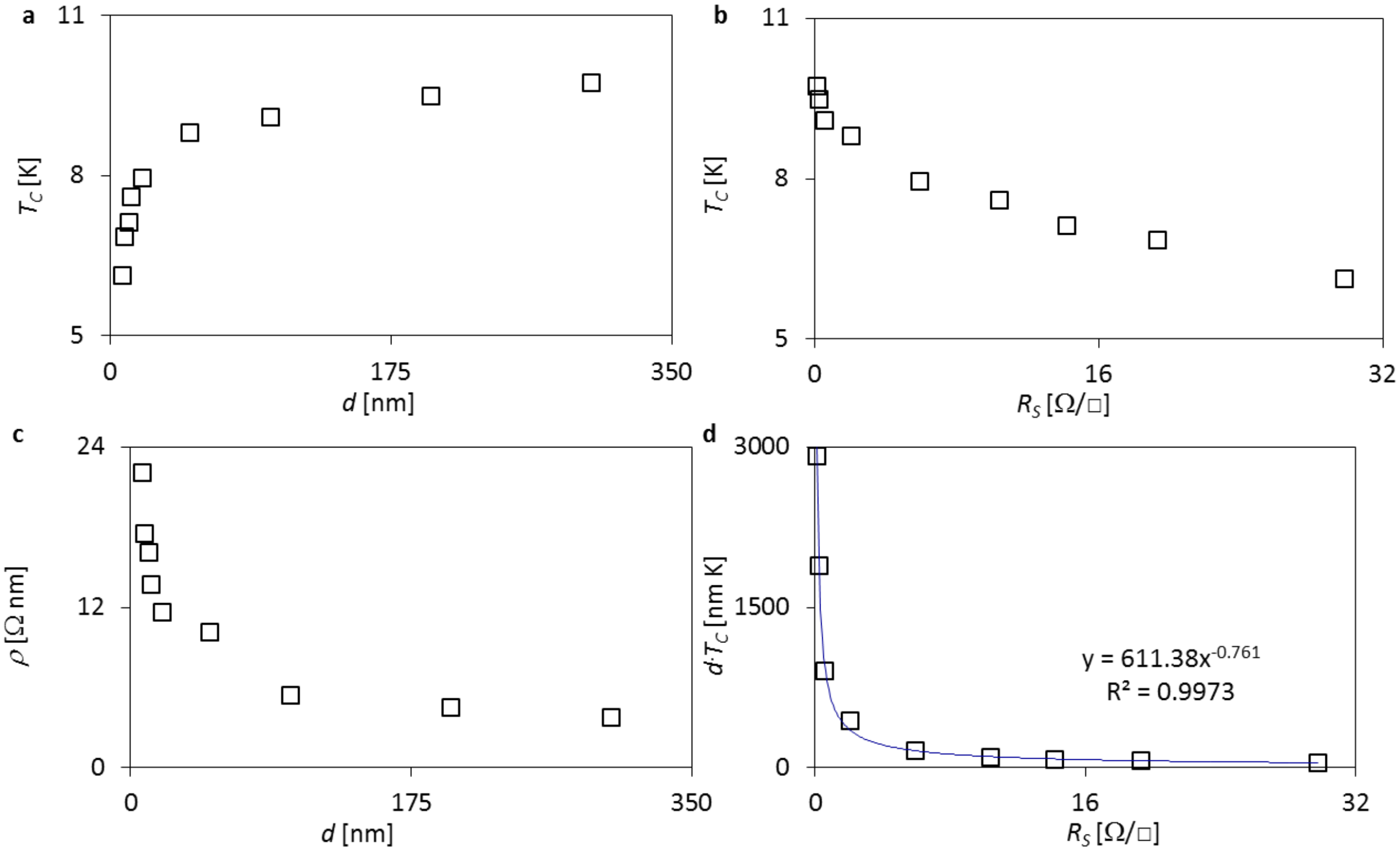

**Figure S8. Superconductivity in Nb films (Gubin *et al.* [24]), ◀ in Fig. 4.** Critical temperature as a function of (**a**) thickness and (**b**) sheet resistance indicates enhancement. (**c**) Resistivity as a function of thickness. (**d**) $d{\cdot}T_c$ vs. $R_s$ fits to Eq. 1 with $A$ = 611.38 and $B$ = 0.761. Data presented in (d) includes he red data point in Table S8.

## 9. Amorphous $Nb_3Ge$- extracted from Kes and Tsuei [25] (▷ in Fig. 4).

Kes and Tsuei studied superconductivity in α-$Nb_3Ge$ films [25]. These films cannot be considered thin, as they are ~0.5-3 μm thick. We found that these films scale with $d{\cdot}T_C$ vs. $R_S$ better than they scale, for instance, with $T_c(R_s)$, or with $T_c(d)$ or $\rho(d)$. Nevertheless, we believe that the reason for this is mainly because the changes in $T_c$ for these films are very small (a 10% maximum difference in $T_c$ between the films, which is comparable to the empirical ±5% typical error of the scaling of Eq. 1). Moreover, in these films, $R_s$ is almost inversely proportional to $d$ rather convincingly. Hence, given the scaling of $R_s$ with $d$, and given that the exponent $B$ of these films is close to unity, one may wonder whether the origin of the observed scaling in this particular case is mainly due to the electrical properties of the films in the normal state. It should be noted that, in most sets of films with $B$ = 1, the scaling of Eq. 1 cannot be explained only with the inverse relations of $R_s(d)$ because the scattering is usually reduced for $d{\cdot}T_c(R_s)$, instead of increasing as it would have been if the reason for this scaling were the inverse relations of $R_s(d)$. In fact, this may still be the case for α-$Nb_3Ge$, but data for thinner films that enable the examination of the scaling are unavailable to us.

**Table S9. Superconductivity in α-$Nb_3Ge$ (Kes and Tsuei** [25]**), ▷ in Fig. 4.** $d$, $T_c$, and $R_s$ of α-$Nb_3Ge$ films extracted from Gubin *et al.* [24] (Fig. 1 therein) as well as the values calculated for $A$, $B$, $T_{c_RC}$, and Error in $T_{c_RC}$%. Since the data points were matched through a common thickness value, the difference in the extracted $d$ is also added ('Error in $d$%').

| | | | | |
|---|---|---|---|---|
| | | | $A$ | 6658.3 |
| | | | $B$ | 1.032 |
| $d$ [nm] | $T_c$ [K] | $R_s$ [Ω/□] | $T_{c_RC}$ | Err $T_{c_RC}$% |
| 2920 | 4.25 | 0.565 | 4.11 | -3.3 |
| 1240 | 3.99 | 1.266 | 4.21 | 5.49 |
| 620 | 3.86 | 2.645 | 3.94 | 1.96 |
| 460 | 4 | 3.609 | 3.85 | -3.76 |

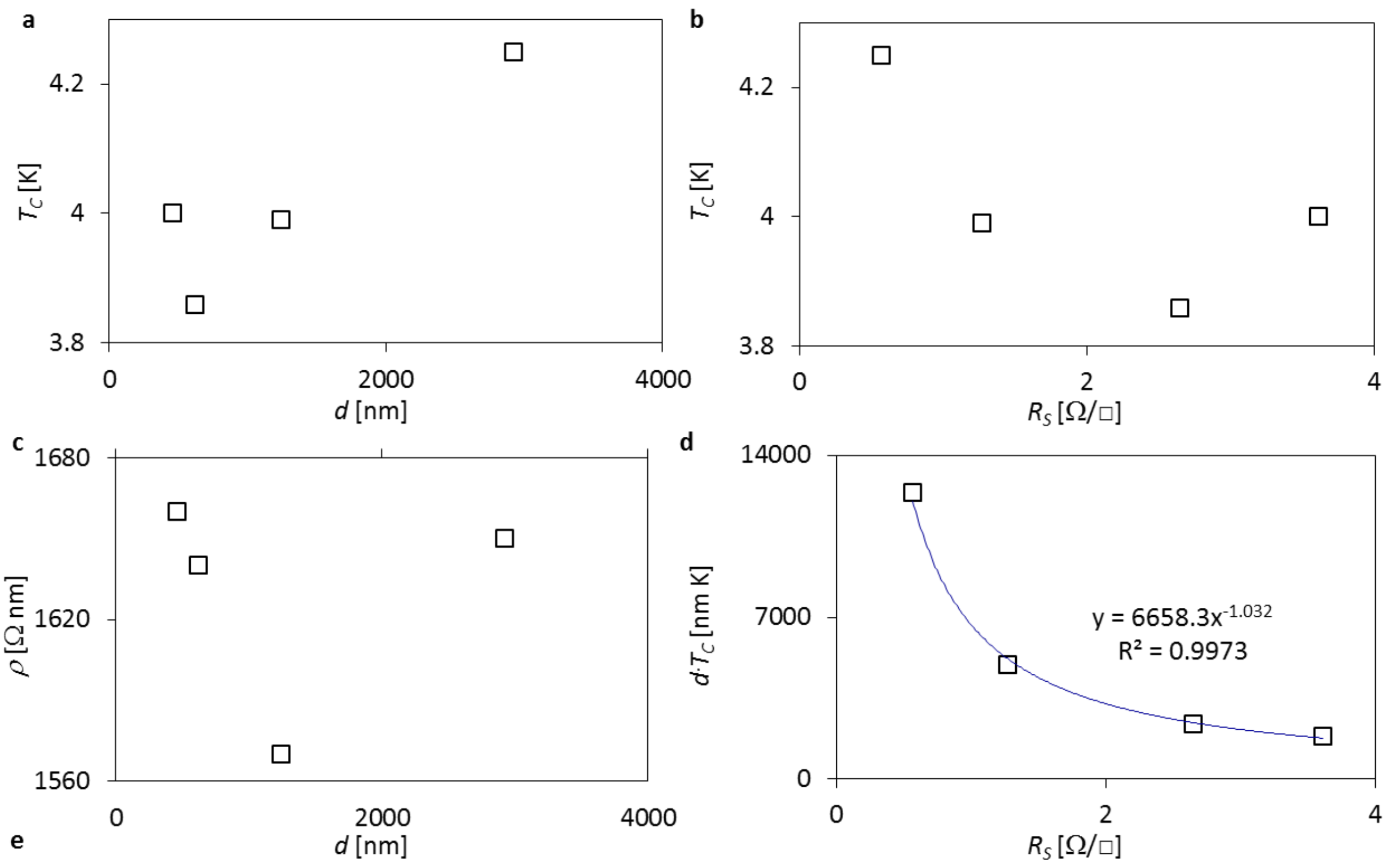

**Figure S9. Superconductivity in α-$Nb_3Ge$ (Kes and Tsuei** [25]**), ▷ in Fig. 4.** Critical temperature as a function of (**a**) thickness and (**b**) sheet resistance. (**c**) Resistivity as a function of thickness demonstrates a decrease in resistivity with decreasing film thickness. (**d**) $d{\cdot}T_c$ vs. $R_s$ fits to Eq. 1 with $A = 6658.3$ and $B = 1.032$.

**10. $Nb_3Sn$- extracted from Orlando *et al.*** [26]**.**

Orlando *et al.* studied superconductivity in what they described as highly damaged or highly defected $Nb_3Sn$ films. This is one of the only examples where the scaling $d{\cdot}T_c$ vs. $R_s$ does not seem to work. We should mention here that, in addition to the defects in the films, the authors also suggested: "since the samples were deposited in a 'compositional phase spread' configuration, the unpatterned samples vary to some degree in composition across the films." [26] In addition, except for one film, all the reported films had a rather constant $T_C$ independent of the thickness, suggesting that they are not in the two-dimensional limit. Lastly, the films were examined over the course of two calendar years, which may have allowed their degradation. Hence, we do not believe that the fact that these films do not agree with the proposed scaling invalidates it. Yet, we present the data for these films here.

**Table S10. Superconductivity in $Nb_3Sb$ (Orlando *et al.*** [26]**).** *d*, *Tc*, and $R_S$ of $V_3Si$ films extracted from Orlando *et al.* [26] (Table 1 therein).

| *d* [nm] | $R_s$ [Ω/□] | $T_S$ [K] |
|---|---|---|
| 210 | 0.762 | 17.9 |
| 730 | 0.121 | 17.9 |
| 260 | 0.654 | 17.8 |
| 510 | 0.704 | 16.1 |

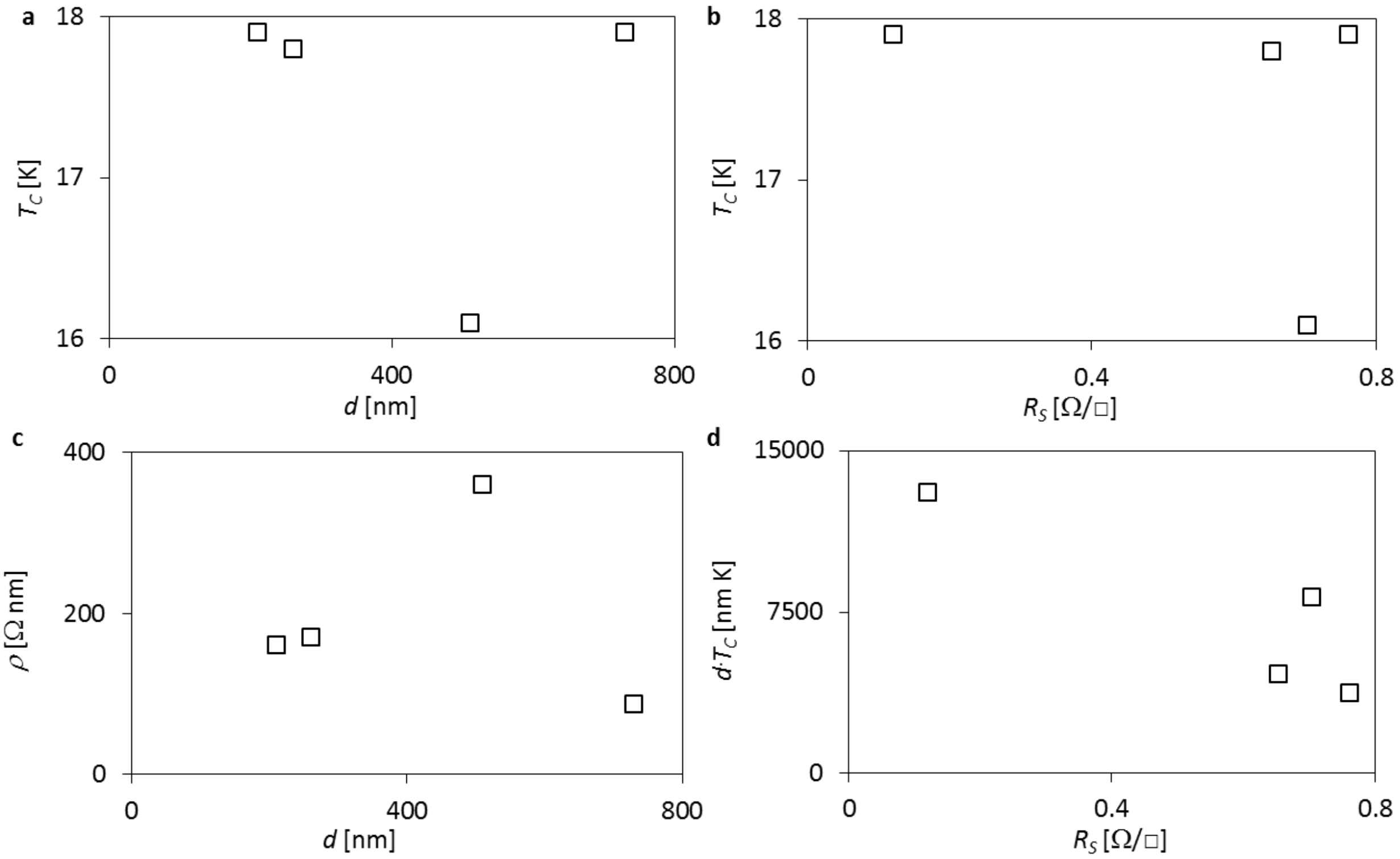

**Figure S10. Superconductivity in $Nb_3Sn$ (Orlando *et al.*** [26]**).** Critical temperature as a function of (**a**) thickness and (**b**) sheet resistance. (**c**) Resistivity as a function of thickness. (**d**) $d{\cdot}T_c$ vs. $R_s$ does not fit the highly-defected and inhomogeneous thick films well.

## 11. NbN.

As mentioned in the main text, superconductivity in NbN is appealing from both technological and fundamental science perspectives. In addition, each of the following datasets is reported to fit a different scaling. Hence, there is also a need to seek a general mechanism that can describe all these datasets. Although the source of the scaling proposed here and that of the empirical power law of Eq. 1 is unknown, demonstrating universal behavior empirically is essential in the process of seeking a general mechanism.

**11.1. NbN- extracted from Wang and co-authors** [27–29] **(■ in Fig. 4).**

We merged data from three different works by Wang and co-workers, in which they grew NbN on MgO substrates [27–29]. In their latest work, the authors found that the suppression of $T_c$ scales with $R_s$ in accordance to Finkel'stein's theory [22]. Indeed, Fig. S11.1b demonstrates a relatively low level of scatter. Yet, we found that these data can also be fitted by Eq. 1 with good agreement. It should be noted that in two of the samples the error in the extracted values was higher than in the others. The authors also reported lower confidence in the thickness values for the thinner films (highlighted in red in Table S11.1). We included these values and made the fitting with and without them, as well as only for the older report [27], only for the second report [28] and for the first and latter works [27,29]. Moreover, we presented the $d{\cdot}T_c$ vs. $R_s$ curves for these three datasets separately in Fig. S11.1. We should mention here that, although an inverse proportional relation between $R_s$ and $d$ was suggested, this by itself cannot explain the fitting to Eq. 1, with which the data was found to be in better agreement.

Finally, it should be noted that, although the authors reported some of the thickness values explicitly, not all the films were identified, so we extracted them from their figures, as detailed in Table S11.1.

**Table S11.1. Superconductivity in NbN (Wang and co-authors [27–29]), ■ in Fig. 4.** *d*, $T_c$, and $R_s$ of NbN films extracted from Wang and co-authors [27–29] (Fig. 3 in the older reference [27], Fig. 3 in [28] and Fig. 1 and 4 in the newer report [29]) as well as the values calculated for *A*, *B*, $T_{c_RC}$, and Error in $T_{c_RC}$%. Since the data points were matched through a common critical temperature value, the difference in the extracted $T_c$ in the second dataset is also added ('Err in $T_c$%'). The data analysis was done with and without the data points for which the authors reported low confidence of their *d* values by the authors (highlighted in red), as well as due to the data extraction process (highlighted in blue), and both calculations are presented. Older data [27] are separated from newer data [29] by a horizontal line, while the report from 2002 [28] is brought below the data points, in the end of the table.

| | | | | | With red values Old and New [27,29] | | Without red values Old and New [27,29] | | Older paper only [27] | |
|---|---|---|---|---|---|---|---|---|---|---|
| | | | | | *A* | 12141 | *A* | 10471 | *A* | 10843 |
| | | | | | *B* | 1.041 | *B* | 0.9935 | *B* | 0.984 |
| *d* [nm] | $T_c$ [K] | $R_s$@20K [Ω/□] | $T_c$[K] | Err $T_c$% | $T_{c_RC}$[K] | Err $T_c$% | $T_{c_RC}$ [K] | Err $T_c$% | $T_{c_RC}$ [K] | Err $T_c$% |
| 5 | 12.5 | 174 | | | 11.29 | -9.68 | 12.45 | -0.4 | 13.54 | 8.32 |
| 9 | 14 | 95.556 | | | 11.71 | -16.36 | 12.54 | -10.43 | 13.56 | -3.14 |
| 17 | 14.5 | 50 | | | 12.17 | -16.07 | 12.64 | -12.83 | 13.58 | -6.34 |
| 40 | 15.5 | 18.75 | | | 14.35 | -7.42 | 14.23 | -8.19 | 15.15 | -2.26 |
| 85 | 15.5 | 8.412 | | | 15.56 | 0.39 | 14.85 | -4.19 | 15.69 | 1.23 |
| 175 | 15.6 | 4 | | | 16.39 | 5.06 | 15.09 | -3.27 | 15.84 | 1.54 |
| 340 | 15.9 | 2.03 | | | 17.09 | 7.48 | 15.25 | -4.09 | 15.89 | -0.06 |
| 700 | 16 | 0.957 | | | 18.15 | 13.44 | 15.62 | -2.38 | 16.17 | 1.06 |
| 19.76 | 14.48 | 34.3 | 14.486 | 0.04 | 15.5 | 7.04 | 15.81 | 9.19 | | |
| 12.39 | 14.013 | 70.3 | 14.044 | 0.22 | 11.71 | -16.43 | 12.36 | -11.8 | | |
| 12.9 | 13.712 | 42.4 | 13.741 | 0.21 | 19.04 | 38.86 | 19.62 | 43.09 | | |
| 9.91 | 13.566 | 78.5 | 13.551 | -0.13 | 13.05 | -3.83 | 13.85 | 2.07 | | |
| 7.43 | 13.126 | 103.1 | 13.135 | 0.06 | 13.11 | -0.12 | 14.09 | 7.34 | | |
| 5.17 | 11.107 | 250.5 | 11.089 | -0.17 | 7.47 | -32.75 | 8.38 | -24.55 | | |
| 3.98 | 10.545 | 258.6 | 10.572 | 0.25 | 9.39 | -10.96 | 10.55 | 0.04 | | |
| 2.5 | 7.822 | 597.9 | 7.868 | 0.59 | 6.25 | -20.1 | 7.3 | -6.67 | | |
| 2.6 | 3.689 | 1056.8 | 3.699 | 0.27 | 3.32 | -10.01 | 3.99 | 8.15 | | |

| | | | | | | | | |
|---|---|---|---|---|---|---|---|---|
| 1.99 | 2.676 | 1035.4 | 2.665 | -0.42 | 4.43 | 65.55 | 5.32 | 98.81 |
| 1.99 | 2.51 | 1189.6 | 2.562 | 2.09 | 3.84 | 53.01 | 4.63 | 84.49 |

| | | | | | All data [27–29] | | All but red points | | From $T_c(d)$ [28] | | From $\rho(d)$ [28] | |
|---|---|---|---|---|---|---|---|---|---|---|---|---|
| | | | | | *A* | 11275 | *A* | 9465.4 | *A* | 7693.9 | *A* | 7908.7 |
| | | | | | *B* | 1.006 | *B* | 0.946 | *B* | 0.889 | *B* | 0.891 |
| *d* [nm] | $T_c$ [K] | $R_s$@20K [Ω/□] | *d*[K] | Err *d*% | $T_{c_RC}$ [K] | Err $T_c$% | $T_{c_RC}$ [K] | Err $T_c$% | $T_{c_RC}$ [K] | Err $T_c$% | $T_{c_RC}$ [K] | Err $T_c$% |
| 2.866 | 9.74 | 639.875 | 2.814 | -1.848 | 5.914 | 39.283 | 7.316 | 24.89 | 8.595 | 11.763 | 8.882 | 8.809 |
| 4.314 | 12.271 | 267.95 | 4.255 | -1.399 | 9.432 | 23.135 | 11.074 | 9.752 | 12.38 | -0.889 | 12.76 | -3.987 |
| 5.784 | 12.69 | 212.665 | 5.665 | -2.095 | 8.877 | 30.051 | 10.278 | 19.004 | 11.34 | 10.639 | 11.774 | 7.219 |
| 7.213 | 13.417 | 140.63 | 7.1 | -1.589 | 10.79 | 19.577 | 12.188 | 9.158 | 13.133 | 2.111 | 13.58 | -1.215 |
| 10.106 | 14.352 | 81.696 | 9.941 | -1.66 | 13.3 | 7.33 | 14.542 | -1.319 | 15.192 | -5.851 | 15.736 | -9.642 |
| 14.416 | 14.524 | 55.868 | 14.23 | -1.308 | 13.666 | 5.912 | 14.604 | -0.551 | 14.93 | -2.798 | 15.423 | -6.192 |
| 21.651 | 14.914 | 35.165 | 21.386 | -1.241 | 14.496 | 2.804 | 15.067 | -1.027 | 15.002 | -0.595 | 15.502 | -3.943 |
| 28.902 | 15.085 | 25.181 | 28.561 | -1.196 | 15.195 | -0.73 | 15.48 | -2.623 | 15.124 | -0.257 | 15.631 | -3.618 |
| 86.909 | 15.727 | 7.547 | 85.77 | -1.328 | 16.982 | -7.983 | 16.095 | -2.34 | 14.68 | 6.656 | 15.229 | 3.1678 |
| 174.02 | 15.777 | 3.593 | 172.66 | -0.789 | 17.895 | -13.425 | 16.221 | -2.815 | 14.182 | 10.1 | 14.656 | 7.106 |

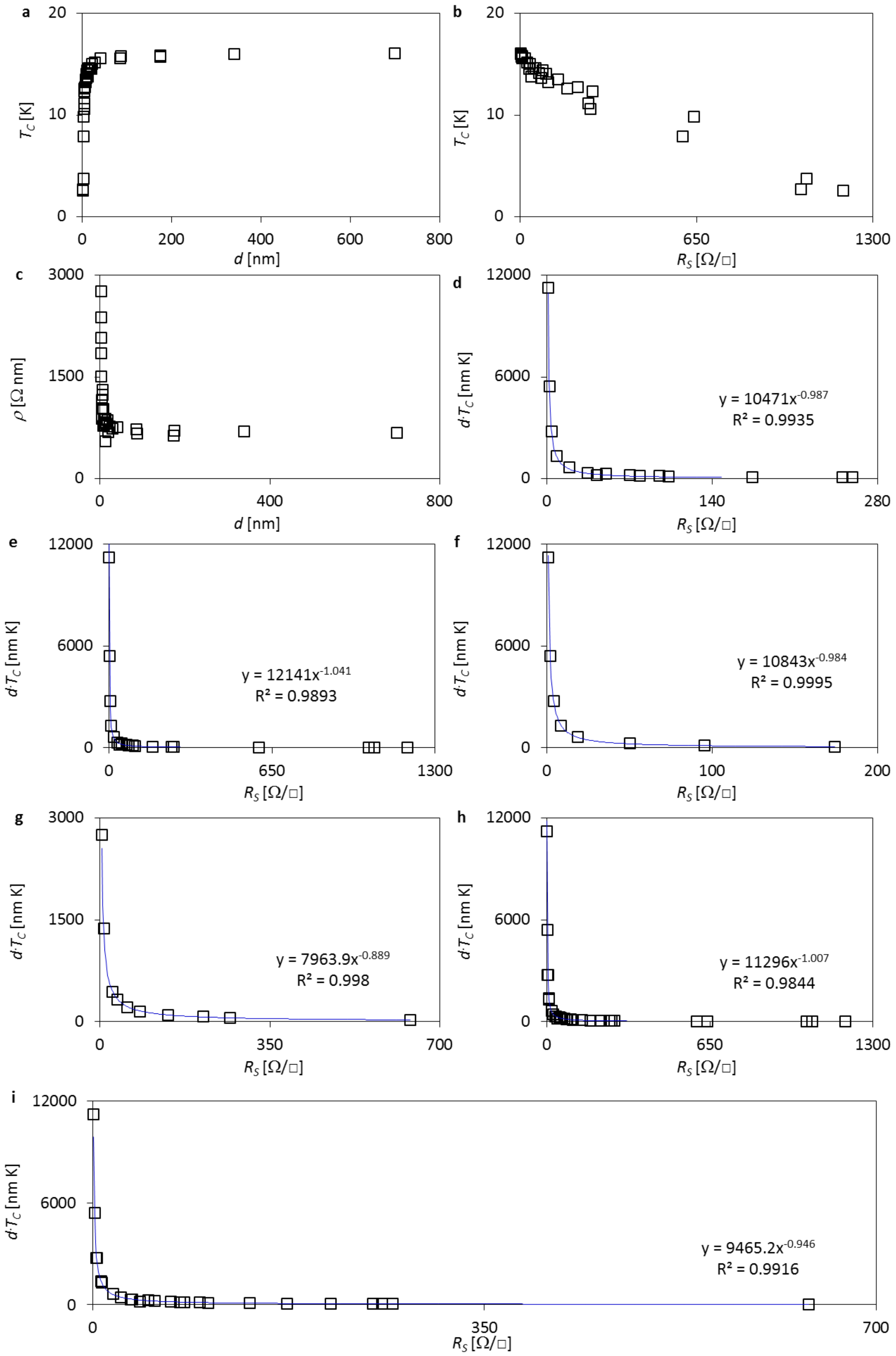

**Figure S11.1. Superconductivity in NbN (Wang and co-authors** [27–29]**), ■ in Fig. 4.** Merged data from Wang *et al.* [27,29] for critical temperature as a function of (**a**) thickness and (**b**) sheet resistance. (**c**) Resistivity as a function of thickness. (**d**) $d{\cdot}T_c$ vs. $R_s$ from these merge data without data points that had a lower level of certainty in their thickness values (highlighted in red in Table S10.1). (**e**) $d{\cdot}T_c$ vs. $R_s$ for all the data presented in Table S10.1. (**f**) $d{\cdot}T_c$ vs. $R_s$ from one dataset only (the older paper by Wang *et al*. [27]). (**g**) $d{\cdot}T_c$ vs. $R_s$ for data only from the 2000 report [28]. (**h**) $d{\cdot}T_c$ vs. $R_s$ for all the three data sets [27–29]. (**i**) $d{\cdot}T_c$ vs. $R_s$ for all the three data sets [27–29], not including the data points marked in red in Table S11.1.

**11.2. NbN- extracted from Semenov *et al.*** [30] **(■ in Fig. 4).**

Semenov *et al*. optimized the conditions for growing NbN films on sapphire while maximizing $T_c$ at the different thicknesses. In addition to variations in growth conditions, some films were of different physical dimensions in the x-y plane. Semenov *et al*. reported good agreement of their data with the proximity effect model, so they found that $T_c$ is a function of $d$. Moreover, they presented two data points for NbN films that were grown with nitrogen deficiency and hence were chemically, crystallography and electronically different than the other superconducting films in this dataset.

We present these films in Table S11.2 and Fig. S11.2 (we included the films of all physical dimensions but took only the large continuous films into consideration in the calculations). One can appreciate that these films fit the empirical power law of Eq. 1 well. Moreover, the films that were grown with nitrogen deficiency were clearly distinguishable from the others (highlighted in red in Table S11.2). This observation strengthens the fact that the scaling of Eq. 1 can be used for controlling and studying the quality of superconducting films, a fact that is significant both experimentally and technologically.

**Table S11.2. Superconductivity in NbN (Semenov *et al*.** [30]) ■ **in Fig. 4.** *d*, $T_c$, and $R_s$ of NbN films extracted from Semenov *et al*. [30] (Tables 1 and 3 therein) as well as the values calculated for *A*, *B*, $T_{c_RC}$, and Error in $T_{c_RC}$%. Films highlighted in red were reported to be grown with nitrogen deficiency, while values highlighted in blue are of films with varying geometries and hence are different than the others.

| | | | *A* | 9544.2 |
|---|---|---|---|---|
| | | | *B* | 0.854 |
| *d* [nm] | $T_c$ [K] | $R_s$@295K [Ω/□] | $T_{c_RC}$ [K] | Err $T_c$% |
| 3.2 | 9.87 | 707 | 10.99 | 11.4 |
| 3.3 | 10.84 | 688 | 10.91 | 0.67 |
| 3.9 | 11.84 | 572 | 10.81 | -8.69 |
| 4.3 | 12.44 | 478 | 11.43 | -8.12 |
| 5.1 | 13.23 | 341 | 12.86 | -2.81 |
| 5.6 | 12.99 | 280 | 13.86 | 6.68 |
| 5.8 | 13.5 | 265 | 14.02 | 3.88 |
| 8 | 13.99 | 191 | 13.45 | -3.88 |
| 8.3 | 14.4 | 165 | 14.69 | 1.99 |
| 11.7 | 15.2 | 105 | 15.33 | 0.84 |
| 14.4 | 15.25 | 84 | 15.07 | -1.2 |
| 5.3 | 11.54 | 261 | 15.55 | 34.73 |
| 6.7 | 13.47 | 145 | 20.32 | 50.83 |
| 3.2 | 10.72 | 940.62 | 8.62 | -19.63 |
| 6 | 14.02 | 235 | 15.02 | 7.14 |
| 12 | 15.17 | 90.83 | 16.91 | 11.49 |

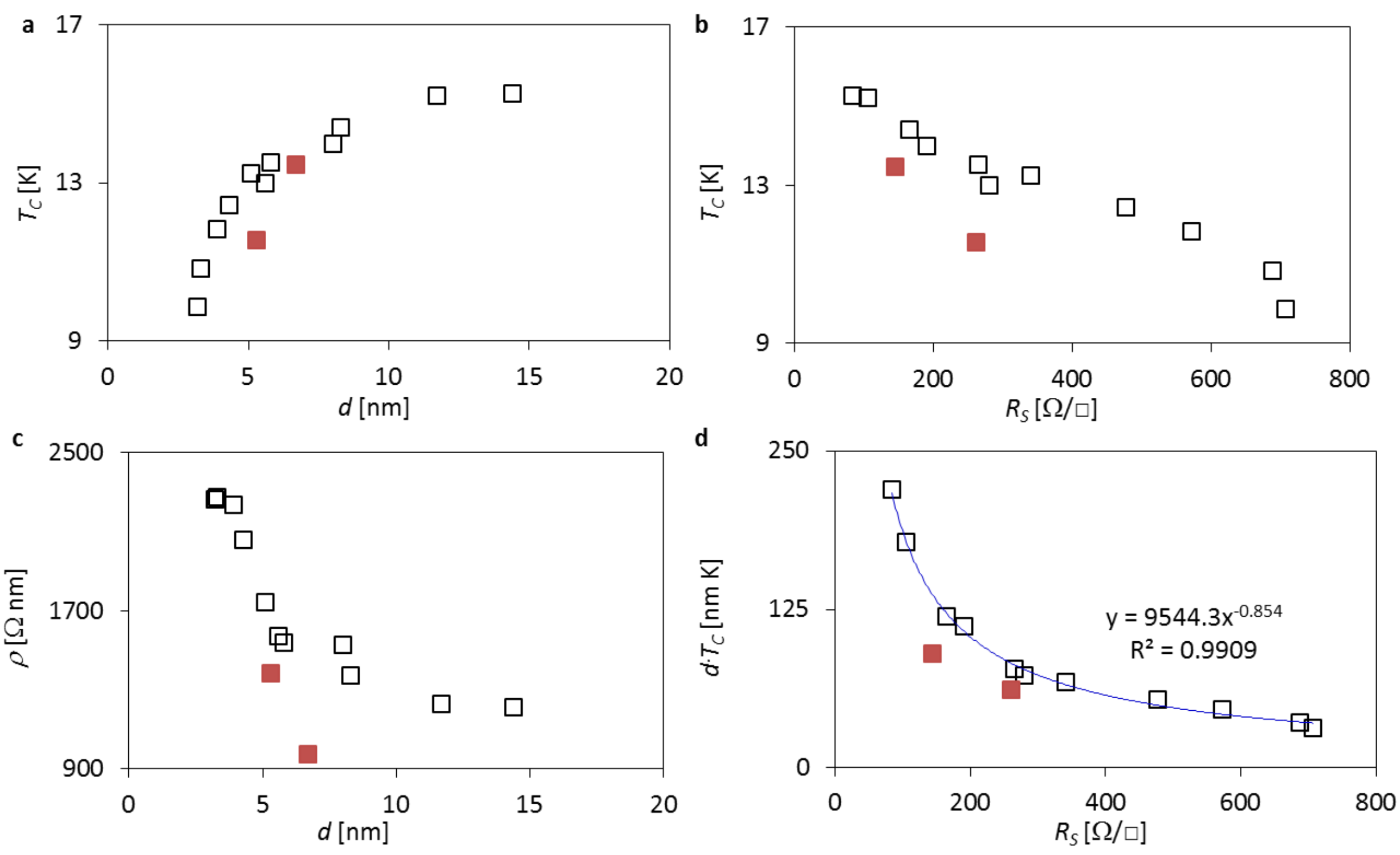

**Figure S11.2. Superconductivity in NbN (Semenov *et al.*** [30]**) ■ in Fig. 4.** Critical temperature as a function of (**a**) thickness and (**b**) sheet resistance. (**c**) Resistivity as a function of thickness. (**d**) $d{\cdot}T_c$ vs. $R_s$ for all films that were prepared under 'normal' conditions (not highlighted in red in Table S11.2). Red data points were reported by Semenov *et al*. to be grown with nitrogen deficiency. These films are not considered for the fitting, but their distance from the trend line in (d) and high values in Err $T_{c_RC}$% suggest such films can be distinguished when plotting $d{\cdot}T_c$ vs. $R_s$. $A$ = 9544.2, $B$ = 0.854.

**10.3. NbN (Kang *et al.*** [31]**, □ in Fig. 4).**

Kang *et al*. supplied a review of some prior works on NbN films. They suggested that, for these films, $T_c$ scales with thickness in accordance with the quantum size effect model ($T_c = T_c(d)$). Moreover, they suggested that the electrical properties of the films change at $d$ = ~5 nm [31]. We found that their data (extracted from Figures 3 and 4, as well as directly from the authors) fit Eq.

1 but not with a great accuracy. A possible reason for that is the reported large error bars in thickness values, mainly for the thinner films.

**Table S11.3. Superconductivity in NbN (Kang *et al.*** [31]**) □ in Fig. 4.** *d*, $T_c$, and $R_s$ of NbN films extracted from Kang *et al*. [31] (Figures 3 and 4 as well as data that were sent directly by the authors) as well as the values calculated for *A*, *B*, $T_{c_RC}$, and Error in $T_{c_RC}$%.

| | | | | |
|---|---|---|---|---|
| | | | *A* | 6583.2 |
| | | | *B* | 0.846 |
| *d* [nm] | $T_c$ [K] | $R_s$ [Ω/□] | $T_{c_RC}$ [K] | Err $T_c$% |
| 3.3 | 11 | 498.435 | 10.42 | -5.3 |
| 4 | 11.8 | 399.431 | 10.36 | -12.16 |
| 5 | 13.2 | 184.642 | 15.93 | 20.66 |
| 7 | 14.1 | 125.353 | 15.79 | 11.97 |
| 9 | 14.5 | 98.8 | 15.02 | 3.58 |
| 13 | 15 | 67.784 | 14.3 | -4.66 |
| 20 | 15.4 | 42.964 | 13.67 | -11.23 |

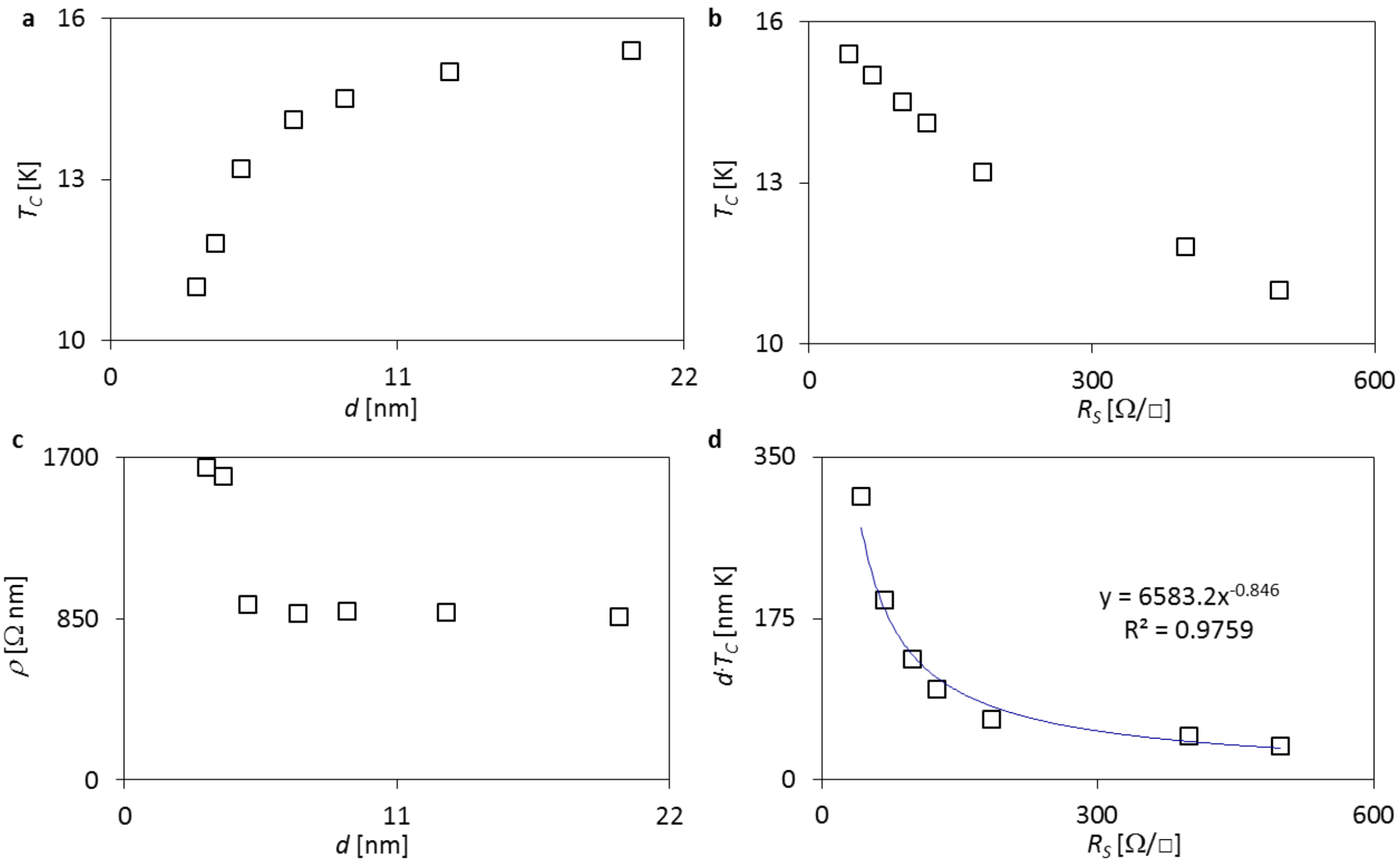

**Figure S11.3. Superconductivity in NbN films (Kang *et al.*** [31]**) □ in Fig. 4.** Critical

temperature as a function of (**a**) thickness and (**b**) sheet resistance. (**c**) Resistivity as a function of thickness. (**d**) $d{\cdot}T_c$ vs. $R_s$. $A$ = 6583.2, $B$ = 0.846.

**<u>11.4 NbN (Delacour *et al.*)</u>** [32].

Delacour *et al*. reported and analyzed superconducting NbN films deposited on sapphire. However, since the data of the reported films that are distributed over a relatively narrow thickness range is scattered over three different graphs (Fig. 5b, Fig. 7 and Fig. 8 in Reference [32]), we could find complete data ($d$, $T_c$ and $R_s$) for only three films. Moreover, the values of these three films are rather close, decreasing the reliability of an analysis with respect to Eq. 1 (*e.g.,* the scale for $d$ is 3-5 nm, while it is 4.4-6.4 K for $T_c$). Hence, we did not have high enough confidence to add these values to Fig. 2. Yet, we present here the relevant data (Table S11.4) and graphs (Fig. S11.4).

**<u>Table 11.4. Superconductivity in NbN (Delacour *et al.*</u>** [32]**).**

$d$, $T_c$, $R_s$@10K, the residual resistance ratio (RRR), and the values of the errors in the extraction process of NbN films extracted from Delacour *et al.* [32] Values for which a match was found are highlighted.

| From Fig. 7 | | From Fig. 8 | | From Fig. 5b | | | | |
|---|---|---|---|---|---|---|---|---|
| $R_s$@10K [Ω/□] | $T_c$ [K] | $T_c$ [K] | $d$ [nm] | $d$ [nm] | RRR^-1 | $R_s$@300K [Ω/□] | Err $T_{c7_8}$% | Err $d_{5b\text{-}8}$% |
| | | 9.203 | 199.084 | 202.072 | 0.028 | | | 1.501 |
| | | 9.092 | 98.629 | 100.455 | 0.031 | | | 1.851 |
| | | 7.588 | 8.84 | 8.925 | 0.414 | | | 0.958 |
| | | 6.608 | 6.492 | 6.523 | 0.775 | | | 0.49 |
| 225.836 | 6.91 | | | | | | | |
| **246.466** | **6.403** | **6.509** | **4.993** | **5.035** | **0.926** | **228.105** | **1.656** | **0.835** |
| | | 6.2 | 4.018 | | | | | |
| **347.184** | **4.901** | **5.005** | **3.344** | **3.344** | **3.233** | **1122.375** | **2.109** | **0.01** |
| **497.513** | **4.411** | **4.5** | **2.982** | **2.986** | **4.344** | **2160.997** | **2.051** | **0.141** |
| | | | | 2.81 | 12.588 | | | |

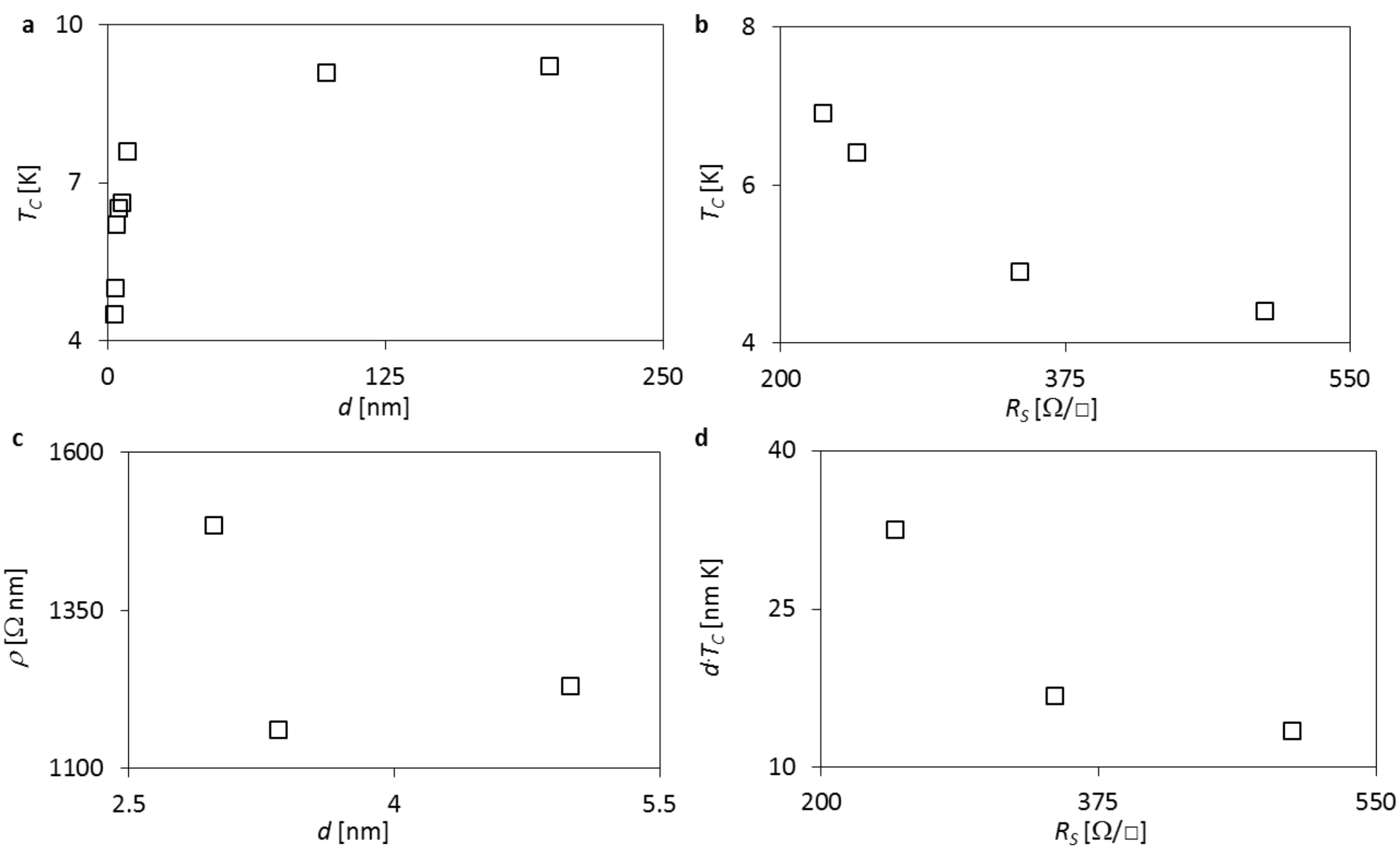

**Figure S11.4. Superconductivity in NbN films (Delacour *et al.* [28]).** Critical temperature as a function of (**a**) thickness and (**b**) sheet resistance. (**c**) Resistivity as a function of thickness. (**d**) $d{\cdot}T_c$ vs. $R_s$. The small number of points in (d) does not allow a quantitative fit to Eq. 1.

**11.5. NbN- Our films (○ in Fig. 4).**

The relations between the different values for the 32 NbN films grown by us were presented in the main text. Here we detail the exact values of the films. The agreement of our films with Eq. 1 was discussed broadly in the main text. The data presented here also include the two films (highlighted in red) whose substrates came into contact with water prior to deposition, which is suspected of influencing their properties (MgO reacts aggressively with water). We found that the best scaling that fits our data is $d{\cdot}T_c$ vs. $R_s$, but we also present the best fit to the power law of Eq. 1.

**Table S11.5. Superconductivity in our NbN films, ○ in Fig. 4.** $d$, $T_c$, and $R_s$ of NbN films grown and characterized by us on MgO substrates as well as the values calculated for $A$, $B$, $T_{c_RC}$, and Error in $T_{c_RC}$%. Data include two films where chemical treatment of the substrate prior to deposition is suspected of influencing their properties (highlighted in red).

| | | | | |
|---|---|---|---|---|
| | | | *A* | 9448.1 |
| | | | *B* | 0.903 |
| *d* [nm] | $R_s$ [Ω/□] | $T_c$ [K] | $T_{c_RC}$ [K] | Err $T_{c_RC}$% |
| 5.67 | 294.53 | 10.7 | 9.82 | -8.22 |
| 5.46 | 303.96 | 10.9 | 9.91 | -9.06 |
| 5.43 | 301.60 | 10.7 | 10.04 | -6.19 |
| 5.43 | 297.39 | 11.3 | 10.17 | -10.04 |
| 5.2 | 275.31 | 11.5 | 11.38 | -1.04 |
| 5.32 | 271.10 | 11.7 | 11.28 | -3.59 |
| 5.18 | 252.99 | 11.5 | 12.33 | 6.81 |
| 5.27 | 259.07 | 11.6 | 11.86 | 2.58 |
| 6.2 | 337.20 | 8.3 | 7.95 | -4.24 |
| 2.88 | 600.79 | 9.9 | 10.16 | 2.6 |
| 3.18 | 396.08 | 11.0 | 13.4 | 21.82 |
| 4.3 | 369.80 | 10.6 | 10.54 | -0.53 |
| 5.7 | 311.51 | 9.6 | 9.29 | -3.26 |
| 5.4 | 316.27 | 9.8 | 9.67 | -1.33 |
| 5.4 | 177.46 | 14.2 | 16.29 | 14.74 |
| 5.7 | 204.23 | 13.5 | 13.6 | 0.72 |
| 5.9 | 274.29 | 11.3 | 10.06 | -10.93 |
| 5.8 | 275.00 | 11.3 | 10.21 | -9.61 |
| 6.5 | 318.65 | 8.7 | 7.98 | -8.29 |
| 6.4 | 321.03 | 8.7 | 8.05 | -7.48 |
| 5.5 | 292.72 | 11.5 | 10.18 | -11.47 |
| 6.2 | 188.41 | 13.2 | 13.44 | 1.85 |
| 4.4 | 297.84 | 12.0 | 12.53 | 4.4 |
| 3.5 | 396.44 | 11.4 | 12.17 | 6.71 |
| 16 | 54.12 | 15.1 | 16.07 | 6.42 |
| 7.9 | 146.90 | 13.4 | 13.21 | -1.42 |
| 9.3 | 103.60 | 14.5 | 15.38 | 6.08 |
| 6.3 | 206.96 | 12.7 | 12.16 | -4.29 |
| 3.6 | 450.01 | 9.2 | 10.55 | 14.65 |
| 2.9 | 542.30 | 9.5 | 11.06 | 16.47 |
| 6.4 | 430.70 | 13.3 | 6.17 | -53.59 |
| 6.5 | 494.21 | 10.5 | 5.37 | -48.87 |

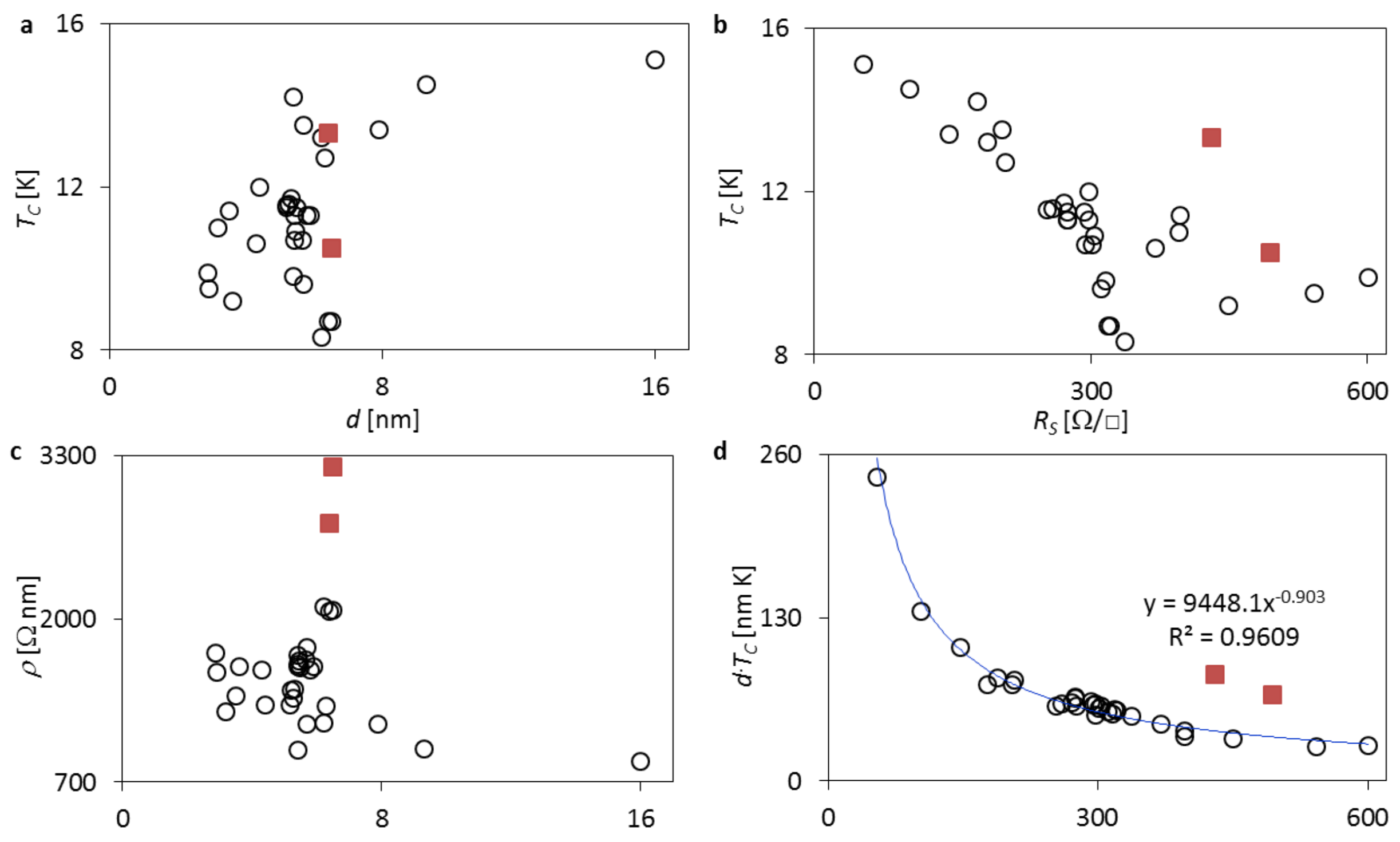

**Figure S11.5. Superconductivity in our NbN films, ○ in Fig. 4.** Critical temperature as a function of (**a**) thickness and (**b**) sheet resistance. (**c**) Resistivity as a function of thickness. (**d**) $d{\cdot}T_c$ vs. $R_s$. Chemical treatment of the substrate prior to deposition is suspected of influencing the properties of the films presented here as red squares. $A$ = 9448.1, $B$ = 0.903.

## 12. Pb.

Similar to bismuth, thin superconducting films of lead were studied by Goldman and co-authors [12], which was a continuation of the work of Strongin *et al*. two decades earlier [6]. Both studies involved cold deposition of Pb films. Strongin *et al*. reported some complete datasets for thin Pb films grown on previously deposited $SiO_2$, while Goldman and co-authors characterized lead films grown on previously deposited Ge films. Here we present all of these datasets.

### 12.1. Pb- extracted from Strongin *et al.* [6] (▲,▼,⌖,•,● and ○ in Fig. 4).

As a part of their study of superconductivity in thin films, Strongin *et al*. reported some 7 complete experiments (datasets) of Pb grown on pre-deposited $SiO_2$ substrates. Six of these datasets were found to fit Eq. 1 very well, while the additional dataset was found to agree with the scaling of $d{\cdot}T_c$ vs. $R_s$, but the quantitative fitting was less successful (an additional dataset, from which we were able to extract only one data point, was also reported). We present all of these datasets here with the same symbols as in both Fig. 4 and the original paper.

**Table S12.1. Superconductivity in Pb (Strongin *et al.* [6]) ▼, ▲, ⌖, ●, • and ○ in Fig. 4.** *d*, $T_c$, and $R_s$ of different sets of Pb films extracted from Strongin *et al*. [6] (Figures 5 and 6 therein). The error in data extraction is evaluated through the error in the extracted $T_c$ values. Values calculated for *A*, *B*, $T_{c_RC}$, and Error in $T_{c_RC}$% are also presented. The symbols representing each dataset in both the original paper [6] and in Fig. 4 are also presented.

| | **From Fig. 5** | | **From Fig. 6** | | | | | | |
|---|---|---|---|---|---|---|---|---|---|
| **Symbol:** | $R_s$ [Ω/□] | $T_c$ [K] | *d* [nm] | $T_c$ [K] | Err $T_c$% | $T_{c_RC}$ [K] | Err $T_{c_RC}$% | | |
| ▼ | 88.959 | 6.753 | 90.1 | 6.646 | -1.58 | 7.27 | 7.67 | *A* | 25803 |
| ▼ | 227.956 | 6.203 | 51.75 | 6.197 | -0.09 | 5.83 | -6.01 | *B* | 0.82 |
| ▼ | 649.151 | 5.195 | 25.77 | 5.4 | 3.94 | 4.94 | -4.87 | | |
| ▼ | 3407.124 | 2.879 | 10.74 | 2.872 | -0.23 | 3.02 | 5.04 | | |
| ▲ | 179.160 | 6.08 | | | | | | | |
| ▲ | 246.512 | 5.905 | 75.99 | 5.889 | -0.26 | 6.22 | 5.4 | *A* | 14299 |
| ▲ | 400.285 | 5.44 | 63.91 | 5.431 | -0.17 | 5.48 | 0.75 | *B* | 0.619 |
| ▲ | 646.182 | 4.998 | 54.38 | 4.978 | -0.4 | 4.79 | -4.17 | | |
| ▲ | 978.841 | 4.39 | 47.8 | 4.379 | -0.24 | 4.21 | -4.01 | | |
| ▲ | 1406.723 | 4.047 | 41.61 | 4.092 | 1.12 | 3.87 | -4.45 | | |
| ▲ | 1935.924 | 3.546 | 37.43 | 3.525 | -0.59 | 3.53 | -0.51 | | |
| ▲ | 2615.306 | 3.034 | 33.25 | 3.043 | 0.29 | 3.3 | 8.64 | | |
| ⌖ | 58.064 | 6.97 | | | | | | | |
| ⌖ | 129.108 | 6.378 | 68.65 | 6.362 | -0.26 | 6.53 | 2.32 | *A* | 45805 |
| ⌖ | 198.665 | 6.124 | 49.78 | 6.145 | 0.34 | 5.97 | -2.5 | *B* | 0.952 |
| ⌖ | 283.058 | 5.978 | 37.12 | 5.907 | -1.19 | 5.72 | -4.37 | | |
| ⌖ | 373.846 | 5.517 | 28.31 | 5.504 | -0.23 | 5.75 | 4.25 | | |
| ⌖ | 544.687 | 5.092 | 22.43 | 5.091 | -0.01 | 5.07 | -0.39 | | |
| ● | 233.689 | 5.648 | 38.08 | 5.626 | -0.39 | 5.29 | -6.28 | *A* | 31806 |
| ● | 327.309 | 5.335 | 28.15 | 5.359 | 0.45 | 5.24 | -1.83 | *B* | 0.928 |
| ● | 524.138 | 4.671 | 18.46 | 4.671 | 0 | 5.16 | 10.5 | | |
| ● | 873.743 | 4.058 | 14.21 | 4.03 | -0.69 | 4.17 | 2.84 | | |
| ● | 2214.532 | 2.667 | 9.895 | 2.636 | -1.15 | 2.53 | -5.24 | | |
| • | 29.428 | 7.128 | | | | | | | |
| • | 88.727 | 6.669 | | | | | | | |
| • | 174.642 | 6.198 | 69.67 | 6.196 | -0.04 | 6.13 | -1.14 | *A* | 47822 |
| • | 282.628 | 5.823 | 46.94 | 5.83 | 0.12 | 5.86 | 0.58 | *B* | 0.914 |
| • | 448.432 | 5.329 | 32.9 | 5.321 | -0.16 | 5.48 | 2.83 | | |
| • | 884.198 | 4.334 | 23.45 | 4.32 | -0.32 | 4.13 | -4.62 | | |
| • | 1726.928 | 3.316 | 15.6 | 3.313 | -0.07 | 3.37 | 1.64 | | |
| ○ | 5.193 | 7.126 | | | | | | | |
| ○ | 31.291 | 6.926 | | | | | | | |
| ○ | 108.621 | 6.854 | 100.7 | 6.847 | -0.1 | 6.22 | -9.25 | *A* | 109742 |
| ○ | 203.292 | 6.045 | 47.61 | 6.046 | 0.01 | 6.59 | 9.08 | *B* | 1.102 |
| ○ | 308.576 | 5.569 | 33.04 | 5.867 | 5.35 | 6 | 7.73 | | |
| ○ | 524.479 | 4.794 | 23.55 | 4.78 | -0.3 | 4.69 | -2.17 | | |
| ○ | 931.595 | 3.951 | 15.59 | 3.932 | -0.48 | 3.76 | -4.8 | | |
| ▽ | 840.466 | 4.289 | | | | | | | |
| ▽ | 1646.527 | 3.154 | | | | | | | |
| ▽ | 1844.345 | 2.847 | | | | | | | |
| ▽ | 1927.64 | 2.304 | 11.1 | 2.322 | 0.76 | 4.68 | 103.2 | | |
| ▽ | 2662.622 | 2.623 | | | | | | | |
| ▽ | 2761.538 | 2.472 | | | | | | | |
| ▽ | 3861.984 | 1.78 | | | | | | | |
| △ | 35.891 | 6.837 | 210.3 | 6.835 | -0.03 | 6.58 | -3.73 | | |
| △ | 88.045 | 6.423 | 173.6 | 6.44 | 0.26 | 3.81 | -40.75 | | |
| △ | 138.09 | 6.122 | 148.9 | 6.104 | -0.29 | 3.06 | -49.98 | | |
| △ | 202.856 | 5.888 | 116.3 | 5.885 | -0.04 | 2.86 | -51.48 | | |
| △ | 317.741 | 5.379 | 95.04 | 5.398 | 0.36 | 2.41 | -55.11 | | |
| △ | 421.107 | 5.085 | 79 | 5.086 | 0.01 | 2.3 | -54.71 | | |
| △ | 524.302 | 4.73 | 70.03 | 4.75 | 0.42 | 2.17 | -54.15 | | |
| △ | 577.377 | 4.648 | 59.04 | 4.632 | -0.34 | 2.38 | -48.88 | | |
| △ | 950.854 | 3.907 | 50.19 | 3.93 | 0.58 | 1.85 | -52.57 | | |
| △ | 1030.484 | 3.791 | 39.85 | 3.791 | 0 | 2.18 | -42.39 | | |
| △ | 1129.304 | 3.605 | 32.61 | 3.599 | -0.16 | 2.48 | -31.35 | | |
| △ | 1296.145 | 3.486 | 28.81 | 3.476 | -0.28 | 2.5 | -28.27 | | |
| △ | 1489.056 | 3.157 | 25.48 | 3.153 | -0.13 | 2.52 | -20.1 | | |
| △ | 1629.329 | 3.067 | 22.46 | 3.05 | -0.56 | 2.66 | -13.4 | | |
| △ | 1781.429 | 2.874 | 20.99 | 2.854 | -0.69 | 2.64 | -8.12 | | |
| △ | 1933.836 | 2.792 | 18.97 | 2.791 | -0.04 | 2.73 | -2.2 | | |

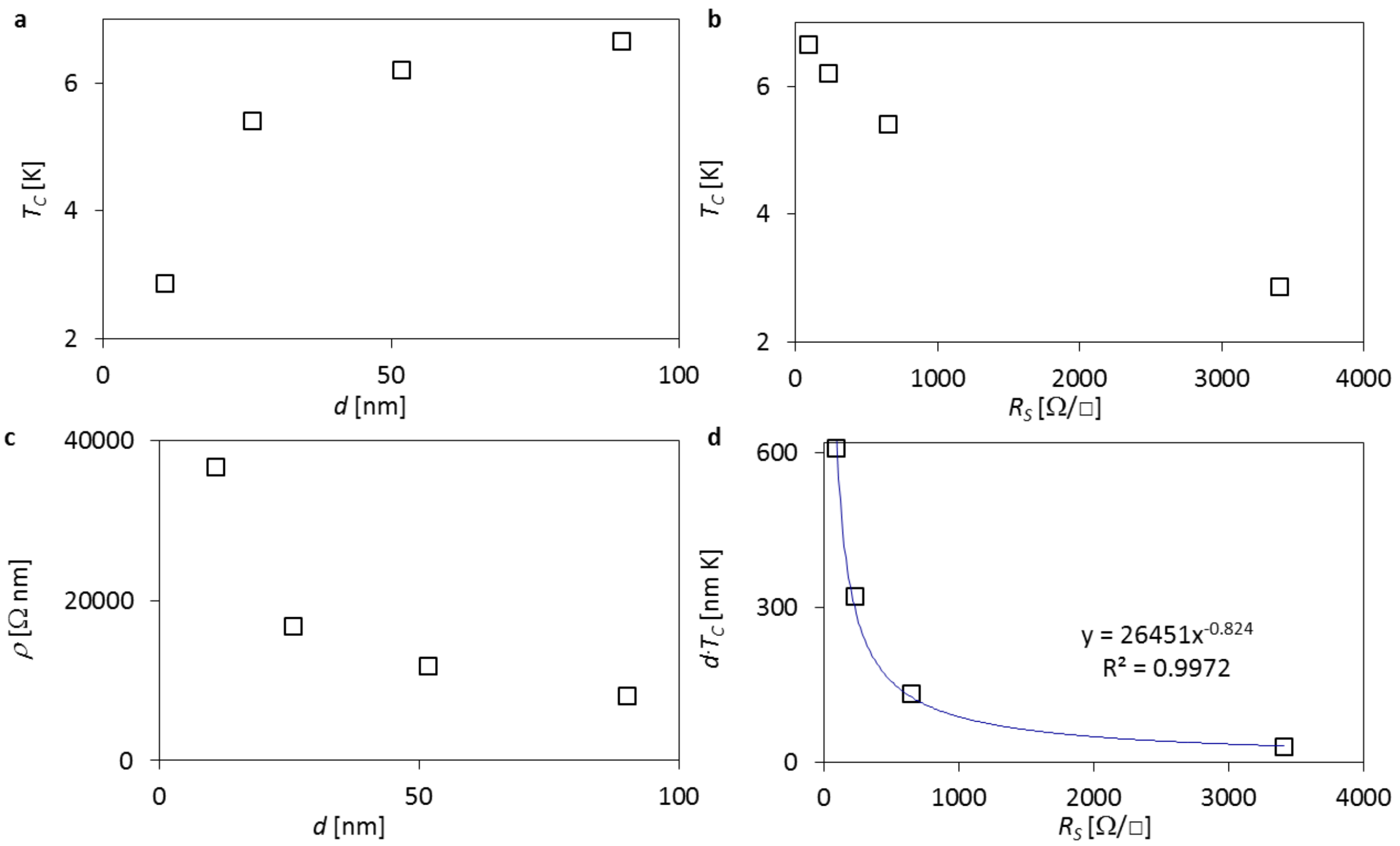

**Fig. S12.1. Superconductivity in Pb (Strongin *et al*.** [6]**), ▼ in Fig. 4.** Critical temperature as a function of (**a**) thickness and (**b**) sheet resistance. (**c**) Resistivity as a function of thickness. (**d**) $d{\cdot}T_c$ vs. $R_s$. Taken from ▼ in Figures 5 and 6 by Strongin *et al*. [6] $A$ = 26451, $B$ = 0.824.

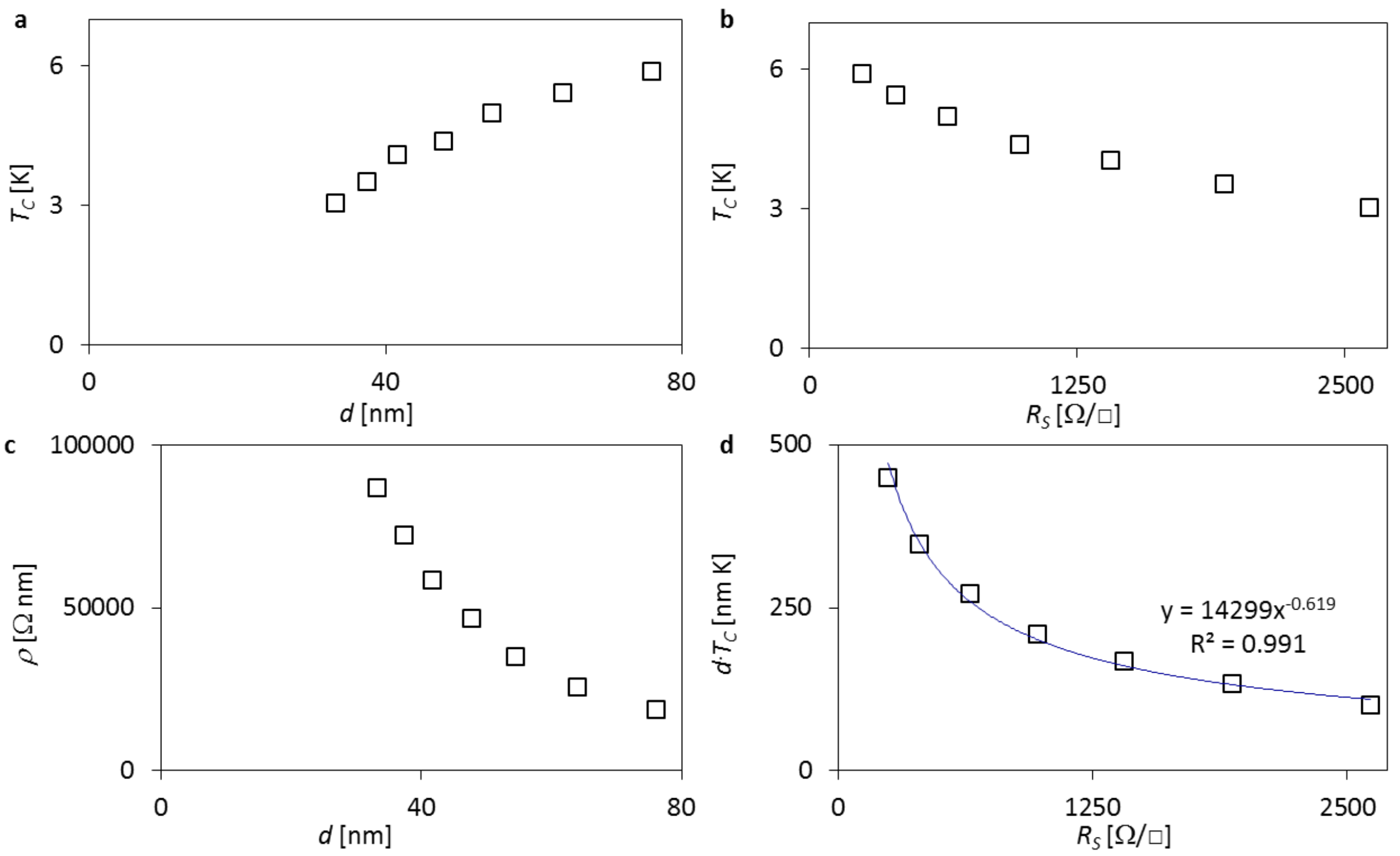

**Fig. S12.2. Superconductivity in Pb (Strongin *et al*.** [6]**), ▲ in Fig. 4.** Critical temperature as a function of (**a**) thickness and (**b**) sheet resistance. (**c**) Resistivity as a function of thickness. (**d**) $d{\cdot}T_c$ vs. $R_s$. Taken from ▲ in Figures 5 and 6 by Strongin *et al*. [6] $A$ = 14299, $B$ = 0.619.

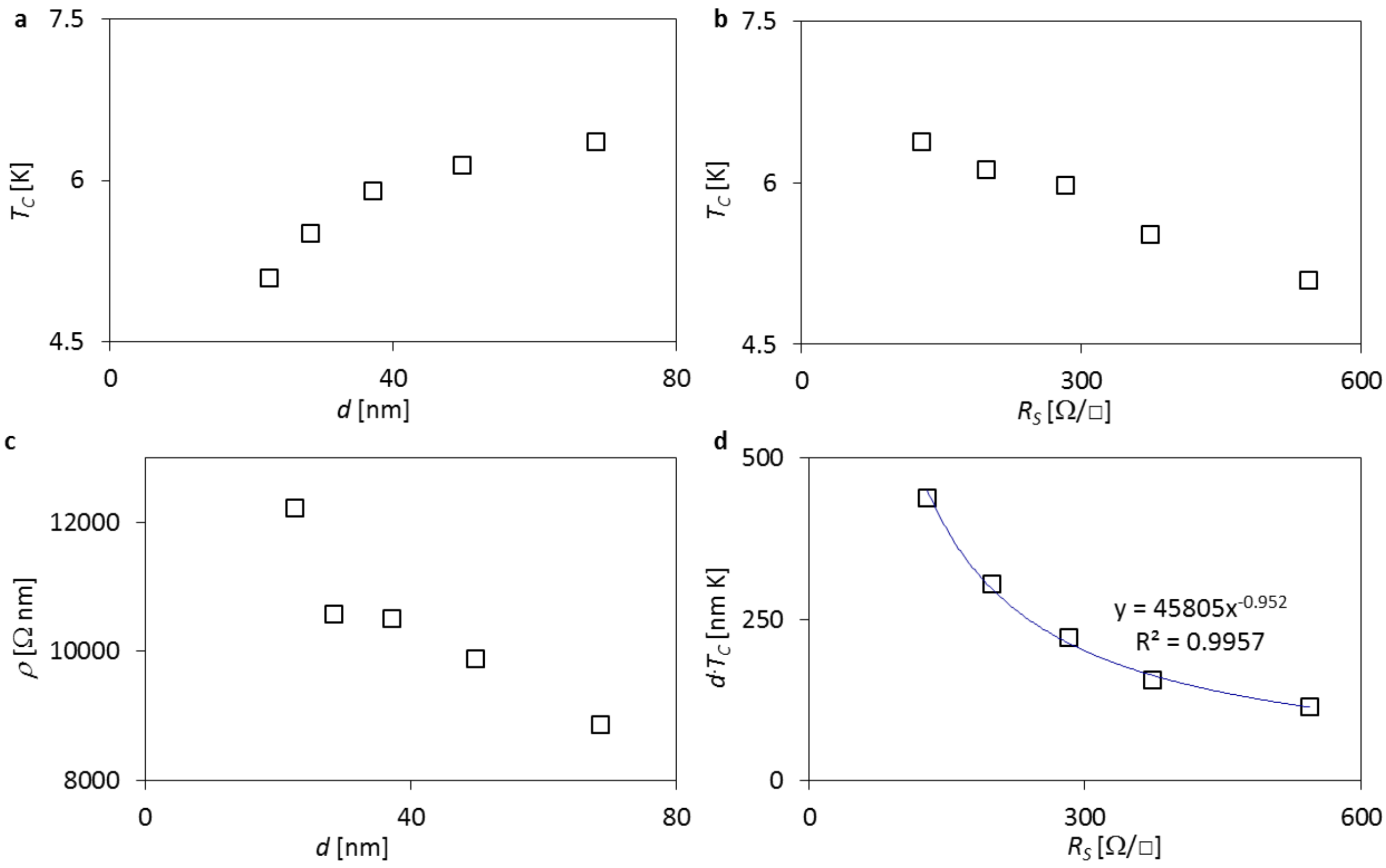

**Fig. S12.3. Superconductivity in Pb (Strongin *et al*.** [6]**), ⌖ in Fig. 4.** Critical temperature as a function of (**a**) thickness and (**b**) sheet resistance. (**c**) Resistivity as a function of thickness. (**d**) $d{\cdot}T_C$ vs. $R_S$. Taken from ⌖ in Figures 5 and 6 by Strongin *et al*. [6] $A$ = 45805, $B$ = 0.952.

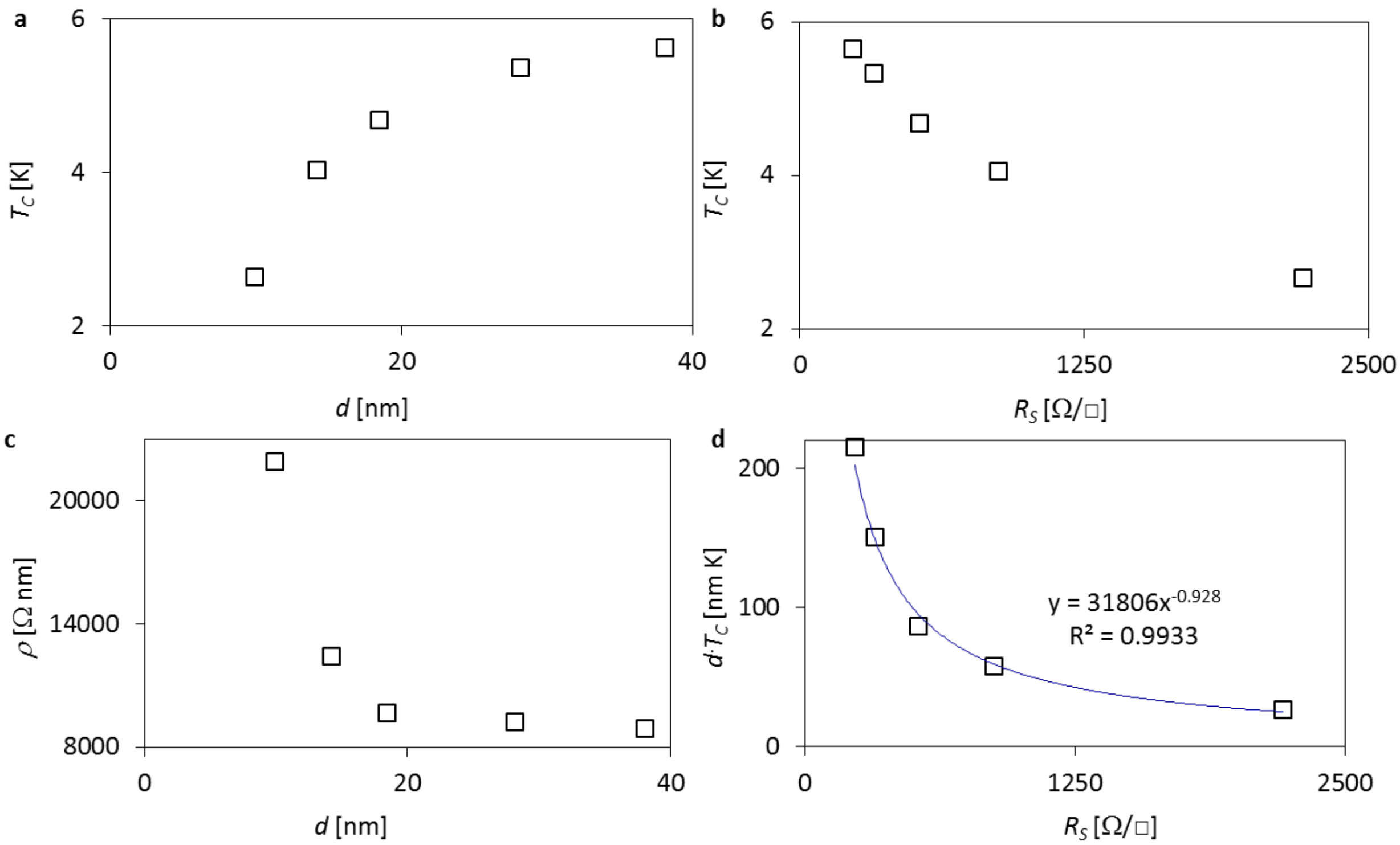

**Fig. S12.4. Superconductivity in Pb (Strongin *et al*.** [6]**), ● in Fig. 4.** Critical temperature as a function of (**a**) thickness and (**b**) sheet resistance. (**c**) Resistivity as a function of thickness. (**d**) $d{\cdot}T_c$ vs. $R_s$. Taken from ● in Figures 5 and 6 by Strongin *et al*. [6] $A$ = 31806, $B$ = 0.928.

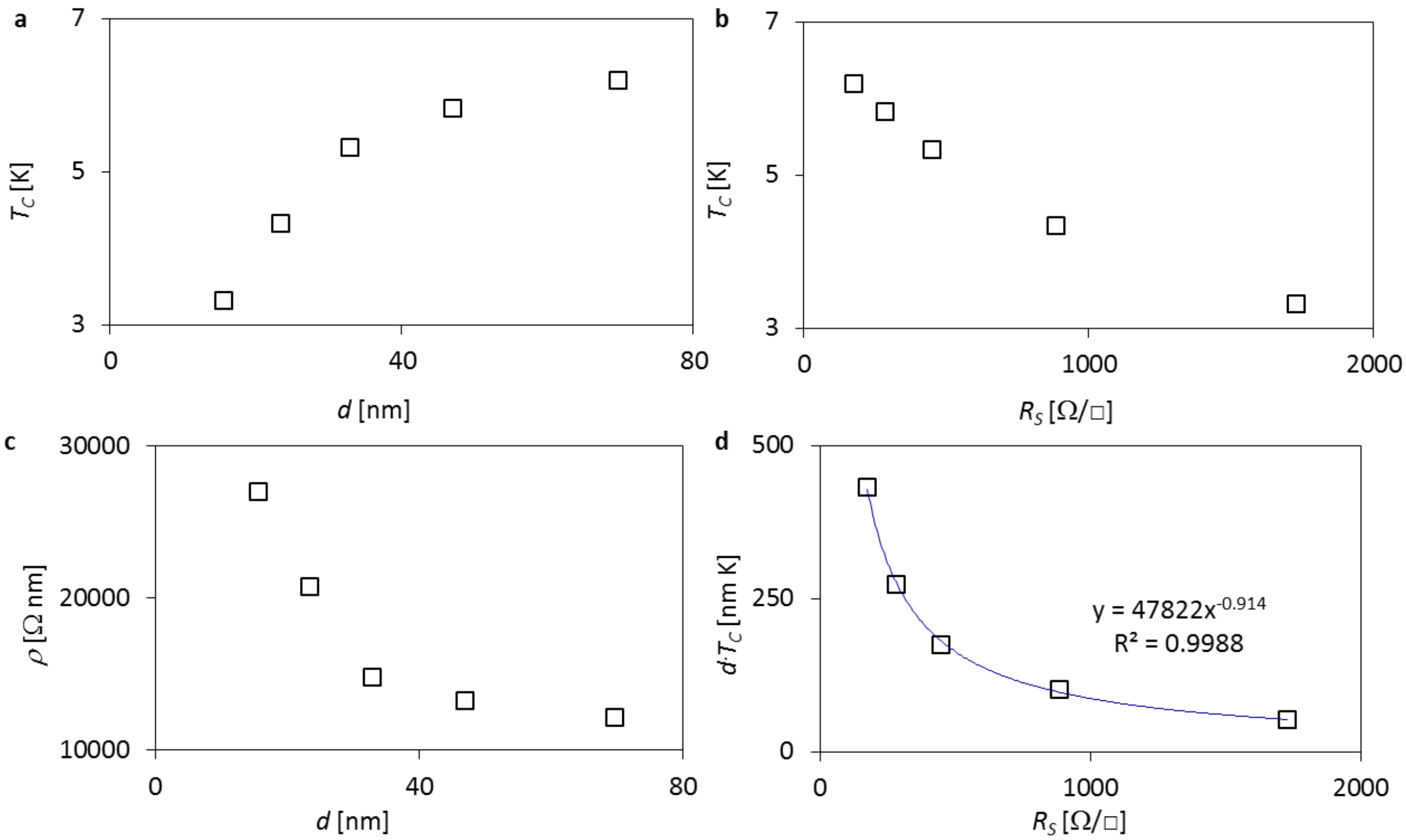

**Fig. S12.5. Superconductivity in Pb (Strongin *et al***. [6]**), • in Fig. 4.** Critical temperature as a function of (**a**) thickness and (**b**) sheet resistance. (**c**) Resistivity as a function of thickness. (**d**) $d{\cdot}T_c$ vs. $R_s$. Taken from • in Figures 5 and 6 by Strongin *et al*. [6] $A$ = 47822, $B$ = 0.914.

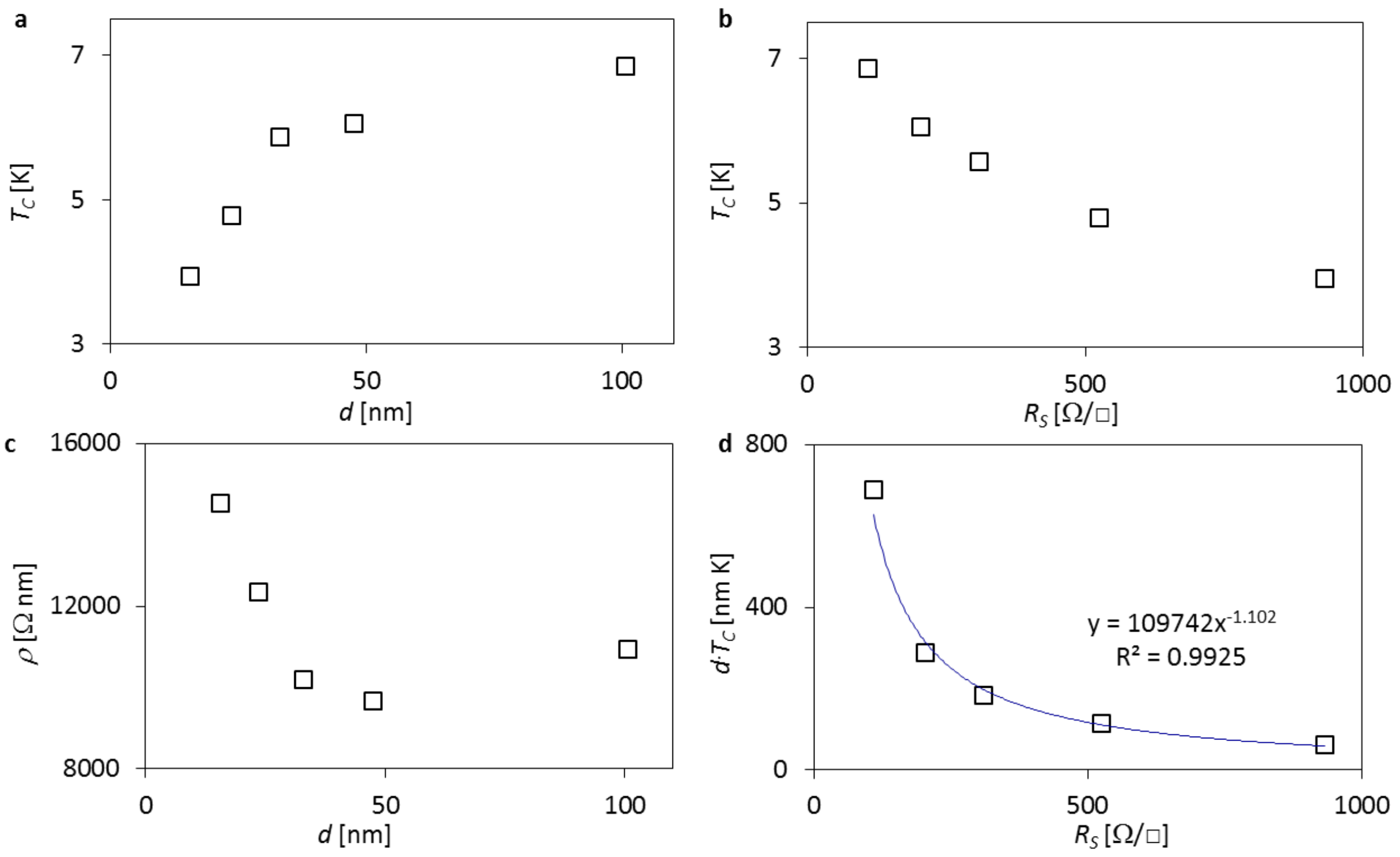

**Fig. S12.6. Superconductivity in Pb (Strongin *et al*.** [6]**), ○ in Fig. 4.** Critical temperature as a function of (**a**) thickness and (**b**) sheet resistance. (**c**) Resistivity as a function of thickness. (**d**) $d{\cdot}T_c$ vs. $R_S$. Taken from ○ in Figures 5 and 6 by Strongin *et al*. [6] $A$ = 109742, $B$ = 1.102.

## 12.7. Pb- extracted from Haviland *et al*. [12] (◁ in Fig. 4).

When Haviland *et al*. studied the quantum transition between superconducting and insulating states (*i.e.*, the 'onset of superconductivity'), they studied superconductivity in thin Pb films in addition to the classic example of Bi films. They found that Pb films exhibit superconductivity for all films thicker than ~0.3 nm. However, since Pb has a typical lattice constant of ~0.49 nm [33], we think one can assume that superconductivity at $d$ < ~0.49 nm cannot be considered as 2D superconductivity of a continuous film. Rather, it is reasonable to assume that films reported to be thinner than 0.49 nm are not homogeneous (or continuous). Hence, we do not expect the mechanism governing superconductivity in these films to be similar to that governing superconductivity in 2D films. Indeed, a log-log scale plot of $d \cdot T_c$ vs. $R_s$ demonstrates linearity of the data for $d$ > ~0.4 nm. Hence, although we present all the data points below to allow the reader to look at the data more carefully, in Fig. 4 we included only the films thicker than 0.4 nm. Moreover, for the quantitative analysis (extracting $A$ and $B$), we used only films with $d \geq 0.49$ nm. For the fitting to Eq. 1 we also excluded several films that had relatively high inconsistency in their values as extracted by us (highlighted in red below). It should be noted that in the thinnest films, superconductivity may have been influenced by the proximization with the Ge substrate, hence changing the trend from a power law to a more complex form. An insightful discussion about superconductivity in these Pb films is presented in Section 17.2.

**Table S12.2. Superconductivity in Pb (Haviland *et al.* [12]), ◁ in Fig. 4.** $d$, $T_c$, and $R_s$ of Pb films extracted from Haviland *et al*. [6] (Figures 2 and 3 therein). The error in data extraction is evaluated through the error in extracted $T_c$ values. Values calculated for $A$, $B$, $T_{c_RC}$, and Error in $T_{c_RC}$% are also presented. Errors in the data extraction larger than 1% are highlighted in red, while films thinner than the nominal lattice constant of Pb are highlighted in blue.

| | | | | | *A* | 1090.9 |
|---|---|---|---|---|---|---|
| From Fig. 2 | | From Fig. 4 | | | *B* | 0.821 |
| $d$ [nm] | $T_c$ [K] | $R_s$ [Ω/□] | $T_c$ [K] | Err $T_c$% | $T_{c_RC}$ [K] | Err $T_{c_RC}$% |
| 2.89 | 6.641 | | | | | |
| 2.725 | 6.612 | 57.476 | 6.546 | 1.02 | | |
| 2.551 | 6.551 | 68.253 | 6.452 | 1.52 | | |
| 2.364 | 6.46 | 53.884 | 6.641 | -2.71 | | |
| 2.212 | 6.299 | 57.476 | 6.609 | -4.69 | | |
| 2.0877 | 6.095 | 107.768 | 6.294 | -3.17 | | |
| 1.9417 | 6.005 | 165.245 | 6.096 | -1.5 | | |
| 1.795 | 5.72 | 280.198 | 5.72 | 0.01 | 5.95 | 3.96 |
| 1.661 | 5.612 | 334.082 | 5.614 | -0.04 | 5.56 | -0.87 |
| 1.506 | 5.482 | 395.15 | 5.482 | 0 | 5.35 | -2.48 |
| 1.376 | 5.357 | 438.258 | 5.36 | -0.06 | 5.38 | 0.35 |
| 1.239 | 5.2 | 510.103 | 5.207 | -0.12 | 5.27 | 1.3 |
| 1.115 | 5.068 | 603.503 | 5.067 | 0.03 | 5.1 | 0.64 |
| 1.001 | 4.854 | 725.64 | 4.847 | 0.13 | 4.88 | 0.61 |
| 0.903 | 4.668 | 872.923 | 4.66 | 0.18 | 4.65 | -0.4 |
| 0.855 | 4.535 | 951.953 | 4.53 | 0.12 | 4.58 | 0.9 |
| 0.812 | 4.4293 | 1070.498 | 4.419 | 0.23 | 4.38 | -1.2 |
| 0.769 | 4.2729 | 1171.082 | 4.266 | 0.17 | 4.29 | 0.47 |
| 0.747 | 4.2119 | 1268.074 | 4.213 | -0.03 | 4.14 | -1.73 |
| 0.712 | 4.0501 | 1386.619 | 4.049 | 0.02 | 4.04 | -0.36 |
| 0.683 | 3.9653 | 1508.756 | 3.97 | -0.12 | 3.92 | -1.05 |
| 0.662 | 3.8486 | 1623.709 | 3.838 | 0.28 | 3.81 | -0.94 |
| 0.639 | 3.7559 | 1774.585 | 3.759 | -0.07 | 3.67 | -2.32 |
| 0.613 | 3.6179 | 1914.683 | 3.621 | -0.1 | 3.59 | -0.71 |
| 0.570 | 3.3528 | 2306.242 | 3.355 | -0.06 | 3.32 | -1.1 |
| 0.545 | 3.1937 | 2539.74 | 3.194 | -0.01 | 3.21 | 0.41 |
| 0.517 | 3.0241 | 2830.714 | 3.02 | 0.14 | 3.09 | 2.3 |
| 0.491 | 2.8226 | 3197.126 | 2.819 | 0.11 | 2.95 | 4.35 |
| 0.469 | 2.6052 | 3635.384 | 2.601 | 0.17 | | |
| 0.445 | 2.3373 | 4174.225 | 2.332 | 0.24 | | |
| 0.426 | 2.1092 | 4774.136 | 2.108 | 0.05 | | |
| 0.407 | 1.9024 | 5309.385 | 1.965 | -0.16 | | |
| 0.4 | 1.7884 | 5560.844 | 1.787 | 0.08 | | |
| 0.392 | 1.6876 | 5841.042 | 1.692 | -0.28 | | |
| 0.386 | 1.6373 | 6056.578 | 1.637 | 0.01 | | |
| 0.38 | 1.4887 | 6387.068 | 1.484 | 0.3 | | |
| 0.372 | 1.3827 | 6555.905 | 1.382 | 0.09 | | |
| 0.36 | 1.128 | 7457.566 | 1.127 | 0.12 | | |
| 0.349 | 0.9663 | 8007.185 | 0.961 | 0.53 | | |
| 0.34 | 0.616 | 9052.537 | 0.611 | 0.75 | | |

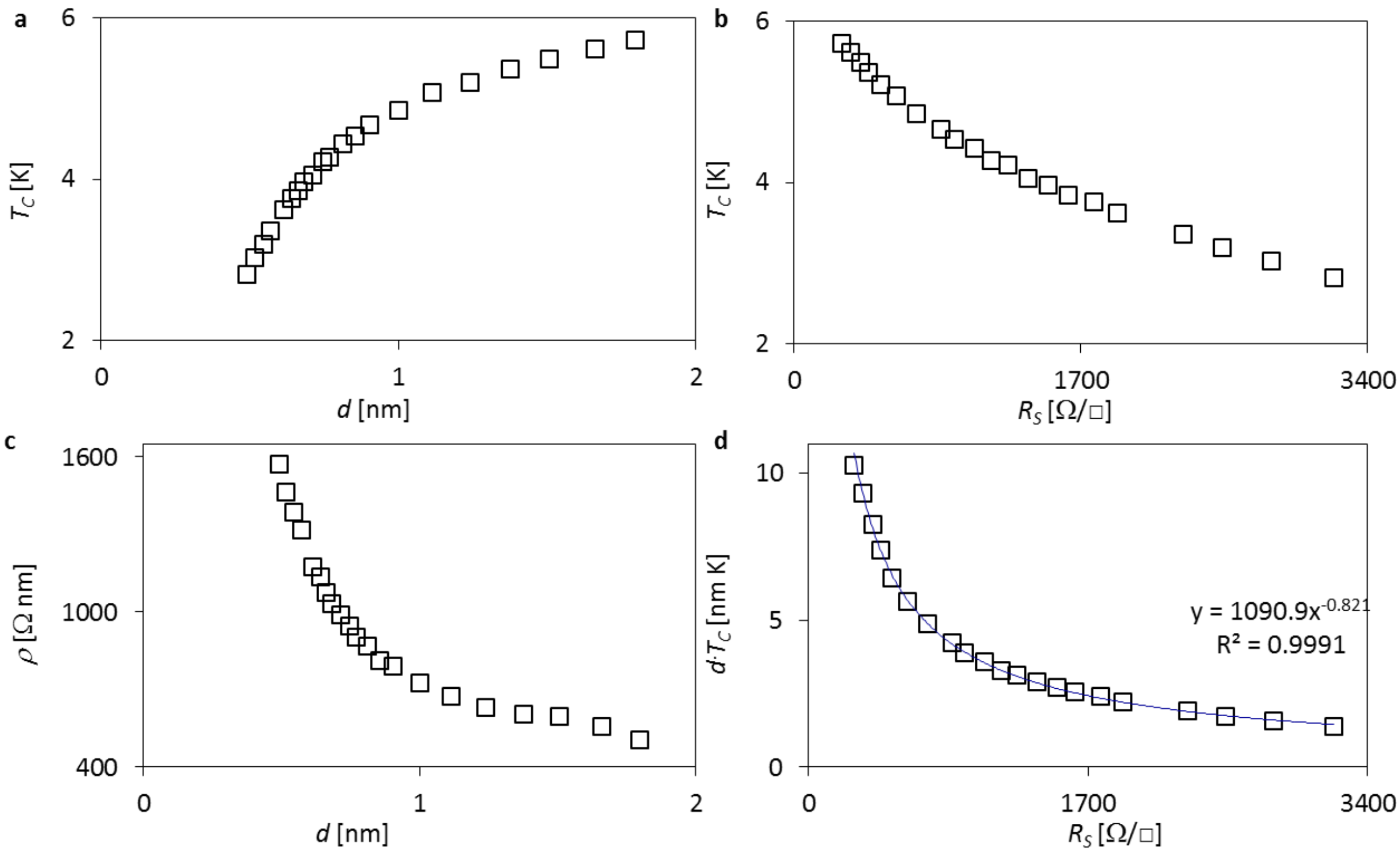

**Fig. S12.7. Superconductivity in Pb (Havliand *et al*.** [12]**) ◁ in Fig. 4.** Critical temperature as a function of (**a**) thickness and (**b**) sheet resistance. (**c**) Resistivity as a function of thickness. (**d**) $d{\cdot}T_c$ vs. $R_S$. Inconsistent data and films thinner than the nominal lattice constant of Pb are not included (see Table S12.2). $A$ = 1090.9, $B$ = 0.81.

**13. α-ReW- extracted from Raffy *et al*.** [34]**, ✱ in Fig. 4.**

Raffy *et al*. reported a set of amorphous ReW films ranging from ~4.5 nm to ~100 nm thick. The different films varied in stoichiometric composition. Similarly to Graybeal and Beasley [17], they deduced that their data fit the Maekawa and Fukuyama model [35]. In this framework, the electrons are localized, resulting in enhanced electron-electron Coulomb interaction, which in turn suppresses $T_c$. Practically, from the perspective of the current report, this means that $T_c$ is a function of the sheet resistance only ($T_c = T_c(R_S)$), similar to Finkel'stein's model [22], which is the successor of Maekawa and Fukayama's model. Raffy *et al*. reported that they believe the properties of their films do not vary by much, despite their different stoichiometric compositions. However, for each stoichiometry, there were not more than ~3 films to allow a more complete examination of this claim with respect to the scaling we propose here. Moreover, the thickness of the grown films was not equally distributed among the collected data points, a fact that does not allow conclusive quantitative examination of the empirical law of Eq. 1. Yet, we also present the quantitative fitting parameters of this data set to Eq. 1. Furthermore, we present this fitting on a log-log scale similar to that of Fig. 4 in the main text. In this respect, it is interesting to note that the parameters for *A* and *B* are in agreement with the linearity in Fig. 5a.

We should mention here that one of the data points was found to behave differently than its sisters. This was noticeable for instance in its resistivity, which is more than 50% higher than the average resistivity of the other films. Hence, although this film is presented below, we did not include it for the quantitative fit.

**Table S13. Superconductivity in α-ReW (Raffy *et al.*** [34]**), ✱ in Fig. 4.** *d*, $T_c$, $R_s$, and the composition of α-ReW films extracted from Raffy *et al.* [34] (Fig. 1 therein). The error in data extraction is evaluated through the error in extracted $T_C$ values, which are also presented. Values calculated for *A*, *B*, $T_{c_RC}$, and Error $T_{c_RC}$% are also presented. Films that not all of their properties are known ($T_c$, *d*, and $R_s$) are highlighted in red. A film that seemed to be different than its sister films is highlighted in blue and was not used for the quantitative fitting.

| | | | | | | *A* | 14545 |
|---|---|---|---|---|---|---|---|
| | From Fig. 1 Top | | From Fig. 1 Bottom | | | *B* | 1.078 |
| Composition | *d* [nm] | $T_c$ [K] | $R_s$ [Ω/□] | $T_c$ [K] | Err $T_c$ % | $T_{c_RC}$ [K] | $T_{c_RC}$% |
| Re70W30 | 106.8296 | 6.811 | 14.828 | 6.732 | -1.15 | 7.44 | 9.24 |
| Re70W30 | 10.26937 | 5.3 | 158.483 | 5.3 | -0.02 | 6.02 | 13.57 |
| Re70W30 | 5.051 | 4.047 | 421.556 | 4.031 | -0.39 | 4.26 | 5.36 |
| Re 65W35 | 9.709 | 5.242 | 187.98 | 5.292 | 0.96 | 5.3 | 1.07 |
| Re 65W35 | 5.051 | 4.371 | 465.7 | 4.327 | -1.02 | 3.83 | -12.39 |
| Re60W40 | 11.691 | 5.343 | 141.357 | 5.383 | 0.76 | 5.98 | 11.96 |
| Re60W40 | 5.051 | 3.771 | 385.566 | 3.789 | 0.46 | 4.69 | 24.47 |
| Re55W45 | 106.974 | 6.368 | 18.7 | 6.256 | -1.75 | 5.79 | -9.14 |
| Re55W45 | 15.392 | 5.587 | 128.179 | 5.588 | 0.03 | 5.05 | -9.63 |
| Re55W45 | 7.092 | 4.77 | 277.806 | 4.722 | -1.01 | 4.76 | -0.22 |
| x Re50W50 | 9.174 | 4.619 | 239.005 | 4.601 | -0.38 | 4.33 | -6.31 |
| X Re50W50 | 4.525 | 3.487 | 545.099 | 3.499 | 0.34 | 3.61 | 3.44 |
| ▽ Re50W50 | 10.142 | 4.7 | 257.292 | 4.669 | -0.66 | 3.62 | -23.08 |
| Re 65W35 | 107.091 | 6.03 | 29.073 | 6.082 | | | |
| Re70W30 | | | 55.753 | 6.021 | | | |
| Re70W30 | | | 335.876 | 4.062 | | | |
| ▽ Re50W50 | 0.02 | 3.492 | | | | | |

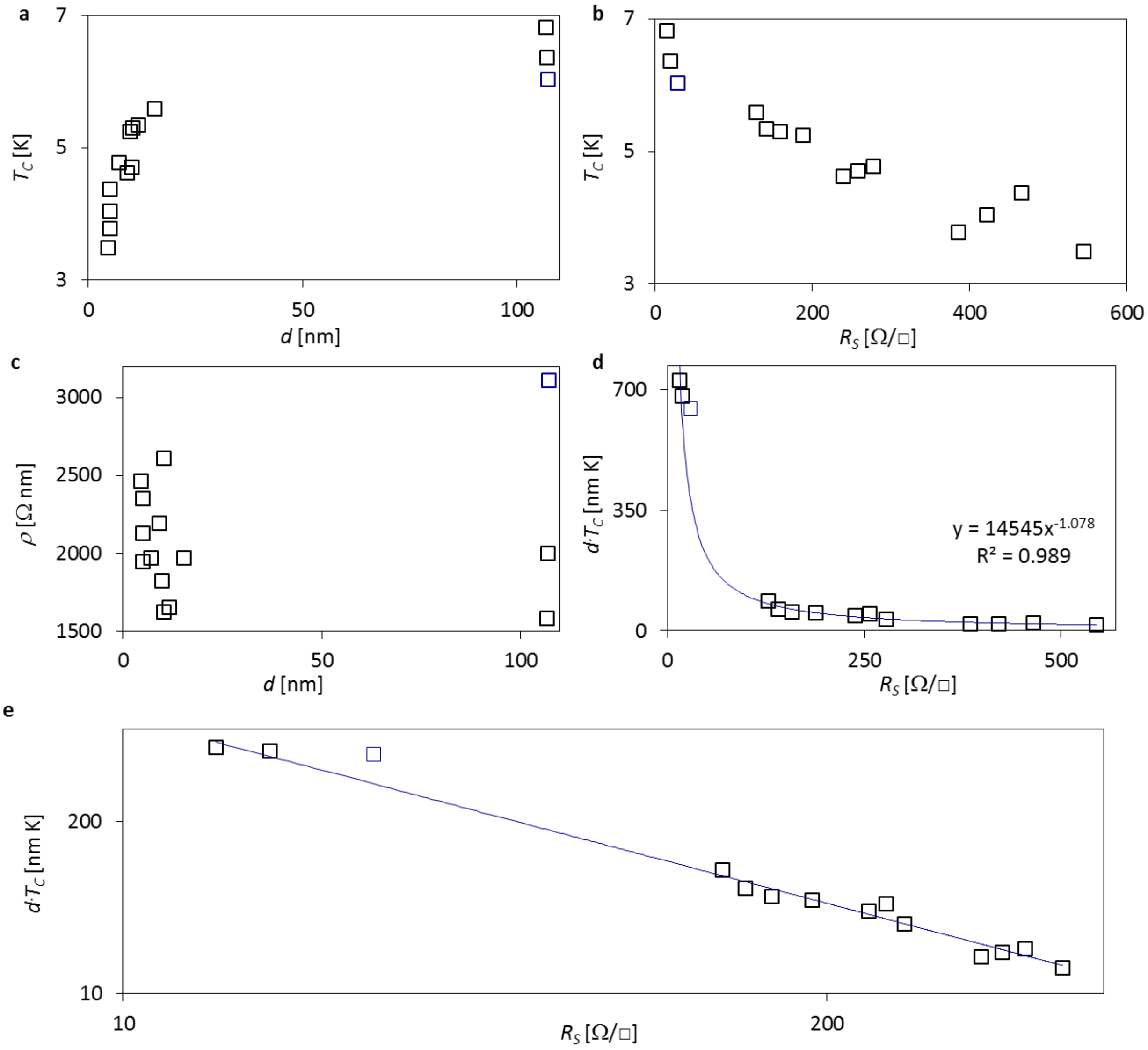

**Fig. S13. Superconductivity in α-ReW (Raffy *et al.* [34]), ✱ in Fig. 4.** Critical temperature as a function of (**a**) thickness and (**b**) sheet resistance. (**c**) Resistivity as a function of thickness. (**d**) $d{\cdot}T_c$ vs. $R_s$ on a linear scale with the best fit to Eq. 1 (blue curve). (**e**) $d{\cdot}T_c$ vs. $R_s$. on a log-log scale

with the best fit to Eq. 1 (blue curve). The point appears in blue in (a-e) was not included in the fitting, as mentioned above. $A$ = 14545, $B$ = 1.078.

## **14.4. Sn- extracted from Strongin *et al.*** [6].

In addition to the Al and Pb set of films that are discussed above, Strongin *et al*. reported thin superconducting films of Sn [6]. Specifically, they observed an increase in $T_c$ when films get thinner. This $T_C$ increase is followed by suppression of superconductivity for films thinner than ~27 nm. Unlike the cases of Al and the various data sets for Pb, the Sn films seem not to agree well with the scaling of Eq. 1. Here we present these data. We should mention that Goldman and co-authors also studied Sn films [14]. However, we could not find the $R_s$ values of these films, which prevented us from examining their agreement with Eq. 1. One possible explanation for this deviation is the role of proximity effect as discussed in Section 17.2.

**Table S14. Superconductivity in Sn films (Strongin *et al.*** [6]**).** $d$, $T_c$, $R_s$ , and the composition of Sn films extracted from Strongin *et al*. [6] (Fig. 2 therein). The error in data extraction is evaluated through the error in the extracted $d$ values, which are also presented. Films for which we could not find the complete data are highlighted in red.

| **From Fig. 2 Top** | | **From Fig. 2 Bottom** | | |
|---|---|---|---|---|
| $d$ [nm] | $T_C$ [K] | $d$ [nm] | $R_S$ [Ω/□] | *Err d%* |
| 16.215 | 4.2833 | 16.128 | 778.531 | -0.54 |
| 19.202 | 4.8223 | 19.538 | 555.91 | 1.75 |
| 22.577 | 5.2652 | 22.489 | 469.144 | -0.39 |
| 27.621 | 5.885 | 27.477 | 309.42 | -0.52 |
| 33.494 | 4.6815 | 33.171 | 194.123 | -0.96 |
| 38.475 | 4.5528 | | | |
| | | 9.9705 | 8773.079 | |

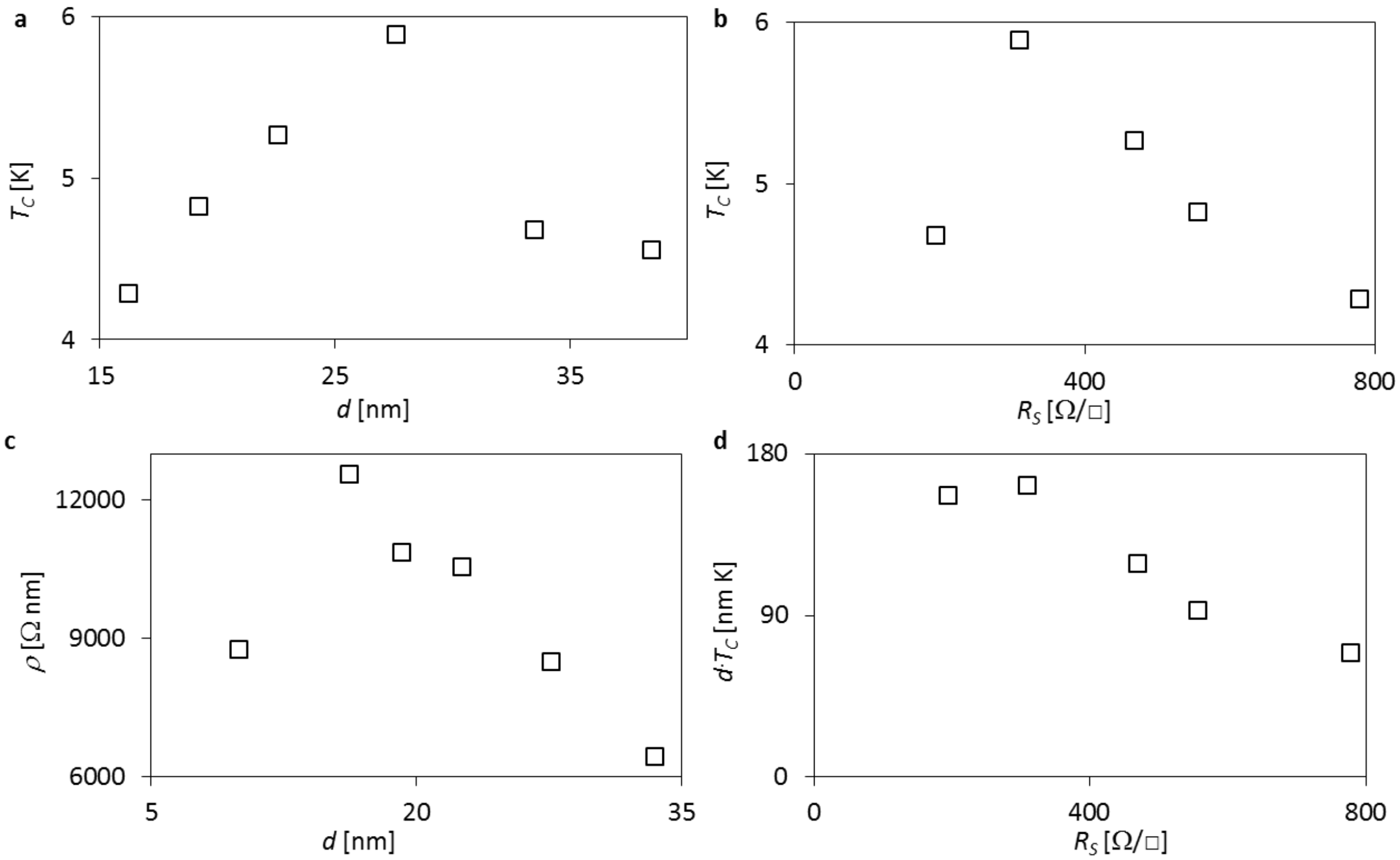

**Fig. S14. Superconductivity in Sn films (Strongin *et al*.** [6]**).** Critical temperature as a function of (**a**) thickness and (**b**) sheet resistance. (**c**) Resistivity as a function of thickness. (**d**) $d{\cdot}T_c$ vs. $R_s$.

## 15. TiN

Thin TiN films are considered highly disordered superconductors. In fact, it is believed that they can undergo a superconducting-to-superinsulating transition [36]. Being 'exotic' disordered superconductors, it is interesting to realize that TiN films grown by different groups agree with the found scaling. In particular, we demonstrated that the discovered scaling describes well the properties of films reported by Klapwijk and co-authors [37,38] and those reported by Baturina and co-authors [39].

Typically, TiN films are grown with the atomic layer deposition method (ALD). In this method, a precursor material is used. For the discussed TiN films, this material is Cl. Due to their chemical properties, the Cl atoms interact with the films and usually some Cl atoms remain in the film and cannot be removed. These Cl atoms give rise to Columbic and magnetic impurities in the system, which in turn affect the inhomogeneity of the films and influence the inherent disorder. The influence of the Cl atoms on the metallic and superconducting properties of the films are not fully know, and it is also not known whether these effects are thickness dependent. Sometimes, the Cl concentration can be measured, but even when it is the case, this concentration cannot be reduced or controlled de-facto significantly. Moreover, it is unknown whether the Cl atoms are distributed in the films homogeneously or not. Yet, the data analyzed below demonstrate that despite the potential effects of the Cl atoms, the sets of films grown under similar conditions follow the discovered scaling $d{\cdot}T_c(R_s)$

**15.1. TiN- extracted from (Klapwijk and co-authors** [37,38]**), ☆ in Fig. 4.**

Thin TiN films are considered as highly-disordered superconductors. As such, Driessen *et al*. reported a set of TiN films grown on SiO substrates, suggesting that their superconducting properties do not comply with the conventional theory [37]. This was followed by a growth of a set of thinner films reported by Coumou *et al*. [38]. In fact, Klapwijk and co-authors *et al*. suggested that the electric properties of these films in the normal state also deviate from conventional theory (*ibid*). Although the authors suggested a heuristic electrodynamic analysis, they claim that their observations are not yet understood and have to be clarified. Hence, we examined the properties of these highly resistive films (ρ = 120-380 μΩ cm) with Eq. 1. Surprisingly, we found that, similar to the other materials examined above, these highly disordered TiN films fit Eq. 1 with a very good agreement. Moreover, by merging the two sets of films we can suggest that they may not be very different with respect to their superconducting and metallic properties. We should note that we believe that one of these films is different in nature than the others. Alternatively, the reported thickness of this film might be thinner than the actual value.

This can be determined from, *e.g.*, the dependence of resistivity on thickness, as presented in Fig. S15c, which suggests that its thickness is ~9.5 nm instead of the nominal 6 nm. Alternatively, a value of $T_c$ higher than the one reported in the paper can also explain the deviation. Hence, for the quantitative analysis, we did not include this film. However, we do present its values quantitatively and graphically in Table S15 (designated in blue) and Fig. S15d (designated in red). Moreover, we can also mention that, given $d$ = ~9.5 nm, as suggested by the $\rho(d)$ curve, this film also agrees with the other data for the $d{\cdot}T_c$ vs. $R_s$ curve (Fig. S14d). We would like to update the reader that prior to publication, after having corresponding with Klapwijk and co-authors, we were informed that indeed, the film that we predicted to have values different than those reported in the literature were re-measured, and indeed, the value of $T_c$ was found to be higher than the value reported in Ref. [37]. Hence, this prediction signifies the usefulness of the model that can be used also to predict the superconducting behavior of thin films.

Lastly, we should mention that Driessen *et al.* also reported highly disordered NbTiN films as well as TiN on a different substrate. However, these sets included too few films to allow an examination of Eq. 1.

**Table S15.1. Superconductivity in highly-disordered TiN films (Klapeijk and co-authors** [37,38]**), ☆ in Fig. 4.** $d$, $T_c$, and $R_s$ of disordered TiN films extracted from Klapijk and co-authors [37,38] (Table 1 in Ref. [37] and Supplemental Material in [38]). A film with a nominal thickness smaller than that we believed is highlighted in blue.

| | | | All from [37,38] but red point | | From [37] without blue point | | From [37] | | All from [37,38] | |
|---|---|---|---|---|---|---|---|---|---|---|
| | | | **A** | **2784.7** | **A** | **2825.6** | **A** | **3889.6** | **A** | **2678.6** |
| | | | **B** | **0.811** | **B** | **0.817** | **B** | **0.906** | **B** | **0.812** |
| $d$ [nm] | $T_C$ [K] | $R_S$ [Ω/□] | $T_{C_RC}$ [K] | Err $T_{C_RC}$% | $T_{C_RC}$ [K] | Err $T_{C_RC}$% | $T_{C_RC}$ [K] | Err $T_{C_RC}$% | $T_{C_RC}$ [K] | Err $T_{C_RC}$% |
| 89 | 3.6 | 13.48315 | 3.79 | 5.12 | 3.79 | 5.4 | 4.14 | 14.98 | 3.64 | 1.12 |
| 45 | 3.2 | 41.55556 | 3.01 | -6.24 | 2.99 | -5.87 | 2.95 | -7.73 | 2.89 | -9.8 |
| 22 | 2.7 | 115 | 2.7 | -0.05 | 2.66 | -0.05 | 2.4 | -11.05 | 2.58 | -4.32 |
| 11 | 2.2 | 323.6364 | 2.33 | 5.66 | 2.29 | 6 | 1.88 | -14.5 | 2.23 | 1.37 |
| **6** | 1.5 | 633.3333 | 2.48 | 39.52 | 2.42 | 65.35 | 1.88 | 25.14 | 2.37 | 58.03 |

| | | | All from [37,38] but red point | | From [38] | | All from [37,38] | |
|---|---|---|---|---|---|---|---|---|
| | | | **A** | **2784.7** | **A** | **8787** | **A** | **2678.6** |
| | | | **B** | **0.811** | **B** | **0.957** | **B** | **0.812** |
| *d* [nm] | $T_C$ [K] | $R_S$ [Ω/□] | $T_{C_RC}$ [K] | Err $T_{C_RC}$% | $T_{C_RC}$ [K] | Err $T_{C_RC}$% | $T_{C_RC}$ [K] | Err $T_{C_RC}$% |
| 4 | 0.7 | 4300 | 0.79 | 11.05 | 0.73 | 4.58 | 0.75 | 7.24 |
| 4.3 | 0.78 | 3700 | 0.83 | 5.68 | 0.79 | 0.81 | 0.79 | 1.15 |
| 4.5 | 0.99 | 3000 | 0.94 | -5.69 | 0.92 | -7.23 | 0.89 | -9.71 |
| 4.8 | 1.3 | 2000 | 1.22 | -6.55 | 1.27 | -2.37 | 1.16 | -10.4 |
| 5 | 1.5 | 1500 | 1.48 | -1.41 | 1.6 | 6.97 | 1.41 | -5.84 |
| 5.5 | 1.6 | 1400 | 1.42 | -12.52 | 1.56 | -2.61 | 1.36 | -15.13 |

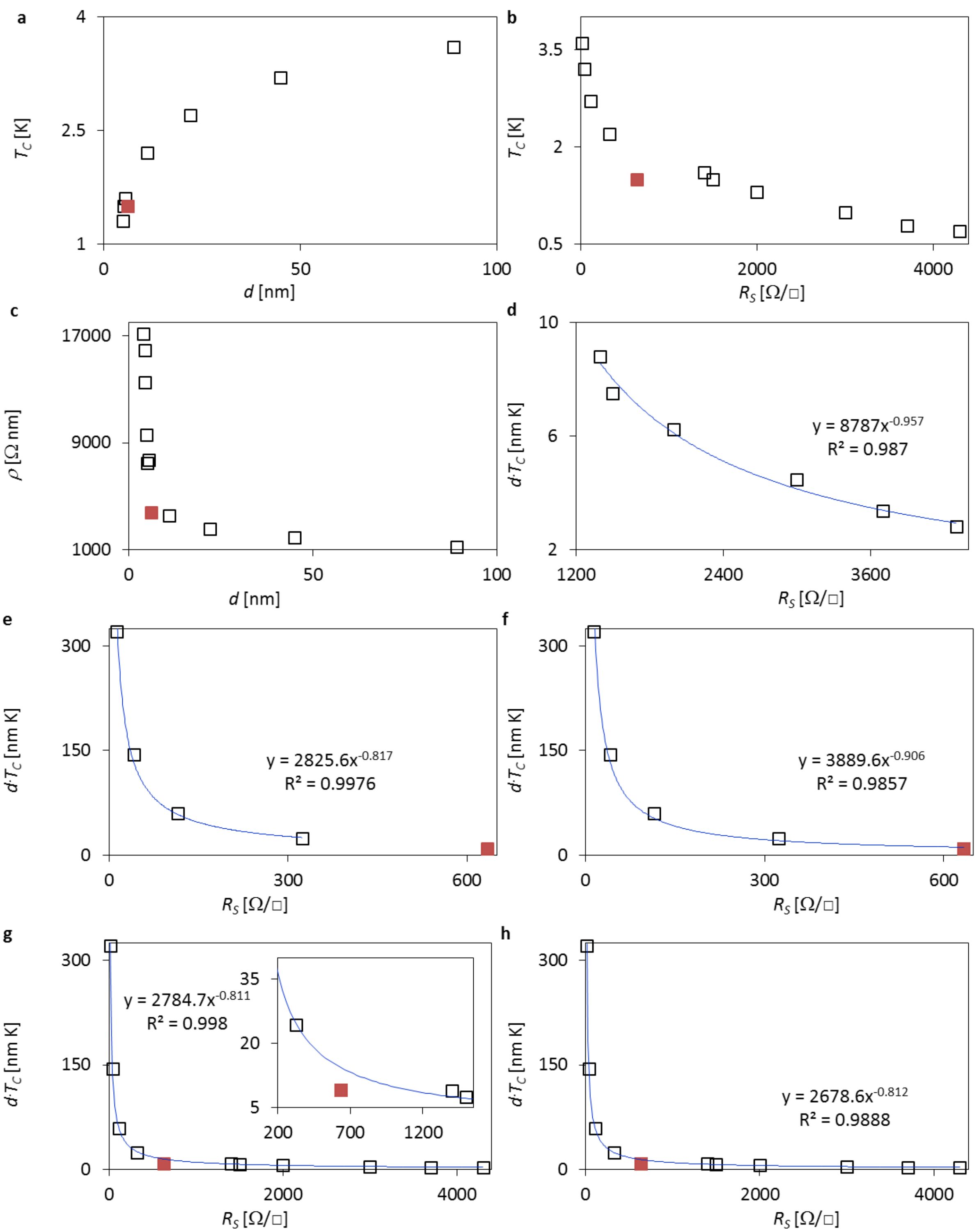

**Fig. S15.1 Superconductivity in in highly-disordered TiN films (Klapwijk and co-authors** [37,38]**), ☆ in Fig. 4.** Critical temperature as a function of (**a**) thickness and (**b**) sheet

resistance. (**c**) Resistivity as a function of thickness. (**d**) $d{\cdot}T_c$ vs. $R_s$ with the best fit of the films from Ref. [38] for Eq. 1. (**e**) Fitting the data from Ref. [37] to Eq. 1 excluding and (**f**) including the red data point that corresponds to the blue values in Table S14 (as discussed in the text). (**g**) Fitting all the data from both Ref. [37,38] to Eq. 1, while excluding the red data point. Inset is a closer look at the area around the red point, emphasizing that this film is different than the others. (**h**) Fitting the entire set of films from Ref. [37,38] to Eq. 1 (including the red data point).

**15.2 TiN- extracted from (Baturina and co-authors [37]), ☆ in Fig. 4.**

Baturina and co-authors have also reported on TiN films. These films were grown by atomic layer deposition (ALD). Since in this method traces of Cl atoms exist in the sample, variation in Cl concentration can derive changes in the homogeneity, disorder and other superconducting- and metallic –related properties. Baturina and co-authors measured the Cl concentration in their films and were able to form a set of films with a constant value of the Cl concentration. We should note that, even though the concentration is the same for the data set, the Cl might not be homogenously distributed in the film and the Cl atoms do expect to affect the measured properties of the films in a way that might, or might not, be thickness dependent. To measure the thickness, Baturina and co-authors imaged the films with transmission electron microscopy, allowing a direct measurement of the film thickness.

Similarly to the case of Al and Sn, the films from Baturina and co-authors seem to exhibit a small enhancement in $T_c$ in the thicker film regime. This increase in $T_c$ cannot be explained in the $T_c(R_s)$ or $T_c(d)$ graphs, but is consistent with the other data points in the $d{\cdot}T_c(R_s)$ scale.

We should note that although the data from Baturina and co-authors and the data from Klapijk and co-authors do not coincide in most other scaling, the grown films looks more similar when are compared on a $d{\cdot}T_c(R_s)$ graph.

**Table S15.2 Superconductivity in highly-disordered TiN films (Batrina and co-authors** [36]**),**

★ **in Fig. 4.** $d$, $T_c$, and $R_s$ of disordered TiN films extracted from Baturina and co-authors [36].

| | | | *A* | 1714.7 |
|---|---|---|---|---|
| | | | *B* | 0.747 |
| *d* [nm] | $T_C$ [K] | $R_S$ [Ω/□] | $T_{C_RC}$ [K] | Err $T_{C_RC}$% |
| 23 | 3.14 | 65 | 3.3 | 4.78 |
| 18 | 3.315 | 90 | 3.3 | -0.32 |
| 12 | 3.25 | 165 | 3.15 | -3.12 |
| 10 | 3.18 | 216 | 3.09 | -2.82 |
| 7 | 3 | 334 | 3.19 | 5.97 |
| 5 | 2.538 | 855 | 2.21 | -14.67 |
| 3.6 | 1.26 | 2520 | 1.37 | 8.1 |

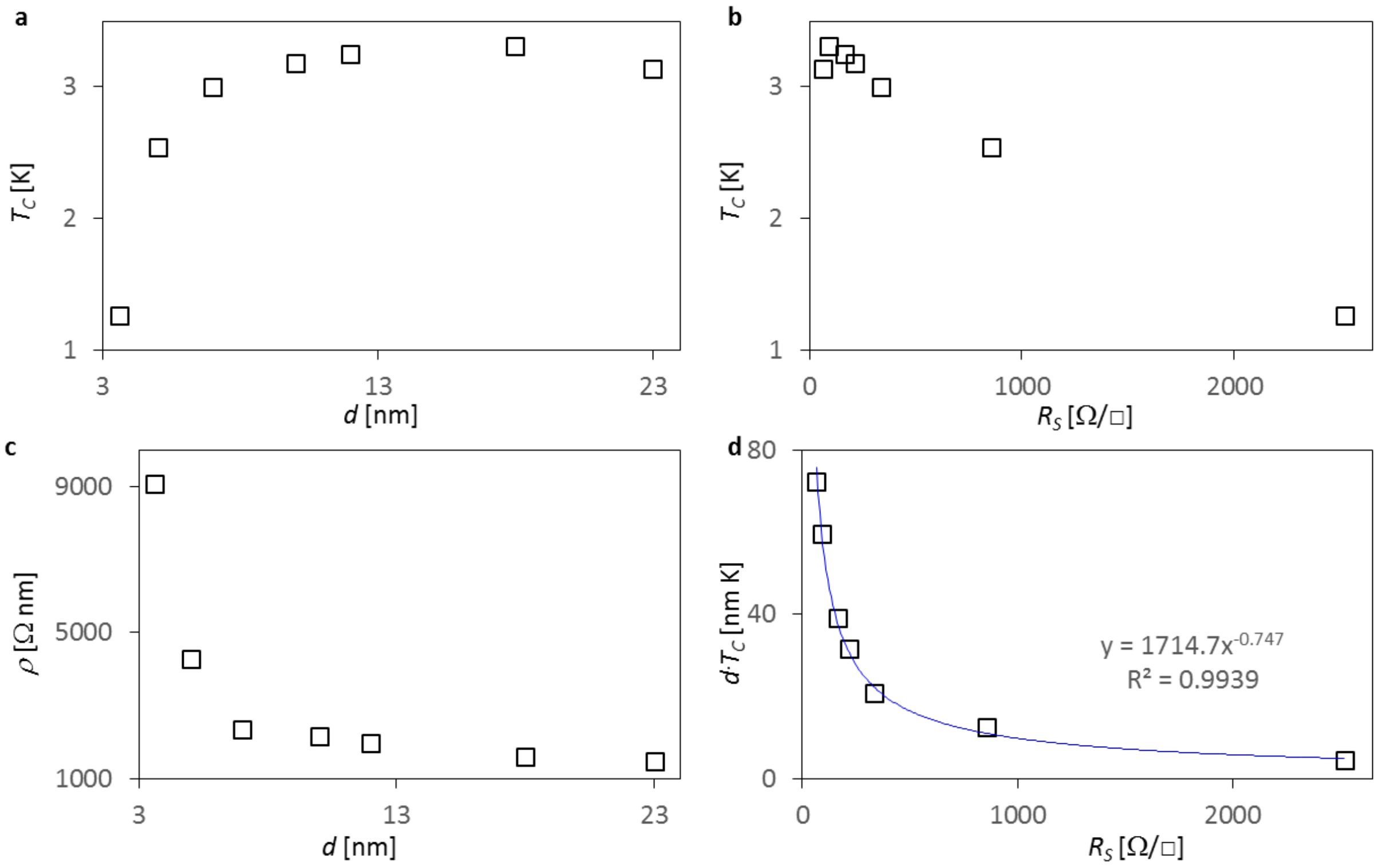

**Fig. S16. Superconductivity in highly-disordered TiN films (Baturina and co-authors** [37]**),**

★ **in Fig. 4.** Critical temperature as a function of (**a**) thickness and (**b**) sheet resistance. (**c**) Resistivity as a function of thickness. (**d**) $d{\cdot}T_c$ vs. $R_s$ with the best fit for Eq. 1.

**16. $V_3Si$- extracted from Orlando *et al.*** [26]**.**

In addition to $Nb_3Sn$ films, Orlando *et al.* also studied superconductivity in highly damaged or highly defected $V_3Sn$ films. The suggestion of the authors that "since the samples were deposited in a 'compositional phase spread' configuration, the unpatterned samples vary to some degree in composition across the films" [26] also addresses $V_3Sn$. In this set of experiments, one film was 1 μm thick, while all the others had $d$ = 300 nm (*i.e.*, small thickness distribution with no 'true' 2D films), while their $R_s$, $T_c$ and ρ values varied (the latter had up to 700% difference in range). Therefore, similar to the $Nb_3Sn$ case, this is one of the only examples where our scaling $d{\cdot}T_c$ vs. $R_s$ does not seem to work. Lastly, these films also were examined over the course of two calendar years, which may have allowed their degradation. Hence, the fact that these films do not agree with the proposed scaling does not necessarily invalidate it. We report the data for these films below.

**Table S17. Superconductivity in $V_3Si$ (Orlando *et al.*** [26]**).** $d$, $T_c$, and $R_s$ of $V_3Si$ films extracted from Orlando *et al.* [26] (Table 1 therein).

| $T_C$ [K] | $R_S$ [Ω/□] | $d$ [nm] |
|---|---|---|
| 16.4 | 0.052 | 1000 |
| 16.1 | 0.24 | 300 |
| 15.7 | 0.503 | 300 |
| 15.6 | 0.203 | 300 |
| 14.8 | 1.04 | 300 |
| 14.3 | 1.263 | 300 |
| 13.9 | 1.317 | 300 |
| 13.8 | 1.42 | 300 |
| 13.6 | 1.44 | 300 |
| 12.9 | 1.42 | 300 |

## 17. Presentation of all unprocessed data.

Above, we discussed each of the data sets for thin superconducting films individually. In particular, we demonstrated how almost all of these data sets agree with the scaling $d{\cdot}T_c(R_s)$ as well as with the power law of Eq. 1. Moreover, whenever applies, we discussed possible sources of error in the data extracted from the literature as well in the film characterization. Hence, we would like to discuss again the entire data as a whole. Specifically, we would like to elaborate on two issues. First, we would like to elaborate on the fact that based on Fig. 5a, the two fitting parameters (*A* and *B*) described in Eq. 1 may be correlated, simplifying the power law. Second, despite the nominated possible sources for errors in the values analyzed in the Supplemental Material, we would like to present the complete, unprocessed data together on one graph. This may supply the reader with the possibility to qualitative estimate the upper limit error of the Eq. 1.

### 17.1. Possible correlation between *A* and *B*.

As mentioned in the main text, the fact that the data presented in Fig. 5a demonstrates a linear trend for more than five orders of magnitude suggests that the parameters *A* and *B* (Eq. 1) are correlated. Linearity on such a log-normal scale indicates an exponential relation between *A* and *B*. We present again the relation between *A* and *B* for the surveyed materials, while we added a linear line (red) to guide the eyes and demonstrate the linearity (Fig. S17.1), suggesting that:

$$\mathrm{Log}(A) = \alpha' + \beta' B \qquad \textbf{(Eq. S1)}$$

where $\alpha'$ and $\beta'$ correspond to the intercept and slope of the red line in Fig. S17.1, respectively.

The exponential dependence of *A* on *B* can be substituted in Eq. 1, while it is more appealing to define $\alpha = 10^{\alpha'}$ and $\beta = e^{\beta' \ln(10)}$ and to substitute Eq. S1 in scaling law as appears in Eq. 2a:

$$T_{\mathrm{c}} = \frac{\alpha}{d} \cdot e^{-B\left(\ln\left(R_{\mathrm{S}}/\beta\right)\right)} \qquad \textbf{(Eq. S2a)}$$

In this way, the only parameter that represents a certain set of films is the parameter $B$, while α and β are universal constants. In particular, the parameter β represents a universal constant for a resistance value.

In fact, the red line in Fig. S17.1 used to guide the eye is also the best fit calculated for the possible exponential dependence of $A$ on $B$, with α' = 1.14 and β' = 2.67 and with the corresponding standard errors of 0.27 and 0.26. That is, Eq. S2a becomes:

$$T_{\mathrm{c}} = \frac{13.7}{d} \cdot e^{-B\left(\ln\left(R_{\mathrm{S}}/464\right)\right)} \qquad \textbf{(Eq. S2b)}$$

where we remind the readers that we use $T_{\mathrm{c}}$, $d$ and $R_{\mathrm{S}}$ in K, nm and Ω/□.

The value β = 464 Ω/□ is very different from the quantum resistance $\hbar/4e$ = 6.45 kΩ. However, this difference is not surprising when bearing in mind that the quantum resistance is not a universal value and that different models use different constants for the sheet resistance [22]. Fig. S17.1 shows that many of the material sets are crowded around $B$ = 1. Although the scatter around the Eq. S2b is smaller at the extreme points along the curve, and larger around $B$ = 1, it worth mentioning that for the case of $B$ equals to unity, Eq. S2b becomes: $T_{\mathrm{c}} = e^{-\ln(R_{\mathrm{S}}/6355)}/d$, while the 6355 Ω/□ in the exponent is when we used again $T_{\mathrm{c}}$, $d$ and $R_{\mathrm{S}}$ in K, nm and Ω/□. We should remind the reader that an examination of the validity of Eq. S2b and a comparison of the fitting made with this equation to Eq. 1 are discussed in Section 7.1 for the case of α-MoGe.

One may identify that one material is rather far from the linear fit (molybdenum, designated by △). Hence, although Fig. S17.1 demonstrates on exponential trend in the dependence of $A$ on $B$ and although the data fit such an exponent quantitatively with a reasonable agreement (Eq. S2b), there is still uncertainty with regard to the limits of the framework of Eq. S2b. Yet, examining the materials at the two extreme points of the linear curve (αMoGe and Al) suggests that the position of the different materials along the curve is determined by its disorder and homogeneity (with respect to stoichiometry, granularity etc). We should note that if we discard the data for Mo, α' and β' almost do not change ($\alpha'_{_noMo}$ = 1.23 and $\beta'_{_noMo}$ = 2.64), the standard errors are reduced significantly from 0.26 and 0.26 to 0.14 and 0.14, respectively.

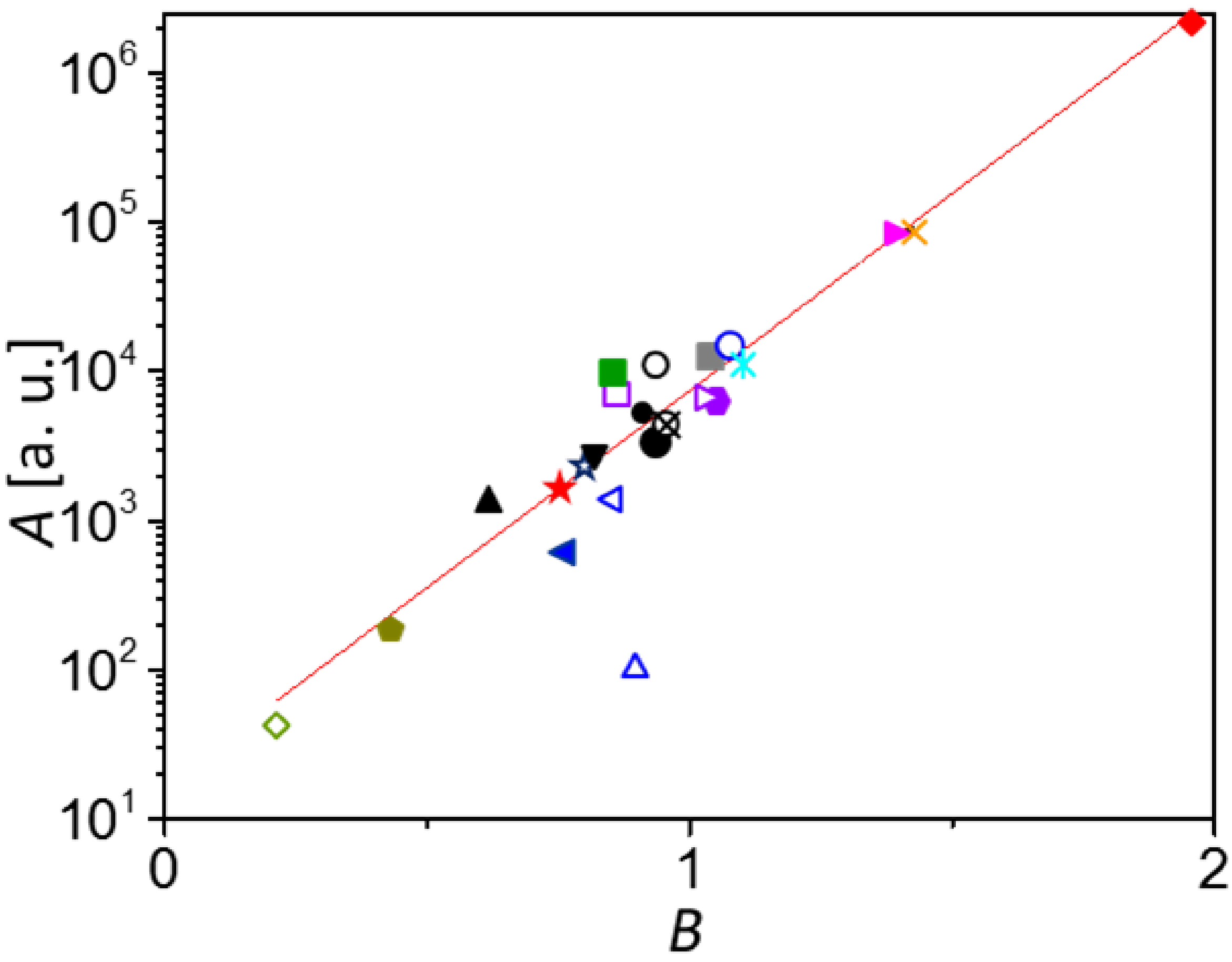

**Fig. S17.1. Intercept versus slope (*A* vs. *B*) of the best fits for the different data sets to Eq. 1 (reproduced from Fig. 5a).** The values for *A* vs. *B* for the different sets of materials as were calculated in 1-16 on log-normal axes. The red line is the best fit to Eq. S1, suggests that the parameters *A* and *B* are correlated with α' = 1.14 and β' = 2.67 and with the corresponding standard errors of 0.26 and 0.26. The symbols used here are similar to those specified in Fig. 4, including the Mo (designated by △) sample that unlike the other

materials, seems to deviate from the linear trend. Without the Mo films, the parameters become: $\alpha'_{_noMo}$ = 1.23 and $\beta'_{_noMo}$ = 2.64, with the standard error of 0.14 for both.

An additional explanation for the dependence of *A* on *B* could be that although the fit to a power law is rather accurate, a different expression is hidden in Eq. 1.In particular, when dealing with universal relations, often, a power law dependence with correlated *A* and *B* may imply that Eq. 1 can be rewritten as: $d{\cdot}T_c = A'{\cdot}n/(R_s{\cdot}\ln^n(R_s))$, where *A*' is a global parameter and $n \ll 1$ is an exponent specific to each material. Such an explanation would be valid only in the limit $B \sim 1$. However, although most of the materials do have indeed an exponent around unity, the scatter of the graph, and hence the deviation from the linear fit in Fig. S17.1 are rather large. Moreover, this cannot explain the materials with $|B\text{-}1| \gg 0$. This means that the merit of such logarithmic approximation is low either due to the scatter of the data, where the mathematical approximation is valid, or due to the invalidity of the approximation, where the data fits well Eq. S2.

**17.2. All unprocessed data – upper limit for the error in the scaling.**

Above, we analyzed superconductivity in each of the materials individually, demonstrating mostly agreement with Eq. 1, while discussing briefly the superconducting characteristic of the specific films. As a part of this analysis, we sometimes had to neglect some of the films due to uncertainty in the values we presented in the above tables. Here we would like to present all the data, as is, without processing it (despite some cases of large errors in the data extraction process). One can see that the linearity of $d{\cdot}T_c$ vs. $R_s$ on a log-log scale (Fig. S17.2) is still convincing. Yet, at the bottom right side of the graph, some of the data sets curve down faster than a power law. A closer look at these materials reveals that this rapid decrease occurs mainly in Al, Bi and Pb films, all were grown by Goldman and co-authors [7,12,14], as well as some of the samples grown by Strongin *et al*. [6]. The immediate direct relation between these films is the fact that they all were grown on conducting substrates. Hence, this may have led to the fact that the superconductivity in these films was affected by the proximity effect, so that Cooper pairs from the superconductor were freely hybridized with electrons from the conductive substrate.

Although Eq. 1 does seem to describe well also sets of films in which the proximity effect is the governing mechanism for the change in $T_c$, *e.g.,* in the case of the discussed Nb films [24], it may be that here, there are two dominant mechanisms that govern $T_c$. That is, for thicker films, the proximity effect may be negligible, and therefore only one mechanism dominates $T_c$, resulting in an agreement of the data with Eq. 1. On the other hand, for the thinner films, the proximity effect may become gradually more significant, competing with the mechanism that is dominant in the thicker films. Such dual dominancy of two competing mechanisms may lead to complex behavior that deviates from the conventional simple form. Indeed, Fig. S17.2 demonstrates that the linearity of the data for these films (on log-log axes) is valid for most of the scale, while $d \cdot T_c$ decays faster than linear beginning from a certain thickness value, which is material dependent. This may suggest that an additional mechanism that changes $T_c$ is introduced at the thinner films. In fact, the Goldman and co-authors indeed reported that for the thinnest films, they suspect that the Ge substrate upon which they grow their films [7,12,14] proximitizes the deposited superconducting films. They used this explanation to support their observation of finite $T_c$ for films measured to be thinner than a single atomic unit cell.

Despite the current discussion, we cannot eliminate the other two potential explanations for the deviation of the thinner films from Eq. 1 in these sets of data: (a) there is a consistent measurement error (this explanation complies also with the observation of finite $T_c$ for films thinner than the unit cell); and (b) Eq. 1 has an unknown limit and is not valid below a certain thickness (such limitation of Eq. 1 cannot explain the existence of superconductivity in films thinner than a single unit cell).

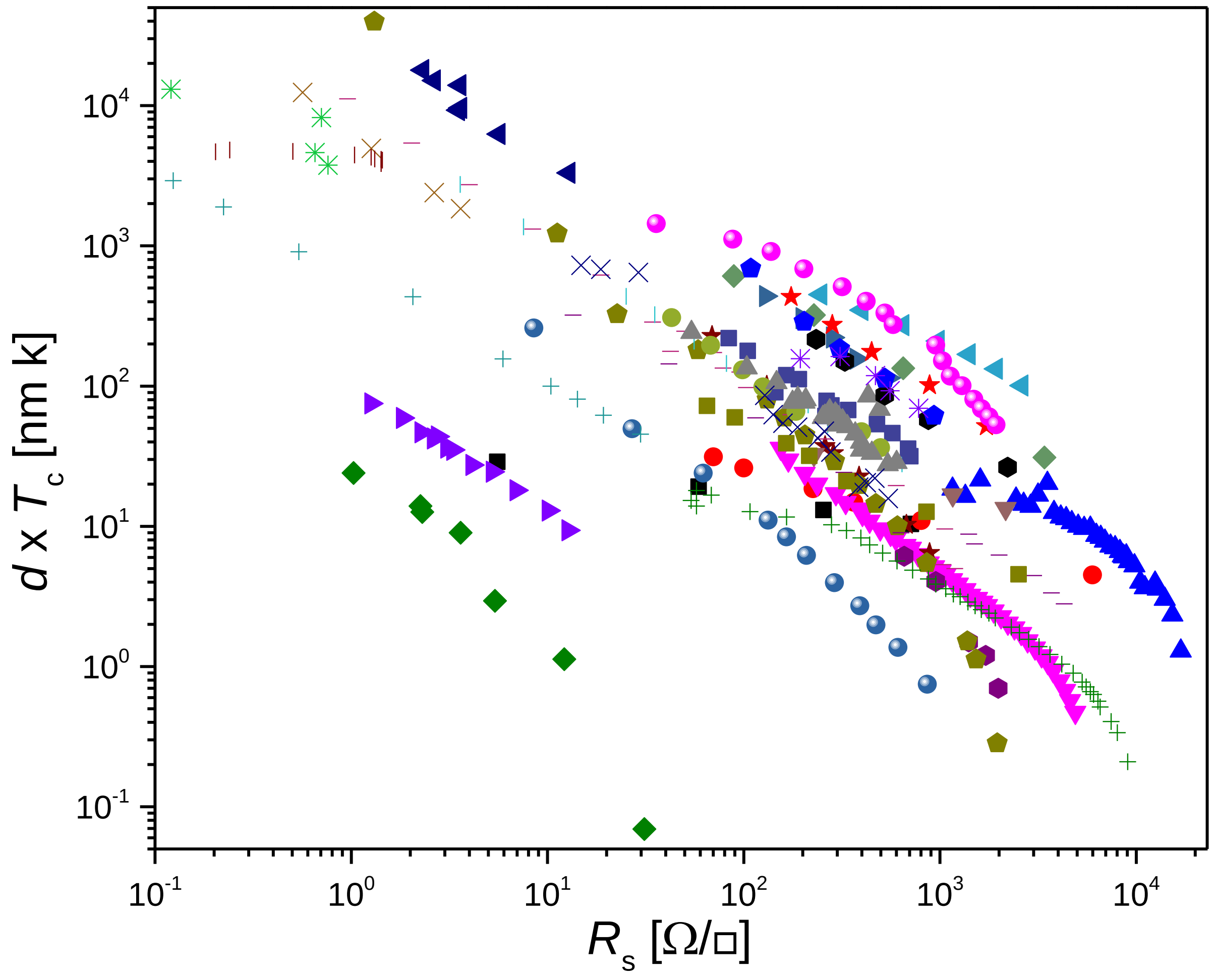

**Fig. S17.2. $d{\cdot}T_c$ vs. $R_s$ for all unprocessed data presented and discussed in the paper.** A collection of all the data presented and discussed above, regardless of the level of certainty in the values, and despite the potential sources of errors that are discussed in the introduction to the Supplemental Material. Yet, universality of Eq. 1 can be realized even here, in comparison, for example, to Fig. 5c-e. A concave down decrease of some of the data sets is likely to be due to measurement error and underestimation of the thickness values or the proximization with conducting substrate upon which the superconducting films was deposited. In this plot, we used the following symbols: ■ Al from Cohen and Abeles [5]; ● Al from in Strongin *et al*. [6]; ▲ Al

from Haviland, Liu and Goldman [12]; ▼ Bi [12]; ◆ $CoSi_2$ [13]; ◀ $MgB_2$ [15]; ▶ Mo [16]; ⬢ αMoGe by Yazdani and Kapitulnik [21]; ★ αMoGe from Graybeal and co-authors [18–20]; ⬟ αMoGe from Graybeal and Beasley [40]; + Nb [24]; × α$Nb_3$Ge [25]; ✳ $Nb_3Sn$ [26]; — NbN from Wang and co-authors [27,29]; | NbN from Miki, Wang and co-authors [28]; ■ NbN by Semenov *et al.* [30]; ● NbN by Kang *et al.* [31]; ▼ NbN by Belacour *et al.* [32]; ▲ our NbN films; ◆, ◀, ▶, ⬢, ★, ⬟ and ● are Pb films by Strongin *et al*. that correspond to triangles facing down, triangles facing up, circles with an 'x', large circles, small circles, empty circles and empty triangles facing up in [6]; + Pb by Haviland, Liu and Goldman [12]; × αReW [34]; ✳ Sn [6]; — disordered TiN by Klapijk and co-authors [37,38], | $V_3Si$ [26], and ■ are disordered TiN by Baturina and co-authors [37].

**References**

[1] D. I. version 1.6, Http://datathief.org B. Tummers, (2006).

[2] C. Strunk, C. Sürgers, U. Paschen, and H. Löhneysen, Phys. Rev. B **49**, 4053 (1994).

[3] W. Siemons, M. Steiner, G. Koster, D. Blank, M. Beasley, and A. Kapitulnik, Phys. Rev. B **77**, 174506 (2008).

[4] B. Matthias, T. Geballe, and V. Compton, Rev. Mod. Phys. **35**, 1 (1963).

[5] R. Cohen and B. Abeles, Phys. Rev. **109**, 444 (1968).

[6] M. Strongin, R. Thompson, O. Kammerer, and J. Crow, Phys. Rev. B **1**, 1078 (1970).

[7] Y. Liu, D. B. Haviland, B. Nease, and A. M. Goldman, Phys. Rev. B **47**, 5931 (1993).

[8] I. L. Landau, D. L. Shapovalov, and I. A. Parshin, JETP Lett. **53**, 263 (1991).

[9] R. P. Silverman, Phys. Rev. B **16**, 2066 (1977).

[10] R. P. Silverman, Phys. Rev. B **19**, 233 (1979).

[11] D. G. Naugle, R. E. Glover III, and W. Moorman, Physica **32**, 250 (1971).

[12] D. B. Haviland, Y. Liu, and A. M. Goldman, Phys. Rev. Lett. **62**, 2180 (1989).

[13] P. A. Badoz, A. Briggs, E. Rosencher, F. A. D'Avitaya, and C. D'Anterroches, Appl. Phys. Lett. **51**, 169 (1987).

[14] H. M. Jaeger, D. B. Haviland, B. G. Orr, and A. M. Goldman, Phys. Rev. B **40**, 182 (1989).

[15] a. V. Pogrebnyakov, J. M. Redwing, J. E. Jones, X. X. Xi, S. Y. Xu, Q. Li, V. Vaithyanathan, and D. G. Schlom, Appl. Phys. Lett. **82**, 4319 (2003).

[16] L. Fàbrega, a Camón, I. Fernández-Martínez, J. Sesé, M. Parra-Borderías, O. Gil, R. González-Arrabal, J. L. Costa-Krämer, and F. Briones, Supercond. Sci. Technol. **24**, 075014 (2011).

[17] J. M. Graybeal and M. R. Beasley, Phys. Rev. B **29**, 4167 (1984).

[18] H. Tashiro, J. Graybeal, D. Tanner, E. Nicol, J. Carbotte, and G. Carr, Phys. Rev. B **78**, 014509 (2008).

[19] S. Turneaure, T. Lemberger, and J. Graybeal, Phys. Rev. B **63**, 174505 (2001).

[20] S. Turneaure, T. Lemberger, and J. Graybeal, Phys. Rev. B **64**, 179901 (2001).

[21] A. Yazdani and A. Kapitulnik, Phys. Rev. Lett. **74**, 3037 (2000).

[22] A. M. Finkel'stein, Phys. B Condens. Matter **197**, 636 (1994).

[23] J. M. Graybeal, Phys. B+C **135**, 113 (1985).

[24] A. Gubin, K. Il'in, S. Vitusevich, M. Siegel, and N. Klein, Phys. Rev. B **72**, 064503 (2005).

[25] H. Kes and C. C. Tsuei, Phys. Rev. B **28**, 5126 (1983).

[26] T. Orlando, E. McNiff, S. Foner, and M. Beasley, Phys. Rev. B **19**, 4545 (1979).

[27] Z. Wang, A. Kawakami, Y. Uzawa, and B. Komiyama, J. Appl. Phys. **79**, 7837 (1996).

[28] S. Miki, Y. Uzawa, A. Kawakami, and Z. Wang, Electron. Commun. Japan (Part II Electronincs) **85**, 77 (2002).

[29] S. Ezaki, K. Makise, B. Shinozaki, T. Odo, T. Asano, H. Terai, T. Yamashita, S. Miki, and Z. Wang, J. Phys. Condens. Matter **24**, 475702 (2012).

[30] a. Semenov, B. Günther, U. Böttger, H.-W. Hübers, H. Bartolf, a. Engel, a. Schilling, K. Ilin, M. Siegel, R. Schneider, D. Gerthsen, and N. Gippius, Phys. Rev. B **80**, 054510 (2009).

[31] L. Kang, B. B. Jin, X. Y. Liu, X. Q. Jia, J. Chen, Z. M. Ji, W. W. Xu, P. H. Wu, S. B. Mi, a. Pimenov, Y. J. Wu, and B. G. Wang, J. Appl. Phys. **109**, 033908 (2011).

[32] C. Delacour, L. Ortega, M. Faucher, T. Crozes, T. Fournier, B. Pannetier, and V. Bouchiat, Phys. Rev. B **83**, 144504 (2011).

[33] B. H. Müller, T. Schmidt, and M. Henzler, Surf. Sci. **376**, 123 (1997).

[34] H. Raffy, R. Laibowitz, P. Chaudhari, and S. Maekawa, Phys. Rev. B **28**, 6607 (1983).

[35] S. Maekawa and H. Fukuyama, J. Phys. Soc. Japan **51**, 1380 (1982).

[36] V. M. Vinokur, T. I. Baturina, M. V Fistul, A. Y. Mironov, M. R. Baklanov, and C. Strunk, Nature **452**, 613 (2008).

[37] E. F. C. Driessen, P. C. J. J. Coumou, R. R. Tromp, P. J. de Visser, and T. M. Klapwijk, Phys. Rev. Lett. **109**, 107003 (2012).

[38] P. C. J. J. Coumou, E. F. C. Driessen, J. Bueno, C. Chapelier, and T. M. Klapwijk, Phys. Rev. B **88**, 180505(R) (2013).

[39] T. I. Baturina and S. V. Postolova, in *Int. Work. Strongly Disord. Supercond. Supercond. Insul. Transit.* (2104), Villard de Lans, France.

[40] J. M. Graybeal, Phys. B+C **135**, 113 (1985).